\documentstyle[aps,prl,amssymb,eqsecnum,epsf]{revtex}

\def\mathbold{\bf}
\def\be{\begin{equation}}
\def\ee{\end{equation}}
\def\bea{\begin{eqnarray}}
\def\eea{\end{eqnarray}}
\def\br{{\mathbold r}}

\def\bq{{\mathbold q}}
\def\ba{{\mathbold a}}
\def\bA{{\mathbold A}}
\def\eps{\epsilon}
\def\Q{{\cal Q}}
\def\W{{\cal W}}
\def\U{{\cal U}}
\def\J{{\cal J}}
\def\C{{\cal C}}
\def\Z{Z}
\def\ZBA{Z}
\newcommand{\corr}[1]{\langle #1\rangle}
\newcommand{\Tr}{\mathop{\rm Tr}}
\newcommand{\tr}{\mathop{\rm tr}}

\renewcommand{\Re}{\mathop{\rm Re}}
\renewcommand{\Im}{\mathop{\rm Im}}
\newsavebox{\sigmaa}

\begin{document}

\title{Keldysh action for disordered superconductors}

\author{M. V. Feigel'man$^1$, A. I. Larkin$^{1,2}$, M. A. Skvortsov$^1$}

\address{
$^1$L. D. Landau Institute for Theoretical Physics, Moscow
117940, RUSSIA \\
$^2$Theoretical Physics Institute, University of Minnesota,
Minneapolis, MN 55455, USA
}

\date{\today}

\maketitle

\begin{abstract}
Keldysh representation
of the functional integral for the interacting electron system with
disorder is used to derive microscopically an
effective action for dirty superconductors.
In the most general case this action is a functional
of the $8\times 8$ matrix $Q(t,t')$ which depends on two time variables,
and on the fluctuating order parameter field and electric potential.
We show that this approach reproduces, without the use of the replica trick,
the well-known result for the Coulomb-induced renormalization of the
electron-electron coupling constant in the Cooper channel.
Turning to the new results,
we calculate the effects of the Coulomb interaction upon: i)
the subgap Andreev conductance between superconductor and 2D dirty
normal metal, and ii) the Josephson proximity coupling between superconductive
islands via such a metal.  These quantities are shown to be strongly
suppressed by the Coulomb interaction at sufficiently low temperatures
due to {\it both} zero-bias anomaly in the density of states
and disorder-enhanced repulsion in the Cooper channel.
\end{abstract}


\tableofcontents

\section{Introduction}

Electron transport in hybrid superconductive-normal systems at low
temperatures is governed by the Andreev reflection.
Both finite-voltage conductance $G_A$
between superconductive and normal electrodes, and Josephson critical
current $I_c$ between two superconductive banks, separated by normal
region,  are determined by the Cooper pair propagation in the normal metal.
The theory of the Andreev conductance (without Coulomb effects) was developed
in, e.~g., \cite{volkov,Nazarov94,beenakker}, whereas the Josephson coupling
was calculated (with the account of short-range electron interaction)
in~\cite{ALO}. When normal conducting region
is made of a dirty metal film, or two-dimensional electron gas with low
density of electrons, Coulomb interaction in the normal region may lead
to strong quantum fluctuations which suppress both the Andreev conductance
and the Josephson proximity effect. Several different kinds of quantum effects
are known to be relevant in low-dimensional conductors at low temperatures.
Quantum corrections to the conductivity of two-dimensional dirty conductors
grow logarithmically as temperature $T$ decreases and
become large at $\ln\frac{1}{T\tau} \sim g$
(where $g=(\hbar/e^2)\sigma$ is the dimensionless conductance, $\sigma$
is the conductance per square, and $\tau$ is the elastic scattering time).
There are two main types of these effects:
weak localization corrections~\cite{band4,GLK},
and interaction-induced  corrections~\cite{AA}.
Other important quantum effects include interaction-induced
suppression of the tunneling conductance (``zero-bias anomaly"~\cite{AA,AAL}),
and disorder-induced
suppression~\cite{fuku,kuroda,finkel1,finkel2,finkel3}
of superconductive transition temperature $T_c$.
These effects are of the relative order of $g^{-1}\ln^2\frac{1}{T\tau}$,
i.~e.\ much stronger than the weak localization and interaction effects.
Earlier studies of all those effects~\cite{GLK,AA,fuku,kuroda}
employed perturbative diagram technique
based on the expansion over semiclassical parameter
$(E_F\tau)^{-1} \sim g^{-1}$. This method has an obvious drawback: the number
and complexity of diagrams grows fast with the order of perturbation
theory, which makes its use very difficult
in the lowest-temperature region where effects are strong.
Some combination of the perturbative diagram technique and functional methods
was used in earlier paper~\cite{ovchina}, where the first attempt to
calculate the effect of long-range Coulomb interaction upon $T_c$ was made.
More advanced functional methods
were then developed in the weak localization theory~\cite{ELK,Efetov},
which made it possible to average the fermionic functional integral
over disorder, and reduce it to an
``effective" form which contains slowly varying diffusive modes only.
Those approaches have used either the replica trick~\cite{ELK} or the
method of supersymmetry~\cite{Efetov}. Whereas the supersymmetry method
was found to be very powerful and convenient for the study of
single-electron effects, it cannot be used in the cases where
quantum corrections due to electron-electron interactions are important.
The replica method was generalized for the interacting systems by Finkelstein
(cf.~\cite{finkel2} for the review); in particular, he has shown that
in dirty films the
superconductive $T_c$ vanishes at $g \sim \ln^2\frac{1}{T_{c0}\tau}$
(here $T_{c0}$ is the bare (BCS) transition temperature).
The drawback of the replica method is that it contains an unphysical
procedure of analytic continuation over the number of replicas
$n\to 0$, and, also, it is difficult to use it for the study
of non-equilibrium phenomena.

Long time ago Keldysh~\cite{Keldysh} proposed an approach which allows
to treat kinetic phenomena in metals with the use, both, of the Green function
technique, and of the kinetic equation for the distribution function.
This approach was found to be especially fruitful in the theory
of superconductivity, where dynamic equations for the Green functions
were derived in the dirty limit~\cite{LO1} (cf.\ also review articles
\cite{rammer,lambert}). In the static limit, these
equations coincide with the Usadel equations~\cite{Usadel}.
Keldysh approach often was found to be the most simple and transparent, even
for the treatment of linear-response problems, since it does not involve
tedious analytic continuation procedures.
It is also the only known method for the treatment of
nonlinear and/or non-equilibrium problems. In some cases nonlinearities
with respect to both external and {\it fluctuating} fields are important,
so one needs either to sum up very large number of diagrams, or to
develop some effective action formalism within the Keldysh approach.
Such an approach was recently developed, for normal metals, by Kamenev
and Andreev~\cite{Kam_Andr}. In many respects we will follow this seminal
paper in our analysis. A similar approach was also recently developed
in~\cite{Ludwig}, where Finkelstein's renormalization group equations for
dirty metals were rederived for the case of short-range electron-electron
interaction. For the earlier approaches to develop
functional integral methods in the Keldysh representation
see~\cite{aron_ios,chou,babich,Horbach}.

In the present paper we will develop the Keldysh functional approach
for dirty superconductors, and will use it for the study of the
Coulomb-interaction effects in the low-temperature
Andreev conductance $G_A$ between superconductor and normal metal and in the
Josephson proximity coupling $E_J$ via dirty 2D metal.
It will be shown that both of the above-mentioned effects,
interaction-induced suppression of the tunneling DOS, and
renormalization of the Cooper-channel interaction, contribute
considerably into the suppression of $G_A$ and $E_J$ in the low-energy
limit.  These two effects differ in the following sense:
specific form of the DOS suppression depends crucially on the long-range
behaviour of the Coulomb potential (and thus can be varied externally
by changing electromagnetic environment), whereas renormalization of the
Cooper-channel interaction depends on the short-distance Coulomb amplitude
only. If long-range Coulomb forces are suppressed (i.~e.\ by placing a nearby
screening electrode), the DOS suppression effect may become weak. In this
case the main effect comes from the presence of short-range repulsion
in the Cooper channel; as a result, both the Andreev conductance $G_A(\omega)$
and the Josephson proximity coupling $E_J(r)$ exhibit anomalous power-law
suppression in the infra-red limit, with exponents of the order of $g^{-1/2}$.
In the case of no static Coulomb screening, the DOS suppression effect
is the strongest one in the asymptotic infra-red limit, it leads
to the ``log-normal"  suppression of $G_A(\omega)$ and $E_J(r)$ as
$(\omega, D/r^2) \to 0$.  The influence of long-range Coulomb interaction on
the Andreev conductance was treated previously~\cite{G_A_Hekking}
in a kind of phenomenological circuit theory. In the asymptotic
limit $(1/g)\ln^2(1/\omega\tau) \gg 1$ our results are in
agreement with those of~\cite{G_A_Hekking} and provide a microscopic
derivation for the effective impedance function used there
 phenomenologically; in the intermediate region
$(1/g)\ln^2(1/\omega\tau) \sim 1$ an additional contribution due to
the Cooper-channel repulsion (not treated in~\cite{G_A_Hekking})
 is shown to be equally important.

The rest of the paper is organized as follows:
  in Sec.~\ref{S:Keldysh} we describe the formalism of the Keldysh-type
functional integral and derive an effective action resulting from
disorder averaging;
  in Sec.~\ref{S:Sigma} we determine the saddle manifold of the above action
and derive a kind of a nonlinear $\sigma$-model action formulated in terms of
an $8\times 8$ matrix $Q$ which depends on two times and one spacial
coordinate, and of fluctuating order parameter and electromagnetic fields.
  In section~\ref{S:Pert} the basics of perturbation theory and diagram
technique for the derived $\sigma$-model are presented.
  In section~\ref{S:Fin}, in order to demonstrate the technique developed,
we rederive Finkelstein's renormalization group equations
and calculate $T_c$ suppression for dirty superconductive films.
  Section~\ref{S:Andreev} is devoted to the calculation of the Andreev
conductance as a function of frequency and/or temperature in the presence
of the Coulomb effects; we consider two different geometries of N-S contact
both in the presence and absence of the DOS suppression effect.
  In section~\ref{S:Josephson} we switch to the calculation of the Josephson
proximity coupling, for the same two geometrical configurations.
  Section~\ref{S:Discussion} contains discussion and conclusions.
  Finally, some technical details are presented in two
Appendixes.

\section{Derivation of the effective action in the Keldysh form}
\label{S:Keldysh}

\subsection{General procedure} \label{SS:Keldysh1}

The Lagrangian of the electron system interacting via electromagnetic field
and subject to disorder potential can be written as
\be
  {\cal L} = {\cal L}^e + {\cal L}^f,
\ee
where
\be
\label{L-e}
  {\cal L}^e  = \int
  \left( \psi_\alpha^*
        \left[ i\frac{\partial}{\partial t} +
        \frac{(\nabla - i\ba)^2}{2m} + \mu - U_{\rm dis}(\br) + \phi \right]
      \psi_\alpha
    + \Delta \psi^*_\uparrow \psi^*_\downarrow
    + \Delta^* \psi_\downarrow \psi_\uparrow
  \right) d\br
\ee
and
\be
\label{L-f}
  {\cal L}^f =
    \int \frac{{\bf E}^2-{\bf H}^2}{8\pi e^2}\, d^3\br +
    \frac{\nu}{\lambda} \int \Delta^*\Delta\, d\br
\ee
describe electron and field contributions respectively.
Here $\psi_\alpha(\br,t)$ is a Grassmann spinor field
($\alpha={\uparrow, \downarrow}$), $\phi(\br,t)$ and $\ba(\br,t)$
are the electromagnetic potentials,
\begin{mathletters}
\label{EH}
\bea
  && {\bf E} = -\nabla\phi - \partial_t \ba, \\
  && {\bf H} = {\rm curl}\, \ba .
\eea
\end{mathletters}%
Space integrals in Eq.~(\ref{L-e}) and in the second term of Eq.~(\ref{L-f})
are taken over the system considered while the free electromagnetic Lagrangian
given by the first term of Eq.~(\ref{L-f}) contains integration over the
whole 3-dimensional space.
$\mu$ is the chemical potential, and
$\nu$ is the density of states per one spin projection at the Fermi level.
A short-range interaction in the Cooper channel
mediated by electron-electron and electron-phonon interactions
with momentum transfer of the order of $p_F$ is decoupled in
the standard way\cite{Delta} by the $\Delta$-field.
Our superconductive coupling constant $\lambda$ coincides with
the Finkelstein's definition~\cite{finkel2} of $\Gamma_c$.
Generally speaking, one should also introduce singlet and triplet
coupling constants, $\Gamma$ and $\Gamma_2$ in Finkelstein's notations,
in the diffusion channel.
In 2-dimensional systems, the correction to them has the relative
order of the weak localization effect, $g^{-1}\ln\frac{1}{\Omega\tau}$,
with $\Omega$ being the relevant energy scale.
Therefore, for the study of the Coulomb effects in the Cooper channel
and in the tunneling density of states, they may be considered constant
and can be incorporated into the Fermi-liquid renormalization of the
parameters of the Lagrangian (\ref{L-e}).
Throughout the paper $\hbar=c=1$.

We are going to construct an effective theory of soft modes in the problem.
For this purpose one has to be able to take all possible channels into account
that is accomplished\cite{Efetov} by introducing a bispinor
\be
\label{bispinor}
  \Psi = \frac1{\sqrt2} \left(
    \begin{array}{c}
      \psi_{\uparrow} \\
      \psi_{\downarrow} \\
      \psi^*_{\downarrow} \\
      -\psi^*_{\uparrow}
    \end{array}
  \right) .
\ee
$\Psi$ is a vector in the 4-dimensional space $\Omega$ which can be
considered as the direct product
$S\otimes T$ of the spin ($\psi_{\uparrow},\psi_{\downarrow}$)
and time-reversal ($\psi,\psi^*$) spaces.
The correlations between different time-reversal components of $\Psi$ are
responsible for the quantum correction to the conductivity in the orthogonal
case\cite{GLK} when the Hamiltonian possesses time-reversal symmetry.
On the other hand, in studying superconducting phenomena, it is convenient
to rearrange components of the bispinor in a different manner,
separating explicitly the Gor'kov-Nambu\cite{Gorkov,Nambu}
($\psi_{\uparrow},\psi^*_{\downarrow}$)
and spin spaces.
Finally, one can think of $\Psi$ as acting in the direct product of
the Nambu ($N$) and time-reversal spaces.
These three representations are equivalent,
$\Omega \simeq S\otimes T \simeq N\otimes S \simeq N\otimes T$;
we will not specify a certain one and will change between them
depending on the problem at hand.
For later reference, we define the Pauli matrices in the spin, time-reversal,
and Nambu spaces as $s_i$, $t_i$, $\tau_i$ respectively ($i=0,1,2,3$).
The action of these matrices on the bispinor (\ref{bispinor})
is equivalent to multiplication by $4\times 4$ matrices;
a few examples of them are listed below:
\be
  s_x \simeq \left( \begin{array}{cccc}
    0&1&0&0\\
    1&0&0&0\\
    0&0&0&-1\\
    0&0&-1&0
  \end{array} \right)
,\quad
  \tau_x \simeq \left( \begin{array}{cccc}
    0&0&1&0\\
    0&0&0&-1\\
    1&0&0&0\\
    0&-1&0&0
  \end{array} \right)
,\quad
  t_y \simeq \left( \begin{array}{cccc}
    0&0&0&i\\
    0&0&-i&0\\
    0&i&0&0\\
    -i&0&0&0
  \end{array} \right) .
\ee
Below we will use the vector $\Psi$ together
with its conjugate vector $\Psi^+$.
These vectors are linearly dependent and related by
\be
  \Psi^+ = (C\Psi)^T =
  \frac1{\sqrt2}
    \left(
      \psi^*_\uparrow\;\psi^*_\downarrow\;-\psi_\downarrow\;\psi_\uparrow
    \right) ,
\ee
where the charge-conjugation matrix
\be
  C = it_y \otimes s_0 \equiv i\tau_y \otimes s_x. 
\ee
In terms of the $\Psi$ field, the electron Lagrangian ${\cal L}^e$
can be rewritten as
\be
  {\cal L}^e = \int d\br\, \Psi^+
        \left[ i\Xi \frac{\partial}{\partial t} +
        \frac{(\Xi\nabla - i\ba)^2}{2m} + \mu - U_{\rm dis}(\br)
        + \phi + \hat\Delta \right]
      \Psi .
\ee
Here
\be
  \hat\Delta = i\tau_y \Re\Delta + i\tau_x \Im\Delta
  = \tau_+ \Delta - \tau_- \Delta^*
  = |\Delta| [ \tau_+ e^{i\theta} - \tau_- e^{-i\theta} ] ,
\ee
$\theta$ is the phase of the order parameter,
$\tau_\pm=(\tau_x \pm i\tau_y)/2$,
and the $4\times4$ matrix $\Xi$ is given by
\be
  \Xi = t_z\otimes s_0 \equiv \tau_z\otimes s_0. 
\ee

Within the Keldysh approach the time-evolution of the system is considered
along the Keldysh contour ${\cal C}$ going from $t=-\infty$ to $t=+\infty$
and then back to $-\infty$. At the initial time, $t=-\infty$,
the system is supposed to be in the thermal equilibrium, with the
interaction and disorder potential being turned off. The latter are
switched on adiabatically during the time-evolution.
The electromagnetic potentials $\phi$ and $\ba$ entering the action
(\ref{action}) consist of fluctuating and external (source) parts:
$\phi =\phi_{fl} + \phi_s$, $\ba = \ba_{fl} + \ba_s$.
The partition function
describing the evolution along the contour can be written in terms of the
functional integral over fermionic fields $\psi$ and $\psi^*$ and
fluctuating electromagnetic field potentials $\phi_{fl}$, $\ba_{fl}$ as
\be
\label{Z-def}
  \Z = \int D\{\psi^*,\psi\} D\{\phi_{fl},\ba_{fl}\} D\{\Delta\} e^{iS} ,
\ee
where the action is given by
\be
  S = \int\limits_{\cal C} {\cal L} dt .
\label{action}
\ee
External fields (introduced through the {\it source terms} in the action)
should not be integrated out in the functional integral (\ref{Z-def}).
If all external fields are {\it classical}, i.~e. identical
for the forward and backward propagation, then the evolution along
the Keldysh contour brings the system back to the initial state. In this
case the partition function is automatically normalized to unity, $\Z=1$.
To get a nontrivial result for physical quantities one has to consider
the generating functional for the source fields having {\it quantum}
components which are different on the upper and lower parts of the contour.
We will discuss the role of the source terms in section~\ref{SS:Andreev}
and will operate meanwhile with the partition function given
by Eq.~(\ref{Z-def}).

Next we divide each dynamical field into two parts residing on the forward
and backward branches of the contour and labeled by the indices 1 and 2
respectively, and combine them into 2-vectors in the Keldysh space:
\be
\label{K-vectors}
  \roarrow{\Psi} = \left( \begin{array}{c}
      \Psi_1 \\
      \Psi_2
    \end{array} \right)
,\quad
  \roarrow{\phi} = \left( \begin{array}{c}
      \phi_1 \\
      \phi_2
    \end{array} \right)
,\quad
  \roarrow{\ba} = \left( \begin{array}{c}
      \ba_1 \\
      \ba_2
    \end{array} \right)
,\quad
  \roarrow{\Delta} = \left( \begin{array}{c}
      \Delta_1 \\
      \Delta_2
    \end{array} \right)
,\quad
  \roarrow{\theta} = \left( \begin{array}{c}
      \theta_1 \\
      \theta_2
    \end{array} \right).
\ee
Then the action can be written as $S = \int_{-\infty}^\infty L dt$
with the Lagrangian given by
\be
  L = {\cal L}_1 - {\cal L}_2, \qquad
  {\cal L}_i \equiv {\cal L}[\psi^*_i,\psi_i,\phi_i,\ba_i,\Delta_i] .
\ee
It is convenient to arrange the components of the Bose-fields
$\roarrow{\phi}$, $\roarrow{\ba}$, $\roarrow{\Delta}$ and $\roarrow{\theta}$
in the matrix form:
\be
\label{tensors}
  \tensor{\phi} = \left( \begin{array}{cc}
      \phi_1 & 0 \\
      0 & \phi_2
    \end{array} \right)
,\quad
  \tensor{\ba} = \left( \begin{array}{cc}
      \ba_1 & 0 \\
      0 & \ba_2
    \end{array} \right)
,\quad
  \tensor{\Delta} = \left( \begin{array}{cc}
      \hat\Delta_1 & 0 \\
      0 & \hat\Delta_2
    \end{array} \right)
,\quad
  \tensor{\theta} = \left( \begin{array}{cc}
      \hat\theta_1 & 0 \\
      0 & \hat\theta_2
    \end{array} \right) .
\ee
Making use of the Keldysh-space matrices (\ref{tensors}) we can
rewrite the electron and field parts of the Lagrangian
in the following concise form:
\be
\label{Le}
  L^e = \int d\br\, \roarrow{\Psi}^+
        \left[ i\Xi \frac{\partial}{\partial t} +
        \frac{(\Xi \nabla - i\tensor{\ba})^2}{2m}
        + \mu - U_{\rm dis}(\br) + \tensor{\phi}
        + \tensor{\Delta} \right] \sigma_z
      \roarrow\Psi ,
\ee
\be
\label{Lf}
  L^f =
    \int
      \frac{\roarrow{\bf E}^T \sigma_z \roarrow{\bf E}
          - \roarrow{\bf H}^T \sigma_z \roarrow{\bf H} }{8\pi e^2}\, d^3\br
  + \frac{\nu}{\lambda} \int \roarrow{\Delta}^+\sigma_z\roarrow{\Delta}\, d\br,
\ee
where $\roarrow{\bf E}$ and $\roarrow{\bf H}$ are expressed in terms of
$\roarrow{\phi}$ and $\roarrow{\ba}$ analogously to
Eqs.~(\ref{EH}), and
$\sigma_i$ denote the Pauli matrices in the Keldysh space.

Now we are in a position to perform disorder averaging.
The Keldysh formalism allows us to average the partition function
directly utilizing its independence of realization
of disorder potential. For the latter we will assume the model of
a Gaussian $\delta$-correlated white noise with the variance
\be
  \corr{U_{\rm dis}(\br) U_{\rm dis}(\br')} =
  \frac{\delta(\br-\br')}{2\pi\nu\tau}.
\ee
Integrating out disorder potential generates a four-fermion term
in the action:
\be
  S_{\rm dis} =
  \frac{i}{4\pi\nu\tau} \int d\br dt dt'
  \bigl[ \roarrow{\Psi}^+(\br,t) \sigma_z \roarrow\Psi(\br,t) \bigr]
  \bigl[ \roarrow{\Psi}^+(\br,t') \sigma_z \roarrow\Psi(\br,t') \bigr].
\ee
The slow part of the resulting non-local in time action can be decoupled
in the standard way (cf.\ \cite{Efetov}) by the Hubbard-Stratonovich
matrix field $\check Q$:
\be
\label{HS}
  e^{iS_{\rm dis}} = \int D\check Q
  \exp \left \{
    - \frac{1}{2\tau} \int d\br dt dt'\,
      \roarrow\Psi^+(\br,t) \check Q(\br,t,t') \sigma_z \roarrow\Psi(\br,t')
    - \frac{\pi\nu}{8\tau} \int d\br dt dt'
      \tr \check Q(\br,t,t') \check Q(\br,t',t)
  \right \} .
\ee
In the Keldysh formalism, $\check Q(\br,t,t')$ is
a matrix in the time space as well as
an $8\times8$ matrix in the $K\otimes \Omega$ space;
it is local in the coordinate space once
the $\delta$-correlated random potential is considered.
Here and in what follows $\tr(\cdots)$ stands for the trace
in the $K\otimes \Omega$ space whereas the complete operator trace
involving integration over space and time indices will be denoted
by $\Tr(\cdots)$.

After the Hubbard-Stratonovich transformation (\ref{HS}), the fermionic part
of the action becomes quadratic and can be written as
\be
  S^e = \Tr \roarrow{\Psi}^+ G^{-1} \roarrow\Psi,
\ee
where the inverse Green function is defined as
\be
\label{G}
  G^{-1} = \left[
   i\Xi \frac{\partial}{\partial t} +
   \frac{(\Xi \nabla - i\tensor{\ba})^2}{2m} + \mu
   + \frac{i}{2\tau} \check Q + \tensor{\phi}
   + \tensor{\Delta} \right] \sigma_z .
\ee
The field part of the action takes the form
\be
  S^f = \Tr
    \frac{\roarrow{\bf E}^T \sigma_z \roarrow{\bf E}
        - \roarrow{\bf H}^T \sigma_z \roarrow{\bf H} }{8\pi e^2}
    + \frac{\nu}{\lambda} \Tr \roarrow{\Delta}^+\sigma_z\roarrow{\Delta}
    + \frac{i\pi\nu}{8\tau} \Tr \check Q^2 .
\label{Sf}
\ee
As mentioned above, the trace operator $\Tr(\cdots)$ includes integration
over space coordinates; in the first term of Eq.~(\ref{Sf}) this
integral goes over the whole 3-dimensional space, whereas in the other terms
the integral is taken over the volume of the system considered. In the
present paper we will consider  thin metallic films only, so the integral
in the second and third terms will be effectively 2-dimensional.
Gaussian integration over $\roarrow\Psi$ can easily be performed
resulting in
\be
  S^e = -\frac i2 \Tr \ln G^{-1} .
\ee

\subsection{Keldysh rotation}

Among four Keldysh subblocks of the Green function (\ref{G}) only three
appear to be linearly independent\cite{Keldysh}.
To simplify its structure, it is convenient to pass to the rotated basis:
\be
  G' = L \sigma_z G L^{-1} ,
\ee
with the unitary matrix $L$ given by\cite{LO1}
\be
  L = \frac{1}{\sqrt2}
    \left( \begin{array}{cc} 1 & -1 \\ 1 & 1 \end{array} \right) .
\ee
After such a rotation $G'$ acquires a triangular form
(provided that source terms have no quantum component):
\be
\label{Gtriang}
  G' =
    \left( \begin{array}{cc} G^R & G^K \\ 0 & G^A \end{array} \right) ,
\ee
where $G^{R(A)}(t,t')=0$ for $t\leq t'$ ($t\geq t'$).

Apart from the Green function, it is also convenient to make a similar
transformation for $Q$:
\be
  \check Q' = L \check Q L^{-1} ,
\ee
and to rotate all 2-vectors defined in Eq.~(\ref{K-vectors}) according to
\be
  \roarrow{\phi}'
  \equiv \left( \begin{array}{c}
      \phi'_1 \\
      \phi'_2
    \end{array} \right)
  = \frac12 \left( \begin{array}{c}
      \phi_1+\phi_2 \\
      \phi_1-\phi_2
    \end{array} \right),
\label{vecrot}
\ee
and analogously for $\roarrow\ba$, $\roarrow\Delta$ and $\roarrow\theta$.
For the reasons discussed above, $\phi'_1$ and $\phi'_2$ will be
referred to as the classical and quantum components of the field.
The matrices (\ref{tensors}) will transform according to
\be
  \tensor\phi' = L \tensor\phi L^{-1}
  = \left( \begin{array}{cc}
      \phi'_1 & \phi'_2 \\ \phi'_2 & \phi'_1 \end{array}
    \right)
  \equiv \phi'_i \gamma^i
\ee
(and analogously for $\tensor\ba$, $\tensor\Delta$ and $\tensor\theta$),
where, following Kamenev and Andreev\cite{Kam_Andr}, we introduced
two vertex matrices
\be
  \gamma^1 \equiv \sigma_0, \quad
  \gamma^2 \equiv \sigma_x.
\ee

In some cases (e.~g., when one considers a uniform superconductor
on time scales much longer than the inverse gap)
it is sufficient to treat the absolute value $|\Delta|$
of the superconductive order parameter as a constant,
while taking into account fluctuations of its phase.
Then, in the rotated basis, the expression for $\tensor\Delta'$
can be written as
$\tensor\Delta' =
|\Delta| [ \tau_+ e^{i\tensor\theta'} - \tau_- e^{-i\tensor\theta'} ]$,
or, in terms of the classical, $\theta_1'$, and quantum, $\theta_2'$,
components:
\begin{mathletters}
\label{Delta'}
\bea
 && \hat\Delta_1' =
    |\Delta| [ \tau_+ e^{i\theta_1'} - \tau_- e^{-i\theta_1'} ] \cos\theta_2',
  \\
 && \hat\Delta_2' =
    |\Delta| [ \tau_+ e^{i\theta_1'} + \tau_- e^{-i\theta_1'} ] i\sin\theta_2'.
\eea
\end{mathletters}%

As a result, the Green function can be written as
\be
\label{G'}
  G'^{-1} =
   i\Xi \frac{\partial}{\partial t} +
   \frac{(\Xi \nabla - i\tensor{\ba}')^2}{2m} + \mu
   + \frac{i}{2\tau}\check Q' + \tensor{\phi}'
   + \tensor{\Delta}' ,
\ee
and the action takes the form
\be
  S = - \frac i2 \Tr \ln G'^{-1}
  + \Tr \left [
     \frac{\roarrow{\bf E}'^T \sigma_x \roarrow{\bf E}'
        - \roarrow{\bf H}'^T \sigma_x \roarrow{\bf H}' }{4\pi e^2}
    + \frac{2\nu}{\lambda} \roarrow{\Delta}'^+ \sigma_x \roarrow{\Delta}'
    + i \frac{\pi\nu}{8\tau}\check Q'^2
  \right ] .
\label{Srot}
\ee
The factor 2 difference between the coefficients in the terms containing
$\roarrow{\bf E}$, $\roarrow{\bf H}$ and $\roarrow\Delta$
in Eqs.~(\ref{Sf}) and (\ref{Srot})
is due to the Jacobian of the transformation (\ref{vecrot}).
In what follows we will omit prime at the designation of
the Keldysh-rotated fields. This cannot lead to an ambiguity
since the original basis will be never used in the subsequent analysis.

\section{$\sigma$-model}
\label{S:Sigma}

In this section we will construct an effective theory that describes
low-energy physics of the action (\ref{Srot}). To start, we subject
the action to the stationary phase analysis. It is a functional of the
matrix field $\check Q$ and the bosonic fields $\roarrow\phi$, $\roarrow\ba$,
$\roarrow\Delta$, and one has to vary the action with respect to
all of them in order to get a set of the saddle point equations.
First of all we note that quantum components of the bosonic fields
are equal to zero in the mean-field approximation. Thus, in this section,
we will designate their classical components without the subscript ``1''
for brevity (i.~e.\ $\phi\equiv\phi_1$, etc.).
Below we will use $\check Q$-matrices defined in the energy domain
according to the relation
\be
\label{FT}
  \check Q_{\eps\eps'} =
  \int\!\!\int dt dt' e^{i\eps t - i\eps't'} \check Q_{tt'}.
\ee
Varying with respect to $\check Q$ yields the saddle point equation
\be
\label{SPE}
  \check Q(\br) = \frac{i}{\pi\nu} G(\br,\br).
\ee
In the absence of quantum components, the Green function has a triangular
form (\ref{Gtriang}) and so does $\check Q$:
\be
\label{Q}
  \check Q
  = \left( \begin{array}{cc} Q^R & Q^K \\ 0 & Q^A \end{array} \right)_K ,
\ee
where $Q^R$, $Q^A$ and $Q^K$ are matrices in the space $\Omega$.
In the stationary case, the solution $\check Q_{tt'}$ depends on the time
difference $t-t'$ only, i.~e.\ in the energy domain we have
$\check Q_{\eps,\eps'} = 2\pi\delta(\eps-\eps') \check Q(\eps)$.
Varying the action with respect to the quantum components $\phi_2$ and $\ba_2$
and setting them to zero one obtains the Maxwell equations.
In the absence of an external magnetic field and/or voltage drops,
the mean-field electromagnetic field vanishes, $\phi=\ba=0$.
In order to have a closed system of equations, one has to supply
Eq.~(\ref{SPE}) with the gap equation. Varying the action (\ref{Srot})
with respect to $\Delta_2^*$ and using Eq.~(\ref{SPE}) we get
the selfconsistency equation for the order parameter:
\be
\label{SPE-Delta}
  \Delta = - \frac{\pi\lambda}{4}
    \int \frac{d\eps}{2\pi} \tr_\Omega Q^K(\eps) \tau_- ,
\ee
where the 4-dimensional space $\Omega$ is defined in Sec.~\ref{SS:Keldysh1}.

To clarify the structure of the saddle point given by Eqs.~(\ref{SPE})
and (\ref{SPE-Delta}), consider first the case of nonsuperconducting
($\Delta=0$) metal. Then it is easy to check that a diagonal in the
energy space matrix
\be
\label{Lambda}
  \Lambda(\eps) = \Lambda_0(\eps) \Xi,
\ee
where
\be
\label{Lambda0}
  \Lambda_0(\eps) =
    \left( \begin{array}{cc} 1 & 2F(\eps) \\ 0 & -1 \end{array} \right)_K ,
\ee
is a solution of Eq.~(\ref{SPE}).
The $4\times 4$ matrix function $F(\eps)$ introduced in Eq.~(\ref{Lambda0})
has the meaning of a generalized distribution function. In the steady
state it reduces  to the single scalar function
$F(\eps)=1-2f(\eps)$ where $f(\eps)$ is the usual Fermi distribution function.
The form of $f(\eps)$ is not fixed by
 the saddle point equation
since $\Lambda(\eps)$ with arbitrary $F(\eps)$ satisfies Eq.~(\ref{SPE}).
The reason for this ambiguity is that any distribution function is allowed in the
absence of interaction. To bring the system into the equilibrium with
\be
\label{Feq}
  F_{\rm eq}(\eps) = \tanh \frac{\eps}{2T},
\ee
either electron-electron or electron-phonon inelastic interactions
must be taken into account.
The Keldysh formalism is suitable for the study of nonequilibrium problems
as well. In this case there is an externally controlled difference of
temperature and/or chemical potential across the system, and the function
$F(\eps,\br)$ should be obtained from Eq.~(\ref{SPE}) with the proper
boundary conditions.

The solution (\ref{Lambda}) captures the eigenvalue structure
of a generic saddle point. All fluctuations of $\check Q$ that alter the
eigenvalues $\pm1$ are massive. The massless modes share the eigenvalue
structure of $\Lambda$ and can be obtained from it by the following
transformation:
\be
  \check Q = U^{-1} \Lambda U ,
\label{ULU}
\ee
where $U$ is some rotation matrix which acts in the $8\times8$ space
$K\otimes\Omega$ as well as in the time domain.
According to Eq.~(\ref{ULU}),
the field $\check Q$ satisfies the nonlinear constraint
\be
  \check Q^2=1
\ee
at the saddle point manifold (SPM). Together with Eq.~(\ref{Q}),
this suggests the following parametrization of the Keldysh block:
\be
\label{RF-FA}
  Q^K = Q^R F - F Q^A,
\ee
where, again, $F$ has the meaning of a generalized distribution function.

Now let us turn to the case of a uniform bulk superconductivity.
Here it is convenient to chose a representation of the space $\Omega$
as a direct product of the Nambu and spin spaces, $\Omega=N\otimes S$;
in this notations $\Xi=\tau_z$. The superconducting saddle point
solution, $\check Q_S$, has the form (\ref{Q}) with\cite{LO1}
\be
\label{Q-SC}
  Q_S^{R,A}(\eps) = \pm \frac{1}{\sqrt{(\eps\pm i0)^2-|\Delta|^2}}
    \left( \begin{array}{cc}
      \eps & \Delta \\ -\Delta^* & -\eps
    \end{array} \right)_N .
\ee
Taken at the saddle point manifold, the matrix $\check Q$ is equivalent to
the Larkin-Ovchinnikov\cite{LO1} quasiclassical Green function $\check g$.
The mean-field value of the order parameter can be obtained form
Eq.~(\ref{SPE-Delta}). Substituting $Q^K$ from Eqs.~(\ref{RF-FA})
and (\ref{Q-SC}) we obtain the standard BCS gap equation
(negative $\lambda$ corresponds to attraction between electrons)
\be
  \Delta = - \lambda \Delta
    \int_{|\Delta|}^{\omega_D}
      \frac{d\eps}{\sqrt{\eps^2-|\Delta|^2}} \tanh \frac{\eps}{2T} .
\ee

The superconducting saddle point, $\check Q_S$, belongs to the metallic SPM
given by Eq.~(\ref{ULU}). However, in the presence of superconductivity,
some excitations on the metallic SPM, having been massless, acquire a gap
proportional to $\Delta$.
A detailed discussion of the hierarchy of gaps in the $\sigma$-model for
N-S systems can be found in the review\cite{alt_sim_tar}.

The next step in the derivation of the $\sigma$-model is to consider
fluctuations of the SPM and to perform the gradient expansion
of the action (\ref{Srot}). Such a procedure is justified in the
dirty limit, $\Delta\tau\ll1$, which will be implied from now on.
It is equivalent to the replacement of the full Eilenberger\cite{Eilen}
equations in the conventional theory of superconductivity
to their approximation proposed by Usadel~\cite{Usadel}.

As was recently suggested in Ref.~\cite{Kam_Andr}, in studying electric
field fluctuations it is convenient to single out the gauge degrees
of freedom in $\check Q$ by the transformation
\be
\label{gauge}
  \check Q_{tt'} = e^{i\tensor{K}(t)\Xi} Q_{tt'} e^{-i\tensor{K}(t')\Xi} ,
\ee
where $\tensor{K}=K_i\gamma^i$ is related to the doublet
$\roarrow{K}=(K_1, K_2)^T$, in analogy with the field $\roarrow\phi$.
After the transformation (\ref{gauge}), the action can still be
written in the form (\ref{Srot}), with the Green function being
substituted by
\be
\label{G2}
  G^{-1} =
   i\Xi \frac{\partial}{\partial t} +
   \frac{(\Xi \nabla - i\tensor{\bA})^2}{2m}
   + \frac{i}{2\tau} Q + \tensor{\Phi}
   + \tensor{\Delta}_K ,
\ee
with
\begin{mathletters}
\bea
 & \tensor{\bA} \equiv \tensor{\ba} - \nabla\tensor{K}, & \\
 & \tensor{\Phi} \equiv \tensor{\phi} - \partial_t \tensor{K}, & \\
 & \tensor{\Delta}_K
   \equiv e^{-i\tensor{K}(t')\Xi}\, \tensor{\Delta}\, e^{i\tensor{K}(t')\Xi} . &
\eea
\end{mathletters}%

Expanding $\Tr\ln G^{-1}$ in the standard way\cite{Efetov,Kam_Andr},
we obtain the following effective action
\be
  S = S_\sigma
  + 2\nu \Tr \roarrow\Phi^T \sigma_x \roarrow{\Phi}
  + \Tr \frac{\roarrow{\bf E}^T \sigma_x \roarrow{\bf E}
        - \roarrow{\bf H}^T \sigma_x \roarrow{\bf H} }{4\pi e^2}
  + \frac{2\nu}{\lambda} \Tr \roarrow{\Delta}^+ \sigma_x \roarrow{\Delta} ,
\label{sigma-model}
\ee
where $S_\sigma$ is the $\sigma$-model action for the matrix field $Q$,
\be
  S_\sigma = \frac{i\pi\nu}{8}
    \Tr \left[
      D (\bbox{\partial} Q)^2
      + 4i \bigl( i\Xi \partial_t + \tensor{\Phi} + \tensor{\Delta}_K \bigr) Q
    \right] .
\label{sm0}
\ee
Here $D$ is the diffusion coefficient and $\bbox{\partial}$ denotes
a long covariant derivative,
\be
  \bbox{\partial} X \equiv
      \nabla X - i \bigl[ \Xi \tensor{\bA}, X \bigr] .
  \label{cov-grad}
\ee

Derivation of the effective action (\ref{sigma-model}) that describes
interacting disordered normal/superconducting electron liquid
is the main result of this section. The model is formulated in terms of the
interacting matter field $Q$ subject to the nonlinear constraint $Q^2=1$,
electromagnetic fields $\roarrow\phi$, $\roarrow\ba$,
and the pairing potential $\roarrow\Delta$.
At the present stage, the phase $\roarrow{K}$ introduced in Eq.~(\ref{gauge})
is left unspecified. It will be expressed in terms of the electromagnetic
potentials in section~\ref{SS:emf}.

In Eq.~(\ref{sm0}), $Q$ is an $8\times8$ matrix in the $K\otimes\Omega$ space.
In what follows we will assume that all interactions are spin-independent.
Then $Q$ is proportional to the unit matrix in the spin space,
and the 4-dimensional space $\Omega=N\otimes S$ collapses
into the 2-dimensional Nambu space. As a result, the theory will be formulated
in terms of the $4\times4$ matrices $Q_{tt'}$ acting in the $K\otimes N$ space.
The corresponding action can be obtained from Eq.~(\ref{sm0}) by taking
the trace over the redundant spin space:
\be
  S_\sigma = \frac{i\pi\nu}{4}
    \Tr \left[
      D (\bbox{\partial} Q)^2
      + 4i \bigl(
        i\tau_z \partial_t + \tensor{\Phi} + \tensor{\Delta}_K
      \bigr) Q
    \right] ,
\label{sm1}
\ee
where the operator $\bbox{\partial}$ is given by Eq.~(\ref{cov-grad})
with $\Xi=\tau_z$.

Varying the action Eq.~(\ref{sm1}) with respect to $Q$ with
the constraint $Q^2=1$ yields the equation
\be
  D \bbox{\partial} (Q \bbox{\partial} Q)
  + i \bigl[ i\tau_z \partial_t + \tensor{\Phi} + \tensor{\Delta}_K, Q \bigr]
  = 0
  ,
\label{Usadel}
\ee
which (for $\tensor{K}=0$) coincides with the dynamical
Usadel equation\cite{LO1}.
In the absence of superconductive coupling, $\roarrow\Delta=0$,
$Q$ is proportional to $\tau_z$: $Q=Q_{KA}\tau_z$, and our action
(\ref{sm1}) reduces to the Kamenev-Andreev action\cite{Kam_Andr}
for the field $Q_{KA}$.

\section{Perturbation theory}
\label{S:Pert}

\subsection{Free metallic diffusons and Cooperons}

In this section we will show how a systematic perturbative expansion
of the $\sigma$-model (\ref{sigma-model}) can be developed. Keeping in
mind further application to N-S devices with relatively weak
proximity-induced coupling (sections~\ref{S:Andreev}, \ref{S:Josephson}),
we will consider fluctuations near the {\em metallic} saddle point
(\ref{Lambda}). For this purpose, it is convenient to parametrize
the rotation matrix $U$ in Eq.~(\ref{ULU}) in terms of another
matrix $W$ subject to the linear constraint
\be
\label{[WL]}
  W \Lambda + \Lambda W = 0 .
\ee
Such a parametrization is not unique and a number of them
are widely used in literature (see, e.~g., \cite{Efetov}).

In the Keldysh formalism, the saddle point (\ref{Q}) is not diagonal
and, consequently, the solution of Eq.~(\ref{[WL]}) for $W$ explicitly depends
on the distribution function $F(\eps)$. As a result, even for a noninteracting
system, intermediate expressions for the Cooperon and diffuson propagators
would depend on the particle distribution. To surmount such an unphysical
complication, we note that at the saddle point (\ref{Q}), $Q$ can be
diagonalized in the Keldysh space by a nonunitary $2\times2$ matrix
\be
\label{u}
  u = u^{-1} =
    \left( \begin{array}{cc} 1 & F \\ 0 & -1 \end{array} \right)_K ,
\ee
that separates the distribution function $F$ from the retarded and advanced
blocks which are determined by the spectral properties only:
\be
\label{uQu0}
  Q = u \left( \begin{array}{cc} Q^R & 0 \\ 0 & Q^A \end{array} \right) u.
\ee
The function $F$ in Eq.~(\ref{u}) is the stationary fermionic
distribution function and we will assume that the system is in
thermal equilibrium, so that $F(\eps)$ is given by Eq.~(\ref{Feq}).

The decomposition (\ref{uQu0}) suggest to pass from the initial
$Q$-representation to a new variable, $\Q$, defined as
\be
\label{uQu}
  \Q = uQu .
\ee
In terms of the new variable $\Q$, the $\sigma$-model
action (\ref{sigma-model}) acquires the form
\be
  S_\sigma = \frac{i\pi\nu}{4}
    \Tr \left[
      D (\bbox{\partial} \Q)^2
      + 4i \bigl( i\tau_z \frac{\partial}{\partial t}
        + u\tensor{\Phi}u + u\tensor{\Delta}_K u \bigr) \Q
    \right],
\label{sm2}
\ee
with the modified definition of the long derivative:
\be
  \bbox{\partial} X \equiv
      \nabla X - i \bigl[ \tau_z u\tensor{\bA}u, X \bigr] .
\ee
Note that the matrix $u$ couples to the interaction terms only.
For the noninteracting case, the distribution function
drops from the $\Q$-action.

At the metallic saddle point for the action (\ref{sm2}),
$\roarrow\Phi = \roarrow\bA = \roarrow\Delta = 0$
and $\Q$ is diagonal, $\Q = u\Lambda_0\tau_z u = \sigma_z \tau_z$.
Gapless fluctuations of $\Q$ can then be parametrized as
\be
  \Q = \U^{-1} \sigma_z \tau_z \, \U,
\label{ULU2}
\ee
where $\U$ is a unitary matrix, for which we adopt the exponential
parametrization,
\be
  \U = e^{\W/2},
\ee
in terms of the matrix $\W$ which anticommutes with $\sigma_z \tau_z$,
\be
  \{ \W, \sigma_z\tau_z \} = 0 .
\label{W-constraint}
\ee
An explicit expression for $\Q$ in terms of $\W$ reads
\be
\label{QW}
  \Q = e^{-\W/2} \sigma_z\tau_z e^{\W/2}
  = \sigma_z\tau_z (1+\W+\W^2/2+\dots).
\ee

The linear matrix constraint (\ref{W-constraint}) can be resolved
by introducing eight scalar variables, $w_i$ and $\overline{w}_i$
with $i=0,x,y,z$, as
\be
\label{W}
  \W = \left( \begin{array}{cc}
     w_x \tau_x + w_y \tau_y
     & w_0 + w_z \tau_z \\
     \overline w_0 + \overline w_z \tau_z &
     \overline w_x \tau_x + \overline w_y \tau_y
  \end{array} \right)_K .
\ee
The diagonal (offdiagonal) in the Nambu space excitations,
$w_i$ and $\overline{w}_i$ with $i=0,z$ ($i=x,y$),
correspond to diffusion (Cooper) modes.
Extracting quadratic in $\W$ part from the $\sigma$-model action (\ref{sm2}),
\be
  iS^{(2)}[\W] = \frac{\pi\nu}{4}
    \Tr \left[
      ( D q^2 - 2i \sigma_z E ) \W(\bq) \W(-\bq)
    \right] ,
\ee
we obtain the following correlators for diffusons:
\be
\label{diffuson}
  \corr{ w_i(\bq;\eps_1,\eps_2) \overline w_i(-\bq;\eps_3,\eps_4) }
  = -\frac{1}{\pi\nu}
    \frac{(2\pi)^2\delta(\eps_1-\eps_4)\delta(\eps_2-\eps_3)}
      {Dq^2 - i(\eps_1-\eps_2)},
  \qquad i=0,z ,
\ee
and Cooperons:
\be
\label{Cooperon}
  \begin{array}{c}
    \displaystyle
    \corr{ w_i(\bq;\eps_1,\eps_2) w_i(-\bq;\eps_3,\eps_4) }
    = -\frac{1}{\pi\nu}
      \frac{(2\pi)^2\delta(\eps_1-\eps_4)\delta(\eps_2-\eps_3)}
        {Dq^2 - i(\eps_1+\eps_2)},
  \\
    \displaystyle
    \corr{ \overline w_i(\bq;\eps_1,\eps_2) \overline w_i(-\bq;\eps_3,\eps_4) }
    = -\frac{1}{\pi\nu}
      \frac{(2\pi)^2\delta(\eps_1-\eps_4)\delta(\eps_2-\eps_3)}
        {Dq^2 + i(\eps_1+\eps_2)},
    \rule{0pt}{20pt}
  \end{array}
  \qquad i=x,y .
\ee

\subsection{Diagrammatic technique} \label{SS:DT}

\begin{figure}
\vspace{-2mm}
\epsfxsize=70mm
\centerline{\epsfbox{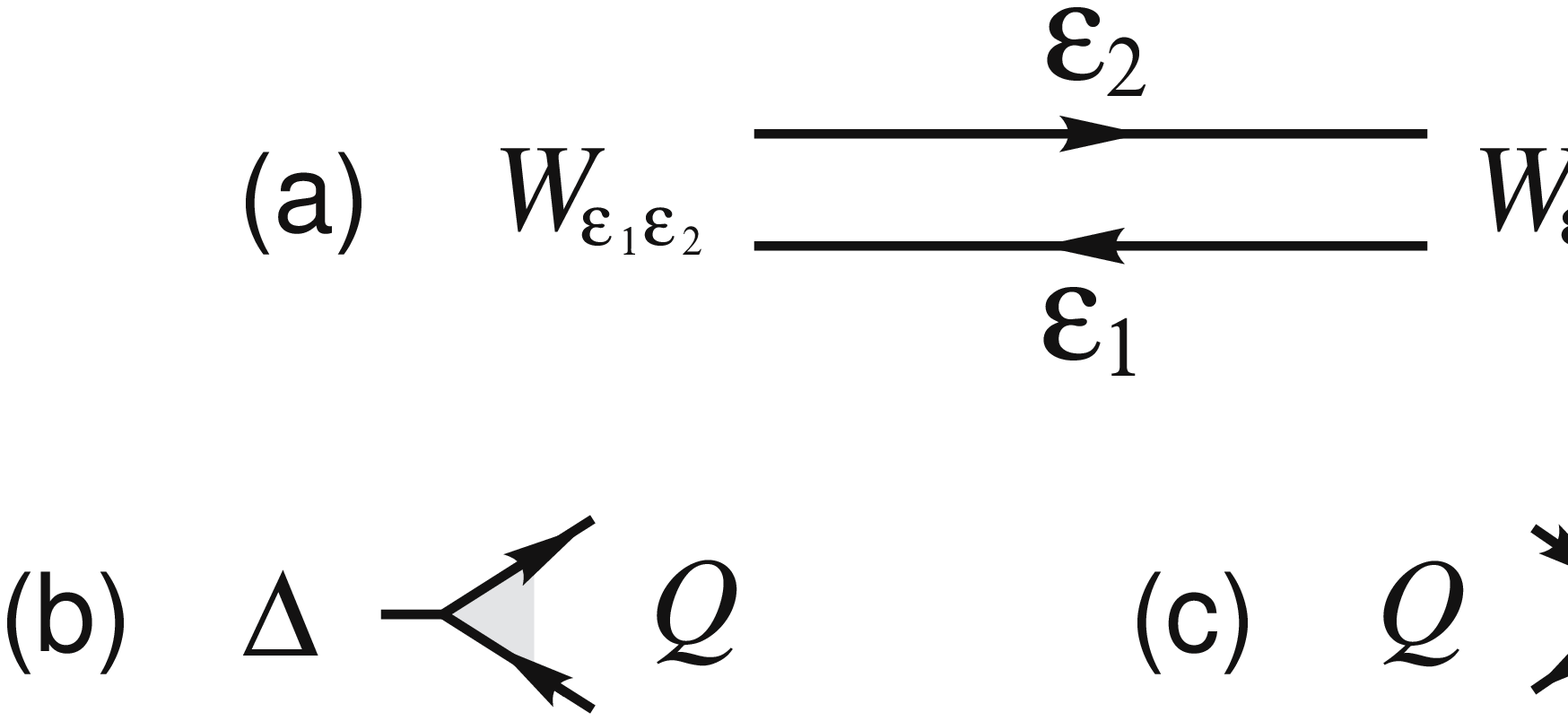}}
\vspace{2mm}
\caption{Basic elements of the diagrammatic technique:
a) diffusion/Cooper propagator;
b) $Q$ -- $\Delta$ interaction vertex;
c) effective interaction $S_\lambda[Q]$ after eliminating $\Delta$ field.
}
\label{F:basic}
\end{figure}

Expression (\ref{QW}) provides a regular way for perturbative expansion
near the metallic saddle point, $\Q = \sigma_z \tau_z$.
Its basic elements are given by the free correlators (\ref{diffuson})
and (\ref{Cooperon}) corresponding to soft diffusion and Cooper modes.
Since $\W_{\eps\eps'}$ is a matrix in the energy space,
its correlators are represented diagrammatically by two parallel lines,
each of them carrying one energy index, see Fig.~\ref{F:basic}a.
Expanding the action (\ref{sm2}) over $\W$ generates nonlinear
vertices describing interaction between diffusion and Cooper modes.
The resulting diagrammatics looks very similar to the standard
cross technique for dirty metals\cite{AGD} where soft modes are
constructed from two Green functions averaged over disorder.
Note however that in the present case both diffusons and Cooperons
are depicted in the same manner, with arrows pointing in the opposite
directions on two lines of the propagator. Such a convention, though
being unusual for Cooperons, is consistent with the definition of the
Fourier-transformed variables (\ref{FT}). Diffusons and Cooperons can
be distinguished by their structure in the Nambu (and Keldysh) space.

The normalization of the functional integral, $\Z=1$, manifests itself
in cancellation of closed loops in the perturbation theory.
In the Keldysh formalism, such a cancellation is related to the
integral over the internal energy (let it be $E$) of a closed loop.
Indeed, all propagators along the loop have poles in one
(upper or lower) half-plane of the complex variable $E$.
Therefore, integrating over $E$ yields to cancellation
of the corresponding diagram.

For future references we present here the contraction rule for averaging
over $\W$:
\bea
  \corr{ \Tr {\cal A}\W \cdot \Tr {\cal B}\W } =
  - \frac{1}{2\pi\nu} \int \frac{d\bq\, d\eps_1 d\eps_2}{(2\pi)^{d+2}}
  \Biggl\{
    \frac{
      \tr ( AB - A\sigma_zB\sigma_z +
        A\tau_zB\tau_z - A\sigma_z\tau_zB\sigma_z\tau_z )
        (Dq^2 + i(\eps_1-\eps_2)\sigma_z)
    }{(Dq^2)^2+(\eps_1-\eps_2)^2}
    + {}
\nonumber
\\ {}
  +
    \frac{
      \tr ( AB + A\sigma_zB\sigma_z -
        A\tau_zB\tau_z - A\sigma_z\tau_zB\sigma_z\tau_z )
        (Dq^2 + i(\eps_1+\eps_2)\sigma_z)
    }{(Dq^2)^2+(\eps_1+\eps_2)^2}
  \Biggr\} ,
\label{WWcontr}
\eea
where $A={\cal A}(\bq,\eps_1,\eps_2)$ and $B={\cal B}(-\bq,\eps_2,\eps_1)$.
The first (second) term corresponds to diffusons (Cooperons).
For $Dq^2\gg\eps_1,\eps_2$, this expression can be simplified as
\be
  \corr{ \Tr {\cal A}\W \cdot \Tr {\cal B}\W } =
  - \frac{1}{\pi\nu} \int \frac{d\bq\, d\eps_1 d\eps_2}{(2\pi)^{d+2}}
    \frac{ \tr ( AB - A\sigma_z\tau_zB\sigma_z\tau_z ) } { Dq^2 } .
\label{WWcontr1}
\ee

Apart from the matter field $\Q$ (or, equivalently, $\W$),
the $\sigma$-model action (\ref{sigma-model}) with $S_\sigma$ given by
Eq.~(\ref{sm1}) contains the electromagnetic
potentials and the field $\roarrow\Delta$.
The former will be considered in the next subsection
while the latter can be easily eliminated by Gaussian integration.
Note that the resulting expression does not depend on the Coulomb
phase $\roarrow{K}$ entering $\tensor\Delta_K$ since it can be
``gauged away'' by the shift of the phase of the integration variable
$\roarrow\Delta$. In other words, $\tensor\Delta_K$ in Eq.~(\ref{sm1})
can be substituted by $\tensor\Delta$ provided that the order parameter
field is to be integrated out.
The resulting contribution to the action reads
\be
  S_\lambda
  = \frac{\pi^2\nu\lambda}4
    \int d\br\, dt \sum\limits_{i=x,y}
      \tr \tau_i Q_{tt} \cdot \tr \tau_i \sigma_x Q_{tt}
  = \frac{\pi^2\nu\lambda}4
    \int d\br\, dt \tr \sigma_x [ Q^2 - (\tau_z Q)^2 ] ,
\label{S_lambda}
\ee
and is shown diagrammatically in Fig.~\ref{F:basic}c.
Note that this term is conveniently expressed
in the original $Q$-representation.

The perturbative expansion near the metallic saddle point is justified
in the case of repulsive ($\lambda>0$) interaction in the Cooper channel
or when the proximity-induced superconducting coherence is weak enough.
Otherwise, deviation from the metallic point is large and one should use
the solution to the Usadel equation as a starting point for the perturbative
analysis (cf.\ a detailed discussion in Ref.~\cite{alt_sim_tar}).

\subsection{Electromagnetic field fluctuations} \label{SS:emf}

In the previous section we have sketched the basic rules of the
diagrammatic perturbation theory in the absence of the Coulomb interaction.
Here we will discuss how fluctuations of the electromagnetic field can
be incorporated into the formalism. As was shown recently
by Kamenev and Andreev\cite{Kam_Andr}, a certain choice of the Coulomb phase
$\roarrow{K}$ introduced in Eq.~(\ref{gauge}) results in a significant
simplification of the theory.
Such a choice essentially depends on the position of the noninteracting
saddle point on the SPM and is quite different for the metallic
(\ref{Lambda}) and superconductive (\ref{Q-SC}) saddle points.
For the metallic (nonsuperconductive) case,
we choose, following Ref.~\cite{Kam_Andr}, $\roarrow{K}$
to be a linear functional of $\roarrow{\phi}$ and $\roarrow{\ba}$
and require the vanishing of the term linear both in $W$ and
$\roarrow\Phi$, $\roarrow\bA$ in the Usadel equation (\ref{Usadel}),
or, equivalently, in the $\sigma$-model action (\ref{sm1}).
The resulting equation reads
\be
\label{eq1}
  D ( \nabla \tensor{\bA} - \Lambda_0 \nabla \tensor{\bA} \Lambda_0 )
  + [ \Lambda_0, \tensor{\Phi} ] = 0 ,
\ee
where the matrix $\Lambda_0$ is introduced in Eq.~(\ref{Lambda0}).
Eq.~(\ref{eq1}) is to be used to express $\roarrow{K}$ in terms of
the electromagnetic field potentials $\roarrow\phi$ and $\roarrow\ba$,
the corresponding relation having the form (in this section we share
most of notations of Ref.~\cite{Kam_Andr})
\be
\label{K-phi}
  {\cal D}^{-1}(\omega,q) \roarrow{K}(\bq,\omega)
  = \Pi^{-1}_\omega \roarrow{\phi}(\bq,\omega)
  + i D \sigma_x \bq \roarrow{\ba}(\bq,\omega) .
\ee
Here
\bea
& \displaystyle
  {\cal D}^{-1}(\omega,q) =
  \left( \begin{array}{cc}
    0 & Dq^2+i\omega \\ Dq^2-i\omega & -2i\omega B_\omega
  \end{array} \right) ,
\\
& \displaystyle
\label{Pi}
  \Pi^{-1}_\omega =
  \left( \begin{array}{cc} 0 & -1 \\ 1 & 2B_\omega \end{array} \right) ,
\eea
and
\be
  B_\omega = \coth \frac{\omega}{2T}
\ee
is the equilibrium bosonic distribution function.

So far our analysis holds for any gauge and any geometry of the sample. From
now on we chose the gauge $\roarrow\ba=0$ neglecting relativistic
effects due to the magnetic field fluctuations. Also we restrict ourselves
to the consideration of 2-dimensional systems. Then one has to integrate
out-of-plane degrees of freedom in the electromagnetic field action
$S_{em} = \Tr \roarrow{\bf E}^T \sigma_x \roarrow{\bf E}/(4\pi e^2)$.
The result depends on the presence or absence of conducting electrodes
that screen the long-range Coulomb interaction (for the sake of simplicity,
we set the dielectric permeability of the medium to unity).
For a single plane, one obtains
\be
  S_{em} = \int dt \int \frac{d\bq}{(2\pi)^2}
   \tr\roarrow{\phi}^T(-\bq) \sigma_x V_0^{-1}(q) \roarrow{\phi}(\bq),
\ee
where $V_0(q)$ is the 2D Coulomb interaction potential,
\be
\label{V0}
  V_0(q) = \int \frac{dk_z}{2\pi} \frac{4\pi e^2}{q^2+k_z^2}
  = \frac{2\pi e^2}{q}.
\ee
If, for example, there is a metallic gate at a distance $b$ from the 2D plane,
then $V_0(q)$ is screened in the long-wavelength limit and we have instead
\be
  V_0^{\rm scr}(q) = \frac{2\pi e^2}{q} \left( 1 - e^{-2bq} \right).
\label{Vscreen}
\ee

Collecting the terms in the action bilinear in $\roarrow\phi$ and $\roarrow{K}$
and making use of the relation (\ref{K-phi}) we get an effective action
for the electromagnetic field propagation in the disordered
metal\cite{Kam_Andr}:
\be
\label{Semeff}
  S_{em}^{eff} = \Tr \roarrow{\phi}^T V^{-1} \roarrow{\phi},
\ee
where $V$ has the meaning of a dynamically screened Coulomb interaction
in the RPA approximation,
\be
  V(q,\omega) = \left( \sigma_x V_0^{-1}(q) + P_0(q,\omega) \right)^{-1},
\ee
where $P_0(q,\omega)$ is the bare density-density correlator.
The matrix $V(q,\omega)$ has the structure of a bosonic propagator
in the Keldysh space\cite{Kam_Andr}:
\be
\label{V}
  V(q,\omega) = \left( \begin{array}{cc}
    V^K(q,\omega) & V^R(q,\omega) \\ V^A(q,\omega) & 0 \end{array}
  \right) ,
\ee
with
\begin{mathletters}
\bea
\label{VR}
&&
  V^{R,A}(q,\omega)
  = \left( V_0^{-1}(q) + \frac{2\nu Dq^2}{Dq^2\mp i\omega} \right)^{-1},
\\
&& V^{K}(q,\omega) = B_\omega \left( V^R(q,\omega) - V^A(q,\omega) \right).
\eea
\end{mathletters}%
An additional factor 2 in Eq.~(\ref{VR}) compared to that
in the Kamenev-Andreev paper~\cite{Kam_Andr} is related
to the fact that they considered spinless electrons.

Eq.~(\ref{Semeff}) determines the bare propagator of the electromagnetic
field:
\be
\label{phi-corr}
  \corr{\phi_i(\bq,\omega) \phi_j(-\bq,-\omega)}
  = \frac{i}{2} V_{ij}(q,\omega).
\ee
The propagator of the field $\roarrow{K}$ can be obtained from
Eq.~(\ref{phi-corr}) with the help of the relation (\ref{K-phi})
and has the form
\be
\label{K-corr}
  \corr{K_i(\bq,\omega) K_j(-\bq,-\omega)}
  = \frac{i}{2} {\cal V}_{ij}(q,\omega),
\ee
where the matrix ${\cal V}$ has the same structure as $V$, Eq.~(\ref{V}),
with
\begin{mathletters}
\label{calV}
\bea
\label{calVR}
&&
  {\cal V}^{R,A}(q,\omega)
  = -\frac{1}{(Dq^2\mp i\omega)^2}
    \left( V_0^{-1}(q) + \frac{2\nu Dq^2}{Dq^2\mp i\omega} \right)^{-1},
\\
&& {\cal V}^{K}(q,\omega)
  = B_\omega \left( {\cal V}^R(q,\omega) - {\cal V}^A(q,\omega) \right).
\eea
\end{mathletters}%
We should note, that in Eqs.~(\ref{VR}), (\ref{calVR}) possible influence of
superconductive pairing upon the dynamic screening of the Coulomb interaction
is neglected; this is safe since we will consider N-S systems in
the limit of weak tunneling only, so the 2D metal is slightly perturbed
by superconductivity.

In a superconductor, the choice of an optimal Coulomb phase $\roarrow{K}$
valid in the whole energy range is a complicated task.
However, it had been shown in Ref.~\cite{NAA} that in the deep subgap limit
($\eps\ll\Delta$) the effect of the electric potential on the quasiclassical
Green function $Q$ is small in the parameter $\eps/\Delta$ and hence
$\roarrow{K}=0$. This result will be used below.

\section{Renormalization of the interaction in the Cooper channel}
\label{S:Fin}

\subsection{Renormalization Group procedure}

In this section we will show how to construct a procedure
of successive eliminating of high-frequency and high-momentum fluctuations of
all interaction modes in the problem:
the matrix field $Q(\bq,\eps,\eps')$,
the order parameter field $\Delta(\bq,\omega)$,
and the electric potential $\phi(\bq,\omega)$.
Elimination of high-energy modes in a dirty 2D metal results in
logarithmic corrections to the parameters entering the action
and governing dynamics of the retained slow modes. This procedure
is known as the Renormalization Group (RG) method. We will closely follow
Finkelstein's approach to RG construction~\cite{finkel1}, with the
necessary modifications due to the presence of the Keldysh space instead of
replicas.

Each elementary RG step consists in elimination of degrees of freedom
in the energy shell from $\Omega_*$ to $\Omega$, where $\Omega$
is the current value of the running ultra-violet cutoff in the problem.
Correspondingly, all fluctuating fields are decomposed into
fast (denoted by a prime) and slow (denoted by a tilde) parts.
For the fields $\roarrow\Delta$, $\roarrow\phi$ and $\roarrow K$
such a representation is trivial:
$\roarrow\Delta = \roarrow{\tilde\Delta} + \roarrow\Delta'$, etc.,
while for the field $\Q$ it must be consistent with the constraint $\Q^2=1$.
To achieve this, we decompose the rotation matrix $\U$ in Eq.~(\ref{ULU2})
into the product of a fast, $\U'=\exp(\W'/2)$, and a slow, $\tilde\U$,
part, so that
\be
  \Q = \, \tilde \U^{-1} \Q' \, \tilde \U,
\ee
where the fast $\Q'$ is expressed in terms of $\W'$ according
to Eq.~(\ref{QW}).
The slow matrix $\tilde\U$ differs from the unit matrix only if all its
arguments are smaller than the new cutoff $\Omega_*$:
\be
  \tilde \U(\bq, \eps, \eps') = \openone, \quad
  \mbox{if $Dq^2$ or $|\eps|$ or $|\eps'| > \Omega_*$} .
\ee
On the other hand, the fast $\W'$ is nonzero if at least one of its arguments
belongs to the energy shell $(\Omega_*,\Omega)$:
\be
  \W(\bq, \eps, \eps') \neq 0, \quad
  \mbox{if $\Omega_*<Dq^2$ or $|\eps|$ or $|\eps'|<\Omega$} .
\ee
After integration over fast variables, $\Omega_*$ becomes a new cutoff
and the whole procedure should be successively repeated.

In a 2D dirty metal, integrating out fast degrees of freedom results
in a relative correction $\propto \ln (\Omega/\Omega_*) = \zeta_* -\zeta$
to the parameters of the effective action (\ref{sigma-model}),
where $\zeta$ is a logarithmic variable defined as
\be
\label{zeta}
  \zeta = \ln\frac{1}{\Omega\tau}.
\ee
The RG near the metallic saddle point $\sigma_z\tau_z$ is justified
provided that the Cooper-channel coupling constant $\lambda\ll1$ while
the dimensionless conductance of the metal $g\gg1$. The latter
is defined as
\be
\label{g}
  g = 2\nu D = \frac{\sigma}{e^2},
\ee
with $\sigma=R_\square^{-1}$ being the conductance per square
(in conventional units, $\sigma = (e^2/\hbar) g$).
Logarithmic corrections to the conductance become large at the
localization scale $\zeta\sim g$. The same is true for the coupling
constants $\Gamma$ and $\Gamma_2$ omitted in the derivation of the
action (\ref{sigma-model}), cf.\ discussion in Sec.~\ref{SS:Keldysh1}.
On the contrary, corrections to $\lambda$ become of the relative order
of unity at much shorter scale, at $\zeta\sim \sqrt{g}$.
Therefore it is possible, at large $g$, to neglect renormalization
of the conductance and consider $g$ as a constant.

\subsection{BCS correction to $\lambda$}

First of all we show how to obtain the standard BCS renormalization of
the Cooper-channel interaction constant $\lambda$ in the present
formalism. The correction originates from the term
$S_{\Delta Q} = -\pi\nu \Tr \tensor{\Delta}Q$ in the $\sigma$-model
action (\ref{sm1}) after eliminating high-frequency fluctuations
of the field $Q$
(here we substitute $\tensor\Delta_K$ by $\tensor\Delta$ as was explained
in section \ref{SS:DT}).
Passing to the rotated $\Q$-representation (\ref{uQu}),
expanding to the first power in $\W'$ according to Eq.~(\ref{QW})
and setting $\tilde\U$ to unity one obtains the relevant interaction vertex
\be
  S_{\rm int} = -\pi\nu \Tr u \tensor{\tilde\Delta} u \sigma_z\tau_z \W'.
\ee
After averaging over the fast $\W'$, this term will generate
the following correction to the action:
\be
  \Delta S = \frac i2 \corr{S_{\rm int}^2}.
\ee

\begin{figure}
\epsfxsize=66mm
\centerline{\epsfbox{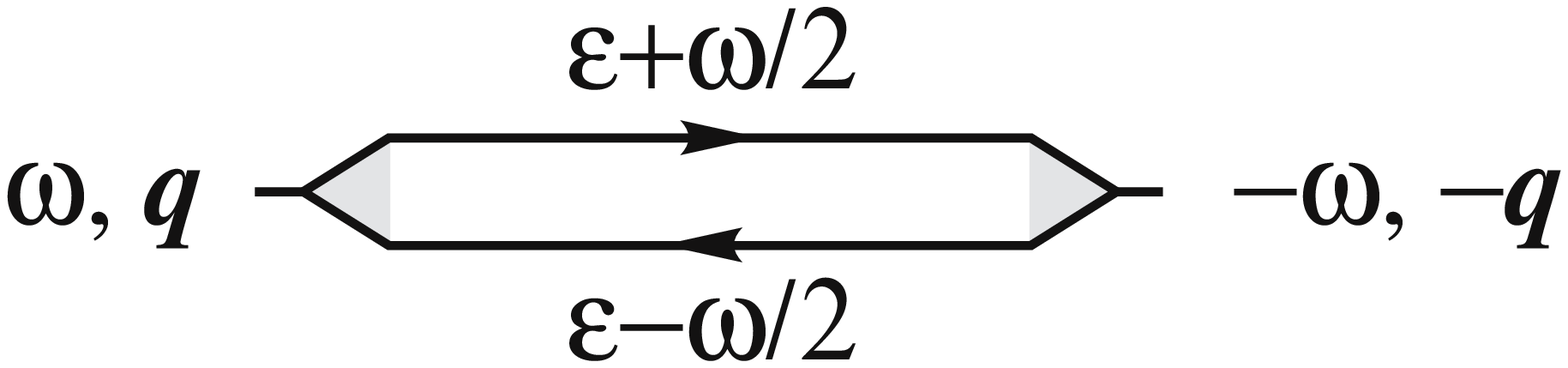}}
\caption{BCS correction to $\lambda$.}
\label{F:d-bcs}
\end{figure}

The corresponding diagram is shown in Fig.~\ref{F:d-bcs}. The only fast
variable is the internal energy of the $\corr{\W'\W'}$ propagator.
Employing then the contraction rule (\ref{WWcontr}) and the
relation $\{\hat\Delta,\tau_z\}=0$, we conclude that the diffuson
contribution vanishes identically while the Cooperon pairing yields
(to logarithmic accuracy)
\be
\label{dSd}
  \Delta S
  = - \frac{\pi\nu}{2}
    \int \frac{d\bq\, d\omega\, d\eps}{(2\pi)^4}
    \, \frac{1}{\eps} \,
    \tr
      \tensor{\tilde\Delta}(\bq,\omega)
      \tensor{\tilde\Delta}(-\bq,-\omega)
      \Lambda_0(\eps) ,
\ee
where the $\eps$-integration is performed over the energy shell
$\Omega_*<|\eps|<\Omega$, while $\bq$ and $\omega$ are restricted to the
domain $Dq^2,|\omega|<\Omega_*$. On deriving Eq.~(\ref{dSd})
we have used the relation $\Lambda_0 = u\sigma_z u$.
Now using the definition of the matrices $\tensor{\Delta}$ and $\Lambda_0$
and omitting the tilde sign over the designation of the slow component of
$\roarrow\Delta$, we transform the trace in the above equation to the form
\be
  \tr (\tau_+ \Delta_i - \tau_- \Delta^*_i)
    (\tau_+ \Delta_j - \tau_- \Delta^*_j)
    \gamma^i \gamma^j \Lambda_0(\eps)
  = - 4 F(\eps) (\Delta^*_1 \Delta_2 + \Delta_1 \Delta^*_2) .
\ee
As a result, Eq.~(\ref{dSd}) can be represented as
\be
  \Delta S
  = 2\nu \Tr \roarrow{\Delta}^+ \sigma_x \roarrow{\Delta}
  \cdot \int_{\Omega_*}^\Omega \frac{d\eps}{\eps} \tanh\frac{\eps}{2T} .
\ee
The temperature $T$ thus determines the infrared cutoff of the RG procedure.
At larger scales, $\Omega\gg T$, one has
\be
\label{dSd2}
  \Delta S
  = 2\nu \, \ln\frac{\Omega}{\Omega_*} \,
  \Tr \roarrow{\Delta}^+ \sigma_x \roarrow{\Delta} .
\ee

Comparing with the last term in the action (\ref{sigma-model}),
we conclude that the correction (\ref{dSd2}) renormalizes $1/\lambda$:
\be
  \frac{\partial(1/\lambda)}{\partial \zeta} = 1 .
\label{bcs}
\ee
For a superfluid Fermi-liquid, the RG equation (\ref{bcs}) was derived
in~\cite{Larkin-Migdal}; in that case only the Cooper channel was
relevant, therefore the RG approach was equivalent to a simple summation
of the standard BCS-theory ladder.

Coulomb interaction in dirty superconductors contributes both to the
Cooper and to the density-density channels. We consider here the range
of parameters where $\ln(1/\Omega\tau) \ll g$ that makes it possible to
neglect the effect of the Cooper channel upon the conductance $g$;
however the effect of the density-density channel upon the Cooper one
has the relative order of $g^{-1}\ln^2(1/\Omega\tau)$ and thus should be taken
into account. This effect can be described in the form of an integral
equation for the energy-dependent Cooper attraction $\lambda(E)$,
as was done by Aleiner and Altshuler~\cite{AAunpub} (the same kind of
equation was derived in~\cite{Vaks-Larkin} for another but similar problem).
For our purpose it will be more convenient to treat the same effect within
the RG procedure, as described in the next subsection.

\subsection{Coulomb correction to the Cooper-channel interaction}
\label{SS:Fin}

In this subsection we calculate the Coulomb-induced correction to the
coupling constant $\lambda$. It appears as a result of eliminating
high-momentum fluctuations of the electric field.
According to Eq.~(\ref{sm2}), the electric field couples to the matter
field $\Q$ by the following terms:
\be
  S_{\rm int} = - \pi\nu
    \Tr \left[
      u(\tensor{\phi} - \partial_t \tensor{K})u \Q +
      \frac12 D \tau_z u\nabla\tensor{K}u [\Q,\nabla \Q] +
      \frac i4 D [\tau_z u\nabla\tensor{K}u, \Q])^2
    \right]
  \equiv S_{\rm int}^a + S_{\rm int}^b + S_{\rm int}^c .
\ee
Here we utilize the relation (\ref{K-phi}) connecting the phase $\roarrow K$
to the field $\roarrow\phi$. The interaction vertices $S_{\rm int}^a$ and
$S_{\rm int}^b$ are linear in $\roarrow\phi$, while the vertex $S_{\rm int}^c$
is quadratic. Then the result of averaging over fast variables can be
written as
\be
\label{Si}
  \Delta S =
  \frac i2 \corr{[S_{\rm int}^a + S_{\rm int}^b]^2} + \corr{S_{\rm int}^c} .
\ee

\begin{figure}[htb]
\epsfxsize=175mm
\centerline{\epsfbox{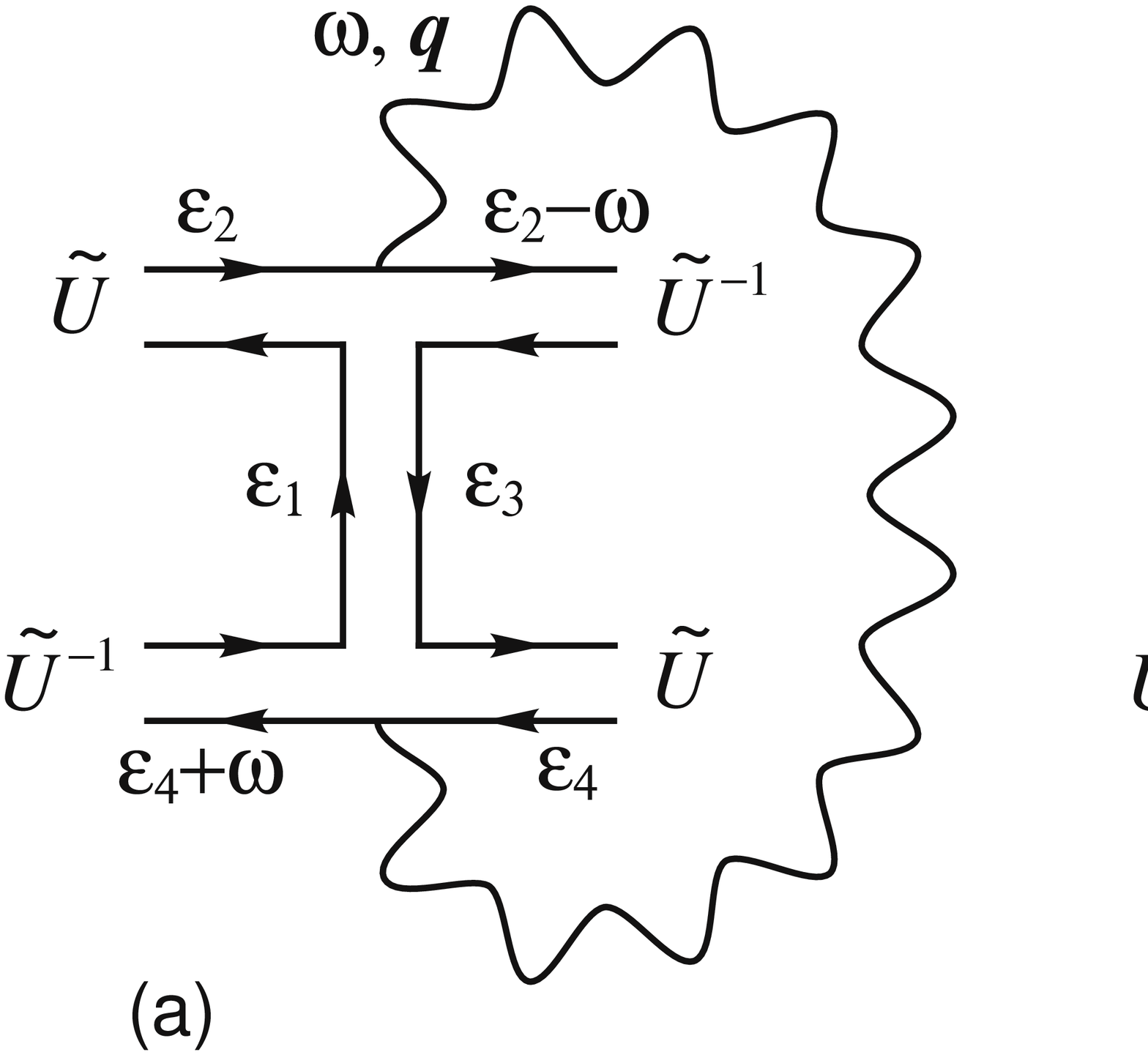}}
\caption{Diagrams for the Coulomb correction to $\lambda$:
a) $\corr{(S_{\rm int}^a)^2}$, wavy line denotes the correlator
$\corr{\phi_i(\bq,\omega) \phi_j(-\bq,-\omega)}$;
b) $\corr{(S_{\rm int}^b)^2}$, zigzag line stands for the correlator
$\corr{K_i(\bq,\omega) K_j(-\bq,-\omega)}$;
c) $\corr{S_{\rm int}^c}$.}
\label{F:d-Fin}
\end{figure}

Consider first the average $\corr{(S_{\rm int}^a)^2}$.
The relevant (i.~e.\ logarithmic) contribution
can be written in the form
\be
  \corr{(S_{\rm int}^a)^2} =
  \pi^2\nu^2
  \Bigl\langle \bigl[
    \Tr
      \, \tilde \U u (\tensor{\phi}'-\partial_t\tensor{K}') u \, \tilde \U^{-1}
      \sigma_z\tau_z \W'
  \bigr]^2 \Bigr\rangle ,
\ee
where the pairings are shown in Fig.~\ref{F:d-Fin}a.
According to this diagram, all energies are coupled to slow variables,
and there is only one fast variable, the internal momentum $\bq$
running over $\corr{\phi'\phi'}$ and $\corr{\W'\W'}$.
Therefore, to logarithmic accuracy we may consider $Dq^2\gg\omega$.
In this limit, one can neglect the term $i\omega \tensor{K}$
compared to $\tensor{\phi}$ as follows from Eq.~(\ref{K-phi})
and use the universal large-$q$ asymptotics of the screened
Coulomb interaction (\ref{V}),
\be
  \corr{\phi_i(\bq,\omega) \phi_j(-\bq,-\omega)}
  \simeq \frac i{4\nu} \sigma_x,
\ee
which is independent of the details of the bare potential $V_0(q)$.
The averaging over $\W'$ can performed with the help of Eq.~(\ref{WWcontr1}),
making use of the inequality $Dq^2 \gg \eps_1$, $\eps_3$.
As a result, one obtains
\be
\label{Siaa1}
  \corr{(S_{\rm int}^a)^2}
  = - \frac{i\pi}{4}
  \int \frac{d\br\, d\eps_1\, d\eps_2\, d\eps_3\, d\eps_4\, d\omega\, d\bq}
    {(2\pi)^7} \frac{1}{Dq^2}
  \sum_{i\neq j} \tr ( A_i B_j - A_i \sigma_z\tau_z B_j \sigma_z\tau_z ) ,
\ee
where the $\bq$-integration is taken over the fast energy shell
$\Omega_* < Dq^2 < \Omega$, the matrices $A_i$ and $B_j$ are given by
\begin{mathletters}
\label{AB}
\bea
 && A_i =
    \, \tilde \U_{\eps_1\eps_2} u_{\eps_2} \gamma^i
    u_{\eps_2-\omega} \, \tilde \U^{-1}_{\eps_2-\omega,\eps_3}
    \sigma_z\tau_z ,
\\
 && B_j =
    \, \tilde \U_{\eps_3,\eps_4} u_{\eps_4} \gamma^j
    u_{\eps_4+\omega} \, \tilde \U^{-1}_{\eps_4+\omega,\eps_1}
    \sigma_z\tau_z ,
\eea
\end{mathletters}%
and all slow matrices $\U$ are taken at a coincident spacial point $\br$.
Performing integration over $\eps_1$ and $\eps_3$ in the first term
under the trace in Eq.~(\ref{Siaa1}), we obtain
\be
  \int \frac{d\eps_1\, d\eps_3}{(2\pi)^2} \sum_{i\neq j} \tr A_i B_j
  = 2 \tr
    \sigma_x
    (u \tilde \Q u)_{\eps_2-\omega,\eps_4}
    (u \tilde \Q u)_{\eps_4+\omega,\eps_2} ,
\ee
where a slow $\tilde\Q$ is defined as
$\tilde\Q = \, \tilde\U^{-1} \sigma_z\tau_z \, \tilde\U$.
In the second term under the trace in Eq.~(\ref{Siaa1}),
the matrices $\sigma_z\tau_z$ cancel with those in Eq.~(\ref{AB}),
integration over $\eps_1$ and $\eps_3$ is equivalent to multiplication
of $\U^{-1}$ and $\U$ that gives the unit matrix, and tracing with
$\sigma_x$ yields zero.

The resulting expression takes a simple form in the initial $Q$-representation
in the time domain. Integrating over $\bq$, one obtains (omitting the tilde
sign)
\be
\label{Siaa}
  \corr{(S_{\rm int}^a)^2}
  = - \frac{i\nu}{4g} \ln \frac{\Omega}{\Omega_*} \,
  \int d\br\, dt \tr \sigma_x Q_{tt}^2(\br).
\ee
The correction to the action $(i/2) \corr{(S_{\rm int}^a)^2}$ is the only one
in the standard gauge with $\roarrow K=0$. Physical quantities should not
depend on the choice of $\roarrow K$, which is a kind of a gauge
transformation upon the matrix $Q$ (cf.\ Eq.~(\ref{gauge})).
For this reason, it would be enough to use Eq.~(\ref{Siaa}) to find
the renormalization of the coupling constants in the effective action.
Below we present, however, the calculation of other terms
in Eq.~(\ref{Si}) related to $\roarrow K$, in order
to show that they really cancel each other.

The diagram for
\be
  \corr{(S_{\rm int}^b)^2} =
  \pi^2\nu^2D^2
  \Bigl\langle \bigl[
    \Tr
      \, \tilde \U \tau_z u \nabla \tensor{K}' u \, \tilde \U^{-1} \nabla \W'
  \bigr]^2 \Bigr\rangle
\ee
is shown in Fig.~\ref{F:d-Fin}b. It looks similar to the one
in Fig.~\ref{F:d-Fin}a, with the wavy line being replaced by the zigzag
line denoting $\corr{K'K'}$ correlator. Repeating the steps that lead
to Eq.~(\ref{Siaa1}) and using the large-$q$ asymptotics of (\ref{calV})
we obtain
\be
\label{Sibb1}
  \corr{(S_{\rm int}^b)^2}
  = \frac{i\pi}{4}
  \int \frac{d\br\, d\eps_1\, d\eps_2\, d\eps_3\, d\eps_4\, d\omega\, d\bq}
    {(2\pi)^7} \frac{1}{Dq^2}
  \sum_{i\neq j} \tr ( M_i N_j - M_i \sigma_z\tau_z N_j \sigma_z\tau_z ) ,
\ee
with
\begin{mathletters}
\label{MN}
\bea
 && M_i =
    \, \tilde \U_{\eps_1\eps_2} \tau_z u_{\eps_2} \gamma^i
    u_{\eps_2-\omega} \, \tilde \U^{-1}_{\eps_2-\omega,\eps_3} ,
\\
 && N_j =
    \, \tilde \U_{\eps_3,\eps_4} \tau_z u_{\eps_4} \gamma^j
    u_{\eps_4+\omega} \, \tilde \U^{-1}_{\eps_4+\omega,\eps_1} .
\eea
\end{mathletters}%
Integrating over $\bq$, $\eps_1$ and $\eps_3$ as described above, we get
\be
\label{Sibb}
  \corr{(S_{\rm int}^b)^2}
  = - \frac{i\nu}{4g} \ln \frac{\Omega}{\Omega_*} \,
  \int d\br\, dt \tr \sigma_x (\tau_z Q_{tt}(\br))^2.
\ee
In the same manner it can be shown that the average
$\corr{S_{\rm int}^aS_{\rm int}^b}=0$ vanishes.

In calculating the average $\corr{S_{\rm int}^c}$ shown diagrammatically
in Fig.~\ref{F:d-Fin}c, all $\Q$'s may be considered slow.
The analytical expression reads
\be
  \corr{S_{\rm int}^c} =
  -\frac{i\pi\nu D}{2}
  \Bigl\langle
    \Tr ( \tau_z \tilde Q \nabla \tensor{K}' )^2
  \Bigr\rangle ,
\ee
where the fast momentum runs over the $\corr{K'K'}$ propagator.
Calculating the corresponding logarithmic integral, one obtains
\be
\label{Sic}
  \corr{S_{\rm int}^c}
  = - \frac{\nu}{8g} \ln \frac{\Omega}{\Omega_*} \,
  \int d\br\, dt \tr \sigma_x (\tau_z Q_{tt}(\br))^2.
\ee

Substituting Eqs.~(\ref{Siaa}), (\ref{Sibb}) and (\ref{Sic})
into Eq.~(\ref{Si}), we see that the contributions from the vertices
describing interaction with the field $\roarrow K$ cancel each other,
and the resulting expression is given by
\be
\label{Si-res}
  \Delta S
  = \frac{\nu}{8g} \ln \frac{\Omega}{\Omega_*} \,
  \int d\br\, dt \tr \sigma_x Q_{tt}^2(\br)
  = \frac{\nu}{16g} \ln \frac{\Omega}{\Omega_*} \,
  \int d\br\, dt \sum_{j=0}^3
    \tr \sigma_x \tau_j Q_{tt}(\br) \cdot \tau_j Q_{tt}(\br).
\ee

Comparing with Eq.~(\ref{S_lambda}), we see that the terms with $j=x,y$
renormalize the Cooper channel coupling $\lambda$, whereas the terms
with $j=0,z$ contribute to the couplings in the diffusion channel.
The latter, $\Gamma$ and $\Gamma_2$, are not taken into account since
corrections to them are of the relative order of $g^{-1}\ln(1/\Omega\tau)$.
As a result, we get the Coulomb contribution to the RG equation for $\lambda$:
\be
  \frac{\partial\lambda}{\partial \zeta} = \frac{1}{4\pi^2 g} ,
\label{fincor}
\ee
which coincides with the Finkelstein's result~\cite{finkel1,finkel2}
in the limit $\Gamma$, $\Gamma_2\to 0$.

\subsection{Solution of the RG equation for $\lambda$ and shift of $T_c$}

Combining Eqs.~(\ref{bcs}) and (\ref{fincor}), we arrive at
the complete renormalization group equation for the Cooper-channel
interaction constant:
\be
  \frac{\partial\lambda}{\partial \zeta}
  = - \lambda^2 + \frac{1}{4\pi^2 g} .
\label{full}
\ee
In the high-energy range $1/\tau \leq E \leq E_F$ the second term
of Eq.~(\ref{full}) is absent. We will use the solution of Eq.~(\ref{bcs})
at $E \approx \tau^{-1}$ as an initial condition for the full equation
(\ref{full}).

The renormalization group equation (\ref{full}) possesses two fixed points,
$\pm\lambda_g$, where
\be
\label{lg}
  \lambda_g = \frac{1}{2\pi\sqrt{g}} .
\ee
The stable fixed point, $+\lambda_g$, is the limiting point of
the RG flow in the metallic region. A trajectory reaches its asymptotic
value $\lambda_g$ at the scale $\zeta\sim\sqrt{g}$.
The unstable fixed point, $-\lambda_g$, separates the regions
of metallic ($\lambda>-\lambda_g$) and superconducting
($\lambda<-\lambda_g$) states.
The solution of Eq.~(\ref{full}) is given by
\be
\label{lambda-res}
  \lambda(\zeta) =
    \frac{\lambda_0+\lambda_g\tanh\lambda_g \zeta}
      {\displaystyle 1+\frac{\lambda_0}{\lambda_g}\tanh\lambda_g \zeta} ,
\ee
where $\lambda_0$ is the bare value of the interaction constant
in the Cooper channel defined at the energy scale $\tau^{-1}$.

To study the superconductor-metal transition we will consider here
the case of an attractive interaction, $\lambda_0<0$.
The superconducting transition temperature $T_c$ is determined by the
position of the pole in $\lambda(\zeta)$, with $\zeta=\ln\frac{1}{T_c\tau}$.
In the clean system ($g\to \infty$),
\be
\label{Tc0}
  T_{c0}\tau = \exp \left( -\frac{1}{|\lambda_0|} \right) .
\ee
For a finite $\lambda_g$, the critical value of $\zeta$ can be easily
obtained from Eq.~(\ref{lambda-res}) and is given by
\be
  \zeta_c = \frac{1}{2\lambda_g} \ln \frac{|\lambda_0|+\lambda_g}{|\lambda_0|-\lambda_g} .
\ee
Consequently, we get for $T_c$:
\be
  T_c\tau =
    \left(
      \frac{|\lambda_0|-\lambda_g}{|\lambda_0|+\lambda_g}
    \right)^{\frac1{2\lambda_g}} .
\ee
Substitution of $\lambda$ in terms of $T_{c0}$ with the help
of Eq.~(\ref{Tc0}) leads to the final result for $T_c$ suppression
by disorder, which coincides (within our accuracy) with that of~\cite{finkel1}:
\be
  T_c\tau =
    \left(
      \frac{1-\frac{1}{2\pi\sqrt{g}}\ln\frac{1}{T_{c0}\tau}}
        {1+\frac{1}{2\pi\sqrt{g}}\ln\frac{1}{T_{c0}\tau}}
    \right)^{\pi\sqrt{g}} .
\label{Tcfin}
\ee

Evaluating expression (\ref{Tcfin}) for $g\gg1$,
we obtain a perturbative reduction of the transition
temperature~\cite{fuku,kuroda}:
\be
  \ln\frac{T_c}{T_{c0}} = - \frac{1}{12\pi^2 g} \ln^3 \frac{1}{T_{c0}\tau} ,
\ee
valid at large conductances.
The critical temperature becomes zero and the superconducting transition
disappears at the critical value of the dimensionless conductance
\be
  g_c = \left(\frac{1}{2\pi}\ln\frac{1}{T_{c0}\tau}\right)^2 .
\label{g_c}
\ee

Note once again, that the result (\ref{Tcfin}) for $T_c(g)$ dependence
is obtained neglecting weak-localization corrections to the conductance $g$,
as well as thermal, quantum and mesoscopic fluctuations.
This is correct provided that the renormalized conductance at the scale $T_c$
is still greater than unity: $g_* = g - (a/\pi^2) \ln \frac1{T_c\tau} \gg 1$,
where the constant $a$ is given by the sum of the usual weak localization
and interaction corrections, and is equal either to $1$ or $1/4$,
depending on the absence or presence of the spin-orbit interaction.

In the above derivation we neglected spin-dependent interactions ($\Gamma_2$
in the notations of~\cite{finkel2}) which may change the numerical coefficient
in Eq.~(\ref{fincor}). On the other hand, it was explained in~\cite{finkel2}
that strong spin-orbit scattering eliminates possible
effect of $\Gamma_2$ upon $T_c(g)$ dependence.

\section{Andreev conductance}
\label{S:Andreev}

\subsection{Tunneling term in the action}

In the previous sections, we considered a uniform 2D system.
In principle, spacial inhomogeneities in the local system's
characteristics such as the conductance and/or the Cooper channel
interaction $\lambda$ can be easily incorporated into the $\sigma$-model
action (\ref{sigma-model}) by a spacial dependence of the parameters
of the $\sigma$-model. Thus, the action (\ref{sigma-model}) is suitable
for a description of N-S interfaces or interfaces between metals with
different conductances. However since the solutions of the Usadel equation
(\ref{Usadel}) are continuous, only interfaces with perfect transmission
$T=1$ can be described in such a manner. In order to be able to deal with
the interfaces of arbitrary transparencies, one has to introduce
a boundary term into the action. Below in this paper we consider the case
of low-transparent interfaces which can be described by means of
the tunneling Hamiltonian approximation.
Then the boundary term in the action can be derived in the second order
over the tunneling Hamiltonian and reads
\be
  S_\gamma = \frac{i\pi}{4} \gamma \Tr\nolimits_\Gamma
  \check Q^{(1)} \check Q^{(2)} .
\label{S_gamma}
\ee
Here $\check Q^{(1)}$ and $\check Q^{(2)}$ refer to different sides
of the interface boundary $\Gamma$;
the notation $\Tr\nolimits_\Gamma$ means that the
space integral is taken over the interface surface,
and $\gamma$ is the (dimensionless)
normal-state tunneling conductance per unit area of the boundary.
In the expression (\ref{S_gamma}) the trace over the spin space
has already been performed.

Variation of the total action $S_{\rm tot} = S_\sigma + S_\gamma$, with
$S_\sigma$ given by Eq.~(\ref{sm1}), with respect to $\check Q$,
reproduces the matrix Usadel equations for $\check Q^{(1)}$
and $\check Q^{(2)}$ together with
the corresponding boundary conditions\cite{Kupriyanov,alt_sim_tar}
(cf.\ similar derivation in\cite{Oreg}):
\be
  g_1 \check Q^{(1)} \nabla_\perp \check Q^{(1)} =
  g_2 \check Q^{(2)} \nabla_\perp \check Q^{(2)} =
  \frac{\gamma}{2} [ \check Q^{(2)}, \check Q^{(1)} ] ,
\label{Usagran}
\ee
where $\nabla_\perp$ stands for the gradient along the normal to the interface
directed from the medium (1) to the medium (2) (Eq.~(\ref{Usagran}) is
written in the form assuming the absence of magnetic field).
It amounts to a straightforward
calculation to show that the action (\ref{S_gamma}) leads to the following
expression for the bare normal-state tunneling conductance $\sigma_T$:
\be
  \sigma_T = \frac{e^2}{\hbar}{\cal A}\gamma ,
\label{sigma_T}
\ee
where $\cal A$ is the area of the tunnel junction.
We omit here such a calculation since it is fairly similar to
the calculation of the Andreev subgap conductance presented in the
next subsection. Similarly, we will not dwell upon Coulomb
interaction-induced corrections to the tunneling
conductance~\cite{AA,LS} which can be derived from the
action (\ref{S_gamma}) by taking into account fluctuations of the fields
$Q$, $\phi$ and $K$ (cf.~\cite{Kam_Andr}). An analogous calculation
of the interaction effects in the Andreev conductance is one of our main
subjects below.

\subsection{Andreev conductance in the effective action formalism}
\label{SS:Andreev}

In this subsection we rederive, within the effective action formalism,
the well-known results for the subgap Andreev conductance $G_A$ between
a superconductor and a dirty normal metal (cf.\ e.~g.~\cite{Nazar_Hekk}).
We start from the simplest situation when $G_A$ does not depend on voltage
and/or frequency. First of all we show that if the effective action
contains the following term
\be
\label{SA1}
  S_A = \frac{i\pi}{16} G_A \Tr (Q_S \Lambda)^2 ,
\ee
then $G_A$ is indeed the dimensionless subgap conductance.
In the second step, we prove that the term of the form (\ref{SA1})
is generated in the second order of expansion over $S_\gamma$.

To begin with, we note that at low energies, $\eps\ll\Delta$, the
superconductive matrix Green function $Q_S$ does not depend on $\eps$
and reduces purely to phase rotations:
\be
\label{QS}
  Q_S = \frac{\tensor\Delta}{i|\Delta|} ,
\ee
where $\tensor\Delta$ can be obtained from Eq.~(\ref{Delta'}).
Therefore we can perform the trace in Eq.~(\ref{SA1})
over energy variables and the Keldysh matrix space,
using Eq.~(\ref{QS}) and the formula\cite{Kam_Andr}
\be
  \int dE \tr_K
    \left[
      \gamma^i \gamma^j -
      \gamma^i \Lambda_0(E+\omega/2) \gamma^j \Lambda_0(E-\omega/2)
    \right]
  = 4 \omega (\Pi^{-1}_\omega)^{ij} ,
\ee
where the matrix $\Pi^{-1}_\omega$ is defined in Eq.~(\ref{Pi}).
As a result, we obtain the effective action as a functional
of the order parameter:
\be
\label{SA2}
  S_A = - \frac{iG_A}{8|\Delta|^2}
    \tr_N \int \frac{d\omega}{2\pi}
      \roarrow{\Delta}^T(-\omega)
        \omega \Pi^{-1}_\omega
      \roarrow{\Delta}(\omega) .
\ee
On deriving this equation we added the constant term $Q_S^2=1$
under the trace in Eq.~(\ref{SA1}).
Employing Eq.~(\ref{Delta'}), one can rewrite (\ref{SA2})
in terms of the phase variables $\theta$ as
\bea
  S_A[\theta_i] = \frac{iG_A}{4}
    \int \frac{d\omega}{2\pi}
    \biggl\{
      i\omega \left[
        (e^{i\theta_1} \cos\theta_2)_{-\omega}
        (e^{-i\theta_1} \sin\theta_2)_{\omega}
      - (e^{-i\theta_1} \cos\theta_2)_{-\omega}
        (e^{i\theta_1} \sin\theta_2)_{\omega}
      \right]
\nonumber \\ {}
      + 2 \omega \coth \frac{\omega}{2T}
        (e^{i\theta_1} \sin\theta_2)_{-\omega}
        (e^{-i\theta_1} \sin\theta_2)_{\omega}
    \biggr\} .
\label{SA3}
\eea
The expression (\ref{SA3}) for the action makes it possible to relate
the coefficient $G_A$ in Eq.~(\ref{SA1}) with the Andreev conductance of
the N-S interface. For this purpose let us suppose that the superconducting
island is biased at some voltage $V(t)$.
Then the phase of the island will rotate with the speed $2eV$.
In the Keldysh formalism this corresponds to the rotation
of its classical component,
\be
\label{2eV}
  \frac{d\theta_1}{dt} = 2eV.
\ee
To find the Andreev current, $I = 2e \frac{dn}{dt}$,
where $n$ is the number of the Cooper pairs on the island,
one may use the fact that $n$ and $\theta$ are canonically
conjugated variables, and thus, $\hat{I} = 2ie \frac{\delta}{\delta\theta}$.
Translating this into the Keldysh formalism, we have,
similar to~\cite{Kam_Andr}:
\be
\label{I}
  \corr{I(t)}
    = ie \left. \frac{\delta \ln\Z}{\delta\theta_2(t)} \right|_{\theta_2=0}
    = ie \left. \frac{\delta \Z}{\delta\theta_2(t)} \right|_{\theta_2=0} ,
\ee
where the last equation follows from the normalization condition $\Z=1$
in the absence of quantum sources.

Transforming Eq.~(\ref{SA3}) with the help of Eq.~(\ref{2eV}), we get
\be
  S_A = - e G_A \int V(t) \theta_2(t) dt + o[\theta_2].
\ee
Substituting this action into Eq.~(\ref{I}) and taking
the functional derivative, we obtain
\be
  \corr{I(t)} = e^2 G_A V(t).
\ee
Hence, the Andreev conductance of the N-S interface
in conventional units is equal to
\be
\label{sigmaA}
\label{AAA}
  \sigma_A = \frac{e^2}{\hbar} G_A.
\ee
This completes the proof of the physical meaning of the term $S_A$,
Eq.~(\ref{SA1}), in the action. An alternative proof is presented
in Appendix A, where we calculate the voltage noise at the N-S barrier
(related to the conductance $\sigma_A$ due to
the fluctuation-dissipation relation).
Now we turn to the derivation of
such a term in the simplest geometry of N-S contact.

\subsection{Rectangular N-S contact} \label{SS:NSbar}

\subsubsection{Semiclassical solution within effective action method}
\label{SSS:NSbar}

\begin{figure}
\vspace{7mm}
\epsfxsize=130pt
\centerline{\epsfbox{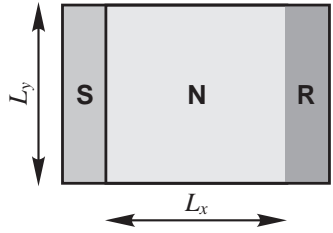}}
\vspace{-2mm}
\caption{%
Rectangular N-S contract.
Superconductor (S) is connected to a normal reservoir (R) via a dirty metal
film (N) of length $L_x$ and width $L_y$.
}
\label{F:NSbar}
\end{figure}

The simplest example of N-S contact is shown in Fig.~\ref{F:NSbar}.
Superconductor (S) is connected to a normal reservoir (R) via a thin film
of dirty metal (N) of length $L_x$ and width $L_y$.
The relevant energy scale in the N region
is determined by the Thouless energy $E_{Th} = D/L_x^2$.
As long as temperature, voltage and frequency of measurement are all
small enough, $\max(eV, T, \omega) \ll E_{Th}$,
the Andreev conductance is just a constant: $\sigma_A = \sigma_T^2 R_D$,
where $R_D = \sigma^{-1} L_x/L_y$ is the resistance of the N
region~\cite{Nazarov94}.  Below we first will show how to get such a result
(described by the term (\ref{SA1})) in the action)
within the effective action method, and later on we turn to its generalizations,
taking into account finite-energy effects as well as effects due to
the interaction in the Cooper channel and due to the zero-bias anomaly.

Consider the boundary action $S_\gamma$ defined in Eq.~(\ref{S_gamma})
in the case when $Q^{(1)}=Q_S$, $Q^{(2)}=Q$. In the present geometry
it is convenient to treat the boundary as a {\it line}
between N and S films, then $\gamma$ becomes the tunneling conductance
{\it per unit length}. To obtain the Andreev
term $S_A$, we need to expand the functional
integral for $\Z$ up to the second order over $S_\gamma$ and average
over fluctuations of the normal-metal matrix $Q$, i.~e.\ over diffusion and
Cooperon modes:
\bea
  S_A = \frac{i}{2} \corr{S_\gamma^2}
    = -\frac{i\pi^2\gamma^2}{32}
      \left\langle \left( \int_0^{L_y} dy \tr Q_S Q(0,y) \right)^2 \right\rangle
\label{SAd1}
\\
    = -\frac{i\pi^2\gamma^2}{32}
    \int_0^{L_y} dy_1 \int_0^{L_y} dy_2 \,
      \corr{\tr \Q_S \sigma_z \tau_z \W(0,y_1)
      \cdot \tr \Q_S \sigma_z \tau_z \W(0,y_2)}
\label{SAd2}
\eea
Proceeding from Eq.~(\ref{SAd1}) to Eq.~(\ref{SAd2}) we switched to the
``rotated'' representation of the $Q$-matrices defined in Eq.~(\ref{uQu})
and used the representation of $\Q$ in terms of the generators $\W$ as defined
in Eq.~(\ref{QW}). Note that we neglect here any possible fluctuations of
the superconductive matrix field $Q_S$.
The next step is to expand the field $\W(x,y)$ over the eigenfunctions
$\psi_{mn}(x,y)$ of the diffusion equation:
\be
  \W(x,y) = \sum_{m,n} \W_{mn} \psi_{mn}(x,y) .
\ee
The choice of the eigenfunctions is determined by the boundary conditions, which
are: vanishing of current at the edges $y=0$, $y= L_y$ and at the N-S
boundary $x = 0$ (the last condition is an approximation valid due
to smallness of $G_T R_D$), and vanishing of the Cooperon amplitude
at the boundary with the normal reservoir.
These conditions result in the following set of eigenfunctions:
\be
\label{eigenpsi}
  \psi_{mn}(x,y) = \frac{2 \cos q_mx \cos k_ny}{\sqrt{(1+\delta_{n,0})L_xL_y}},
  \qquad q_m=\frac{\pi}{L_x} \left( m+\frac12 \right),
  \quad  k_n=\frac{\pi}{L_y} n,
  \quad m,n=0,1,\dots
\ee
After the integration over $y$ in Eq.~(\ref{SAd2}), only zero mode $n=0$
survives.
Using Eq.~(\ref{WWcontr}) for the contraction rule and the fact that
$\{\Q_S,\tau_z\}=0$, we conclude that the diffusion
pairing in (\ref{WWcontr1}) does not contribute, whereas the Cooperon one
does, so we get
\bea
  S_A = \frac{i\pi(\gamma L_y)^2}{16\nu L_xL_y}
    \int \frac{d\eps_1 d\eps_2}{(2\pi)^2} \sum_m
    \frac{
        \tr ( \Q_S(\eps_1,\eps_2) \sigma_z \tau_z
              \Q_S(\eps_2,\eps_1) \sigma_z \tau_z
            - \Q_S(\eps_1,\eps_2) \Q_S(\eps_2,\eps_1) )
        (Dq_m^2 + i(\eps_1+\eps_2)\sigma_z)
    }{(Dq_m^2)^2+(\eps_1+\eps_2)^2}
\nonumber
\\
  = \frac{i\pi(\gamma L_y)^2}{16\nu L_xL_y}
  \int \frac{dE\, d\omega}{(2\pi)^2} \sum_m
    \frac{
      \tr [ Q_S(\omega) \Lambda(E_-)
            Q_S(-\omega) \Lambda(E_+)
          - Q_S(\omega) Q_S(-\omega) ]
        (Dq_m^2 + 2iE\Lambda_0(E_+))
    }{(Dq_m^2)^2+4E^2} ,
\label{SQ_square}
\eea
where $E_\pm=E\pm\frac{\omega}{2}$.
Using the property $Q_S^2=1$ one can verify that the contribution
of the term with $i E \Lambda_0(E_+)$ in the numerator behaves as
$\omega^2$ for small $\omega$. For this reason it can be neglected
compared to the $\omega$-proportional kernel in Eq.~(\ref{SA2}).
However, we will see below that the term of this structure determines the
amplitude of the Josephson proximity coupling.
Then, the term with $Dq_m^2 Q_S(\omega) Q_S(-\omega)$,
being integrated over $\omega$, produces an unimportant
constant that can be omitted. As a result, one has
\be
  S_A
  = \frac{i\pi(\gamma L_y)^2}{16\nu L_xL_y}
  \int \frac{dE\, d\omega}{(2\pi)^2} \sum_m
    \frac{ Dq_m^2
      \tr [ Q_S(\omega) \Lambda(E_-)
            Q_S(-\omega) \Lambda(E_+) ]
    }{(Dq_m^2)^2+4E^2} .
\ee

In the limit $E \ll E_{Th} = D/L_x^2$
the sum over $m$ can be easily calculated:
\be
  \sum_m \frac{1}{Dq_m^2}
  = \frac{L_x^2}{\pi^2D} \sum_m \frac{1}{\left( m+\frac12 \right)^2}
  = \frac{L_x^2}{2D} .
\label{sum_m}
\ee
The region of energies $E$ which are relevant
for the integral in Eq.~(\ref{SQ_square}) is given by
$E \sim \max(\omega,T)$. Thus, as long as
the condition ${\rm max}(\omega,T) \ll E_{Th}$
is fulfilled, we can, using Eq.~(\ref{sum_m}),
 convert (\ref{SQ_square}) into the foreseen expression of the form
(\ref{SA1}):
\be
  S = \frac{i\pi}{16} \frac{G_T^2}{g} \frac{L_x}{L_y} \Tr (Q_S\Lambda)^2,
\ee
where $G_T=\gamma L_y$ is the total normal-state conductance
of the interface, according to Eq.~(\ref{sigma_T}).
Now, comparing with Eq.~(\ref{SA1}), one finds
for the dimensionless Andreev conductance
\be
  G_A^{(0)} = \frac{G_T^2}{g} \frac{L_x}{L_y} .
\label{ga1}
\ee
The above relation coincides with the known~\cite{Nazarov94} relation
$\sigma_A = \sigma_T^2 R_D$, since in the present geometry
$R_D= \sigma^{-1}L_x/L_y$.

At higher temperatures or frequencies the $E$-dependence of the denominator
in Eq.~(\ref{SQ_square}) becomes important and the simple invariant
representation (\ref{SA1}) is not valid anymore. However, one can still
present the Andreev term in the action as a slight modification
of Eq.~(\ref{SA2}):
\be
\label{SA22}
  S_A = - \frac{i}{8|\Delta|^2}
    \tr_N \int \frac{d\omega}{2\pi}G_A(\omega)
      \roarrow{\Delta}^T(-\omega)
        \omega \Pi^{-1}_\omega
      \roarrow{\Delta}(\omega) .
\ee
The function $G_A(\omega)$ is given by
\be
  G_A^{(0)}(\omega) = \frac{G_T^2}{g} \frac{L_{\rm eff}(L_x,\omega,T)}{L_y} ,
\label{ga2}
\ee
where
\be
\label{Leff}
  L_{\rm eff}(L_x,\omega,T)
  = \frac{2D}{L_x \omega}
    \int_0^\infty dE \left[ \tanh\frac{E_+}{2T} - \tanh\frac{E_-}{2T} \right]
    \sum_m \frac{Dq_m^2}{(Dq_m^2)^2+4E^2} .
\ee
In the limit $\max(\omega,T) \ll E_{Th}$ the result $L_{\rm eff}=L_x$
is uncovered, whereas in the opposite case, $\max(\omega,T) \gg E_{Th}$,
one gets
\be
  L_{\rm eff}(\omega,T) = \frac{\sqrt{D}}{2\omega}
    \int_0^\infty \left[ \tanh\frac{E_+}{2T} - \tanh\frac{E_-}{2T} \right]
    \frac{dE}{\sqrt{E}}
  = \cases{
    \sqrt{2D/|\omega|}, & if $T\ll \omega$; \cr
    0.95\sqrt{D/2T},    & if $\omega\ll T$.
  }
\ee
The number $0.95$ in the above equation is the approximate
value for $(1-2^{-3/2})\pi^{-1/2}\zeta(3/2)$.

Representation (\ref{SA22}) can also be used in order to derive an
expression for the {\it nonlinear} Andreev current $I_A(V)$.
We present this derivation in Appendix B, the result is that
{\it dc} subgap current
$I_A(V) = V G_A(2eV)$ where the function $G_A(\omega)$ is defined
in Eq.~(\ref{ga2}). This relation between frequency-dependent linear
response and static nonlinear response is due to the fact that
static voltage $V$ applied to the superconductor leads to
the oscillation of its order parameter $|\Delta| e^{i\theta}$
with the frequency $2eV$.

To summarize this subsubsection, we emphasize that within the usual
semiclassical approximation the term describing Andreev conductance
is seen as a result of Gaussian integration over noninteracting
Cooperon modes in the N metal. Below we will
take into account Cooperon nonlinearities which are due to
interactions in the Cooper and direct Coulomb channels.
Basically where are two different kinds of the interaction effects:
the first one is due to the presence of electron-electron interaction in the
Cooper channel (which is itself renormalized in a way dependent
on the degree of disorder, as shown by Finkelstein), whereas the second one
is of the same nature as the zero-bias anomaly in the usual tunneling
conductance~\cite{AA}.  These two effects are intrinsically different,
as the first one is determined by diffusion modes with low frequencies
$\omega \ll Dq^2$ and does not depend on the long-range tail of the
Coulomb potential, whereas the second one comes from relatively
high-frequency fluctuations with $\omega \gg Dq^2$ and {\it does depend} on
the actual behaviour of the Coulomb potential at large space scales (and thus
it depends on the sample geometry as well). Below we start from the
study of the first (``Finkelstein's") effect and later on will include
the effect of the zero-bias anomaly.
The latter effects may indeed be
strongly suppressed if a good conductor is placed near a dirty metal
film, so that the Coulomb interaction in the film becomes screened on a
relatively short distance (cf.\ Eq.~(\ref{Vscreen}) and Refs.~\cite{AA,LS}).

\subsubsection{The effect of interactions in the Cooper channel}
\label{SSS:NSbarC}

In this subsection we will show that the Cooper-channel interaction
in the normal metal leads to a logarithmic (or weak power law)
dependence of the Andreev conductance on frequency.
To take it into account, one has to sum all inclusion of the
vertex $S_\lambda$ into the Cooperon propagator.
Such a summation can be effectively expressed as a renormalization
of the tunneling action (\ref{S_gamma}) due to
the presence of the Cooper repulsion $\lambda$.
As a result, the tunneling conductance $\gamma$ gets renormalized
down to $\gamma(\zeta)$ which should then be substituted into
Eq.~(\ref{ga2}) for $G_A$.

\begin{figure}
\vspace{3mm}
\epsfxsize=63mm
\centerline{\epsfbox{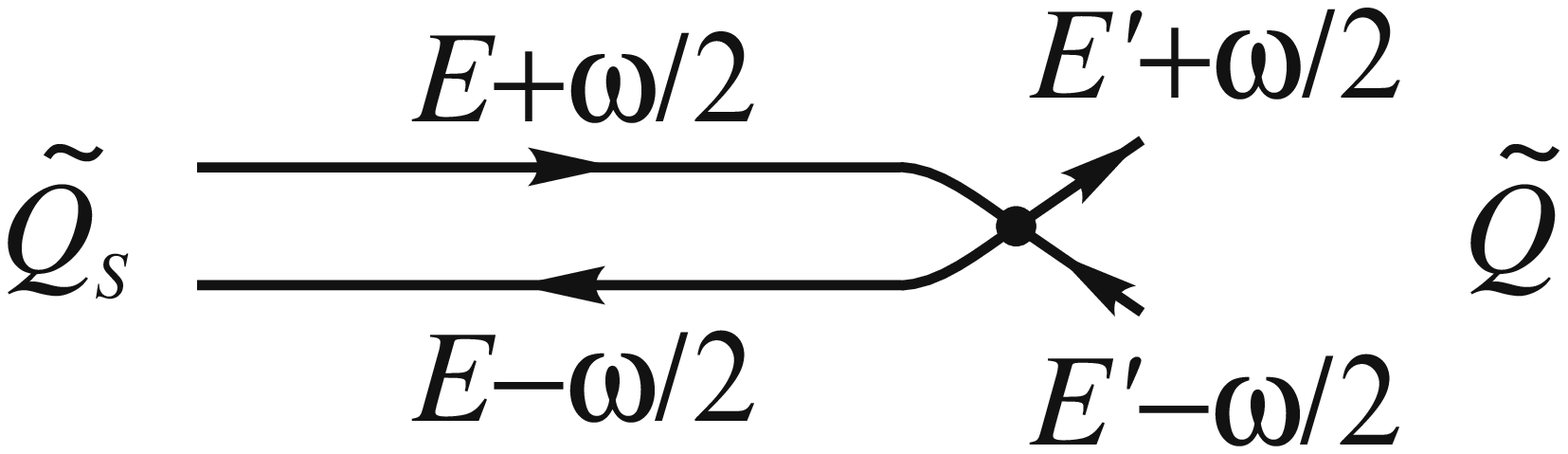}}
\vspace{-2mm}
\caption{Correction to the interface transparency $\gamma$ due to interaction
in the Cooper channel.}
\label{F:d-gamma}
\end{figure}

Logarithmic corrections to the boundary term (\ref{S_gamma}) originate
from pairing with the Cooper nonlinearity vertex $S_\lambda$ given by
Eq.~(\ref{S_lambda}):
\be
\label{gam0}
  \Delta S_\gamma = i \corr{S_\gamma S_\lambda} .
\ee
The diagram is shown in Fig.~\ref{F:d-gamma},
it contains one fast Cooperon mode connecting $S_\gamma$ and $S_\lambda$:
\be
  \corr{S_\gamma S_\lambda} =
  \frac{i\pi^3\nu\gamma\lambda}{16}
  \Bigl\langle
    \Tr\nolimits_\Gamma \tilde\Q_S \sigma_z\tau_z \W' \cdot
    \Tr \{ \sigma_x, \tilde Q \} u\sigma_z [\tau_z, \W']u
  \Bigr\rangle ,
\label{gam1}
\ee
where $\Q_S = u Q_S u$, according to the general rule (\ref{uQu}).

Consider first the spacial structure of Eq.~(\ref{gam1}).
The first trace corresponding to $S_\gamma$ is taken along
the N-S interface $\Gamma$, whereas the second one
corresponding to $S_\lambda$ is taken in the bulk of the N region
at a distance not larger than $\sqrt{D/\Omega_*}$ from the interface.
That is, from the point of view of slow variables,
both $S_\gamma$ and $S_\lambda$ are taken
at the same point on the boundary $\Gamma$.
In virtue of the  coordinate integration in $S_\lambda$,
the correlator $\corr{\W'\W'}$ must be taken at zero momentum,
so that the energy $E$ becomes the only fast variable involved.
Furthermore, due to the presence of the commutator of $\W'$ with $\tau_z$
under the second trace in Eq.~(\ref{gam1}), only Cooperon modes give
a nonzero contribution (cf.\ Eq.~(\ref{W})). Then, from the first trace
one concludes that the whole average (\ref{gam1}) does not vanish only
in the sub-gap region, $\Omega<\Delta$. In other words, this reflects
a trivial fact that the metallic tunneling conductance is not renormalized
by interaction with Cooperons. In the deep sub-gap limit, the superconductive
$Q_S$ is given by Eq.~(\ref{QS}), it is independent of the ``center-of-mass''
energy $E$ and depends on $\omega$ through the rotating phase $\theta(t)$.

Employing the Cooperon-related part of the contraction rule (\ref{WWcontr}),
one obtains
\be
\label{gam2}
  \corr{S_\gamma S_\lambda} =
  \frac{\pi^2\gamma\lambda}{32}
  \int_\Gamma d\br
  \int \frac{d\omega\, dE\, dE'}{(2\pi)^3} \frac{1}{E}
  \tr ( AB + A\sigma_zB\sigma_z -
    A\tau_zB\tau_z - A\sigma_z\tau_zB\sigma_z\tau_z ) \sigma_z ,
\ee
where $\Omega_*<E<\Omega$, whereas $\omega$, $E'<\Omega_*$,
and the matrices
\begin{mathletters}
\bea
 && A = u_E Q_S(\omega) u_E \sigma_z\tau_z , \\
 && B = u_E \{ \sigma_x, Q_{E'-\omega/2,E'+\omega/2} \} u_E \sigma_z\tau_z
\eea
\end{mathletters}%
are taken at the same point at the interface
(to logarithmic accuracy, all fast energies are equal to $E$).

Using the property $\{Q_S,\tau_z\}=0$ and the relation $\Lambda_0=u\sigma_zu$,
we transform Eq.~(\ref{gam2}) as
\be
  \corr{S_\gamma S_\lambda} =
  - \frac{\pi^2\gamma\lambda}{16}
  \int_\Gamma d\br
  \int \frac{d\omega\, dE\, dE'}{(2\pi)^3} \frac{1}{E}
    \tr \{ Q_S, \{ \sigma_x, Q \} \} \Lambda_0(E) .
\label{gam25}
\ee
The matrix $\Lambda_0(E)$ in Eq.~(\ref{gam25}) can be replaced by
$2\,\sigma_+ \tanh\frac{E}{2T} $, since contributions of its
 other components vanish by parity after integration over $E$.
Employing the relations $Q_S \propto 1_K$ and $\{\sigma_+,\sigma_x\}=1$,
one arrives at
\be
\label{gam3}
  \corr{S_\gamma S_\lambda} =
  - \frac{\pi\gamma\lambda}{4} \ln \frac{\Omega}{\Omega_*}
  \Tr\nolimits_\Gamma Q_S Q .
\ee
This term effectively renormalizes the interface transparency;
substituting Eq.~(\ref{gam3}) into Eq.~(\ref{gam0}),
we obtain the RG equation for $\gamma$:
\be
  \frac{\partial\gamma}{\partial \zeta} = - \gamma \lambda .
\label{gamma_rg}
\ee

A few comments are in order concerning this equation.
First of all, we remind that it is valid in the sub-gap region
$\zeta>\zeta_{\Delta}$, where $\zeta_{\Delta} = \ln1/\Delta\tau$;
for higher energies $\gamma$ remains unaffected.
In the derivation we assumed that both $\omega$ and $T$ are small
compared to the running cutoff $\Omega$, otherwise the RG should stop
at the scale $\max(\omega,T)$.

Eq.~(\ref{gamma_rg}) contains the running coupling constant
$\lambda(\zeta)$ which is determined by Eq.~(\ref{lambda-res})
with $\lambda_0 < 0$ replaced by $\lambda_n > -\lambda_g$,
a Cooper-channel repulsion in the N metal,
defined at the energy scale $\tau^{-1}$.
Looking at Eq.~(\ref{gamma_rg}) one can think that it is valid for
any geometry of the system since it arises as a result of integration
over energy. This is however not the case and the RG exists in 2D only.
The point is that the coupling constant $\lambda(\zeta)$ logarithmically
depends on the scale
in 2D case only (cf.\ section \ref{SS:Fin}). Recently a method to treat
$T_c$ suppression for dirty SC films in the 1D-2D crossover region was
developed~\cite{Oreg2}, which does not employ the existence of
 logarithmic RG equations.

Substituting $\lambda(\zeta)$ from Eq.~(\ref{full}),
we write down the solution of Eq.~(\ref{gamma_rg})
with the boundary condition $\gamma(\zeta_\Delta)=\gamma_0$:
\be
\label{gamma-res}
  \gamma(\zeta) =
    \frac{\gamma_0}
      {\displaystyle
      \left( 1+\frac{\lambda_n}{\lambda_g}\tanh\lambda_g (\zeta-\zeta_{\Delta})
 \right) \cosh\lambda_g (\zeta-\zeta_{\Delta})} .
\ee
Now this equation can be used to find the Andreev conductance
modified by the interaction in the Cooper channel.
In this regard we note that all integrals in Eq.~(\ref{SQ_square})
are not logarithmic. Consequently, with logarithmic accuracy,
it is sufficient to use the semiclassical expression for $G_A$, but with
the renormalized barrier transparency:
\be
\label{GAren}
  G_A = \frac{\gamma^2(\ln\frac1{\Omega\tau})}{\gamma^2_0} G_A^{(0)} ,
\ee
where the semiclassical value of $G_A^{(0)}$ is given by Eq.~(\ref{ga2})
and the low-energy RG cutoff
$\Omega = \max(D/L_x^2,\omega,eV,T,\tau_\phi^{-1})$,
with $\tau_\phi$ being the electron decoherence time in the N metal
(all these quantities are assumed to be much below $\Delta$).
Substituting $\gamma(\zeta)$ from Eq.~(\ref{gamma-res}) one obtains
\be
  G_A =
    \frac{G_T^2}{g} \, \frac{L_{\rm eff}(L_x,\omega,T)}{L_y} \,
    \frac{ 4 (\Omega/\Delta)^{2\lambda_g} }
      { \left[ (1 + \lambda_n/\lambda_g)
      + (1 - \lambda_n/\lambda_g) (\Omega/\Delta)^{2\lambda_g} \right]^2 }
  ,
\label{GaFin}
\ee
where $\lambda_g$ is defined in Eq.~(\ref{lg}).
This equation is valid provided that the problem is effectively
two-dimensional, cf.\ discussion after Eq.~(\ref{gamma_rg}).
This condition is satisfied as long as $L_\Omega\leq L_y$,
where $L_\Omega=\sqrt{D/\Omega}$.
According to Eq.~(\ref{GaFin}), in the low-$\Omega$ limit
the Andreev conductance acquires an anomalous power-law suppression
with an exponent $2\lambda_g$.
The total power-law exponent, describing growth
of $G_A (\Omega)$ with the $\Omega$ decrease, is equal to
\be
  x_A = \frac{1}{2} - 2\lambda_g = \frac{1}{2}
    \left( 1 - \sqrt{\frac{4e^2}{\pi^2\hbar} R_\square } \ \right) .
\label{x_A}
\ee

\subsection{Small SC island in contact with 2D film} \label{SS:NSisland}

\begin{figure}
\vspace{8mm}
\epsfxsize=107pt
\centerline{\epsfbox{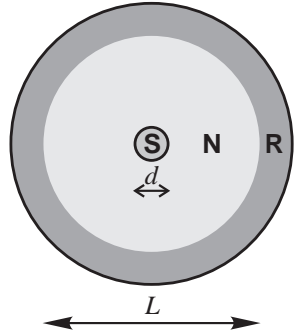}}
\vspace{-4mm}
\caption{%
Small superconductive island (S) of size $d$ connected to
a reservoir (R) through a
dirty normal film (N) of size $L \gg d$.
}
\label{F:NSisland}
\end{figure}

Here we consider another example of a N-S contact, namely, a contact
of a small superconductive island of size $d$
with a thin dirty normal film of linear size $L \gg d$,
shown in Fig.~\ref{F:NSisland}.
The edge of the film is connected to a normal reservoir.
The semiclassical result for the Andreev conductance in this geometry
can be inferred from the general relation $\sigma_A^{\rm cl} = \sigma_T^2R_D$,
where the metal resistance $R_D$ is now a ``spreading" resistance
which grows logarithmically with the film size:
$R_D = \frac{1}{2\pi\sigma}\ln\frac{L}{d} $.
If temperature or frequency is larger than the Thouless energy
scale $D/L^2$, the shortest of the effective lengths
$L_{T,\omega} = \sqrt\frac{D}{T,\omega}$ should be used instead of $L$.
Logarithmic growth of the Andreev conductance with the space scale
implies that $G_A$ itself becomes a running coupling constant
subject to the RG procedure. Below we will derive and solve the RG equation
for the Andreev conductance in the presence of the Cooper interaction
renormalized by the Coulomb repulsion (still we will not touch here
the effect of the tunneling DOS suppression; it will be considered below,
in the next subsection).

\begin{figure}
\vspace{-3mm}
\epsfxsize=63mm
\centerline{\epsfbox{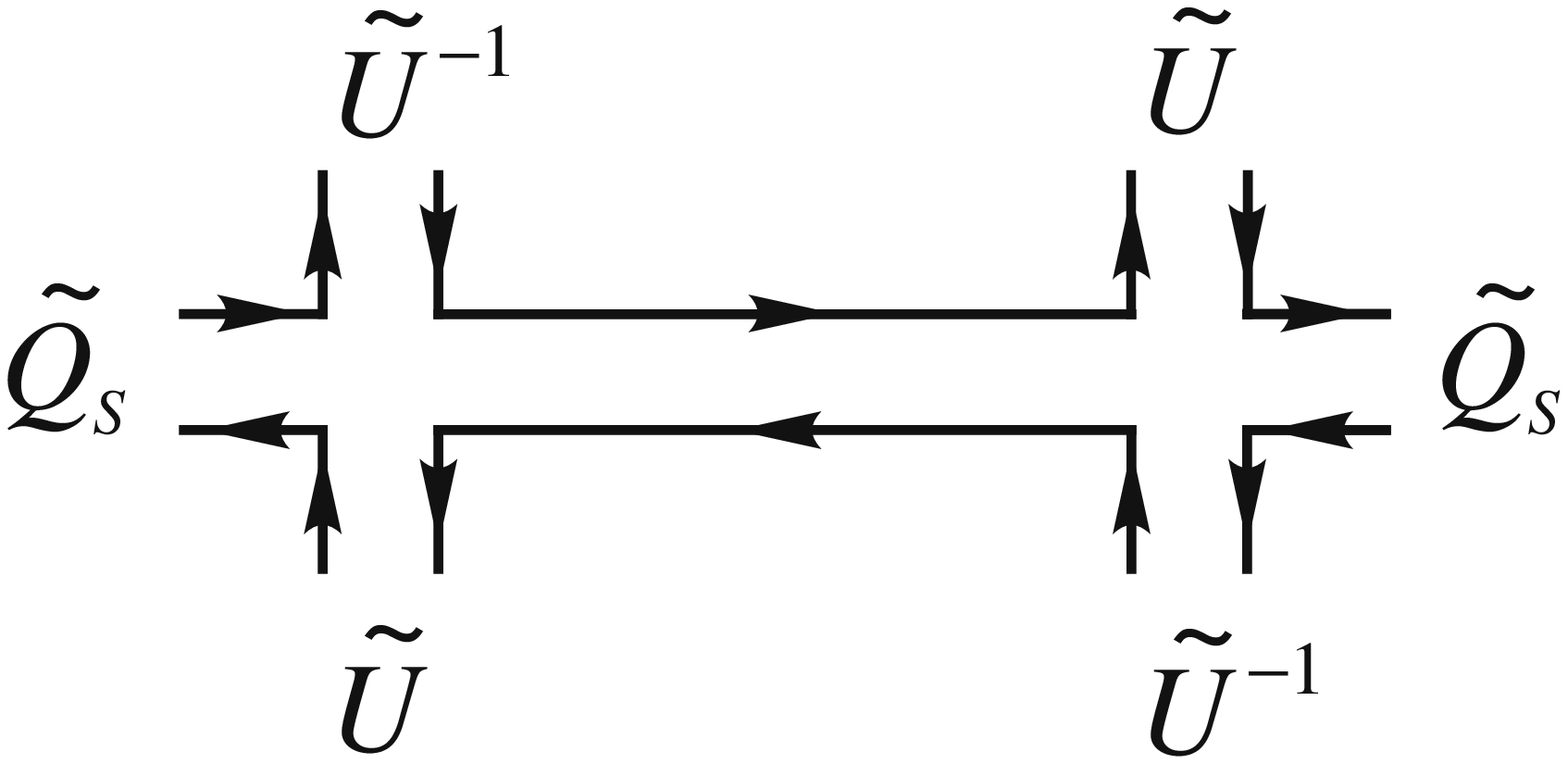}}
\vspace{-3mm}
\caption{Correction to the Andreev term $S_A$}
\label{F:d-GA}
\end{figure}

Similarly to Eq.~(\ref{SAd1}), the correction to the Andreev term (\ref{SA1})
in the action comes from the averaging of two boundary vertices
$S_\gamma$:
\be
  \Delta S_A = \frac{i}{2} \corr{S_\gamma^2} .
\label{gar1}
\ee
At scales larger than its size, $d$, the SC island can be considered
as a point object. Then the boundary term (\ref{S_gamma}) can be written
in a local form: $S_\gamma = \frac{i\pi}{4} G_T \Tr Q_S Q(0)$,
with $G_T = {\cal A} \gamma$, where ${\cal A}$ is the
area of the contact between the island and the film.
The relevant pairing in Eq.~(\ref{gar1}),
\be
  \corr{S_\gamma^2} =
  - \frac{\pi^2G_T^2}{16}
  \Bigl\langle \bigl[
    \Tr
      \, \tilde \U \Q_S \, \tilde \U^{-1}
      \sigma_z\tau_z \W'
  \bigr]^2 \Bigr\rangle ,
\label{gar2}
\ee
is shown diagrammatically in Fig.~\ref{F:d-GA}.
Here all energies are coupled to the slow matrices $\,\tilde\U$,
and the internal Cooperon momentum is the only fast variable.
Integrating out fast modes with the help of the contraction rule
(\ref{WWcontr1}), one obtains
\be
\label{gar3}
  \corr{S_\gamma^2} =
  \frac{\pi G_T^2}{16\nu}
  \int \frac{d\eps_1 d\eps_2\, d\bq}{(2\pi)^4} \frac{1}{Dq^2}
  \Tr ( AB - A\sigma_z\tau_zB\sigma_z\tau_z ) ,
\ee
where (all fields are taken at the point $\br=0$)
\begin{mathletters}
\bea
 && A = (\, \tilde \U \Q_S \, \tilde \U^{-1})_{\eps_1,\eps_2} \sigma_z\tau_z ,
\\
 && B = (\, \tilde \U \Q_S \, \tilde \U^{-1})_{\eps_2,\eps_1} \sigma_z\tau_z .
\eea
\end{mathletters}%
Only the first term under the trace in Eq.~(\ref{gar3}) is important.
Evaluating the logarithmic integral over $\bq$, one finds
\be
  \corr{S_\gamma^2}
  = \frac{G_T^2}{8g} \ln\frac{\Omega}{\Omega_*}
  \Tr (\Q_S \Q)^2 .
\label{SA2d}
\ee
Eq.~(\ref{SA2d}) is valid at space scales $L$ larger than the size
of the island, $d$, i.~e. for energies $\Omega \leq \omega_d = D/d^2$.
It shows that the RG procedure generates the term in the action of the form
\be
\label{SA'}
  S_A = \frac{i\pi}{16} G_A \Tr (Q_S Q)^2 ,
\ee
which reduces to Eq.~(\ref{SA1}) when all fast modes of $Q$ are
integrated out and $Q$ is replaced by $\Lambda$.
The running constant $G_A$ obeys the following RG equation:
\be
  \frac{\partial G_A}{\partial \zeta} = \frac{G_T^2}{4\pi g} ,
\label{GAdot}
\ee
valid for $\zeta > \zeta_d = \ln 1/\omega_d\tau$.

In the semiclassical limit, $\gamma\equiv G_T/{\cal A}$ is constant
and integrating Eq.~(\ref{GAdot}) over $\zeta = 2\ln(L/l)$ one reveals
the above-mentioned result, $G_A^{\rm cl} = \frac{G_T^2}{2\pi g} \ln \frac Ld$.
Beyond semiclassics, Eq.~(\ref{GAdot}) becomes nontrivial since
fluctuations which renormalize the tunneling conductance
$\gamma(\zeta)$ must be taken into account.
Below we will assume that $\omega_d \sim \Delta$ and thereby neglect
small corrections of the order of $\lambda_g|\zeta_{\Delta}-\zeta_d|$.
Then, substituting $\gamma(\zeta)$ from Eq.~(\ref{gamma-res})
and integrating Eq.~(\ref{GAdot}) over $\zeta$ between $\zeta_d$
and $\ln\frac{1}{\Omega\tau}$, we obtain the result for $G_A(\Omega)$:
\be
  G_A(\Omega) =
    \frac{G_T^2}{4\pi g} \,
    \frac{1-(\Omega/\Delta)^{2\lambda_g}}
      { (\lambda_g + \lambda_n)
      + (\lambda_g - \lambda_n) (\Omega/\Delta)^{2\lambda_g} } .
\label{GAdot-res1}
\ee
A general feature of the expression (\ref{GAdot-res1})
is that $G_A(\Omega)$ approaches a constant value,
$\frac{G_T^2}{4\pi g} \frac{1}{\lambda_n+\lambda_g}$,
in the low-$\Omega$ limit.
The semiclassical expression~\cite{Nazar_Hekk} for the Andreev conductance,
$G_A^{\rm cl}(\Omega) = \frac{G_T^2}{4\pi g} \ln \frac{\omega_d}{\Omega}$,
predicts growth of $G_A(\Omega)$ with the increase of the relevant
space scale $L_{\Omega} = \sqrt{D/\Omega}$, due to the growth of the region
where electron and Andreev-reflected hole interfere constructively.
This expression for $G_A^{\rm cl}(\Omega)$ follows from Eq.~(\ref{GAdot-res1})
in the limit $\lambda_n \to 0$ and $\ln\frac{\Delta}{\Omega} \ll \sqrt{g}$.
Our result (\ref{GAdot-res1}) demonstrates, that
(in the present geometry when the current flow is effectively 2-dimensional)
the Cooper-channel repulsion acts as if it imposes an upper limit
for the time duration of such a constructive interference.
The same qualitative behaviour would be seen in the absence of
Finkelstein's corrections as well: taking first the limit
$g \gg \ln^2\frac{\Delta}{\Omega}$ in Eq.~(\ref{GAdot-res1}), one would find
$G_A \approx G_T^2/4\pi g\lambda_n$ as $\Omega \to 0$
(cf.\ Ref.~\cite{FeigLar} where a similar expression was used for $G_A$;
note however that a numerical mistake was made in~\cite{FeigLar}, which
lead to the overestimation of $G_A$ by a factor of 2).
Below we will see that the effect of the Coulomb-blockade
suppression of tunneling is even more drastic,
as it leads to the decrease of the Andreev
conductance at $\Omega\to 0$.

\subsection{Effect of the zero-bias anomaly} \label{SS:ZBA}

\subsubsection{General treatment} \label{SSS:ZBA0}

In the preceding consideration of the Andreev conductivity we neglected
the effect of high frequency ($Dq^2\ll\omega$) Coulomb fluctuations.
They are responsible for the zero-bias anomaly (ZBA) in the usual tunneling
conductance (calculated originally by Altshuler and Aronov~\cite{AA}),
which, in 2D, leads to a suppression of tunneling by
the relative order of $g^{-1}\ln^21/\Omega\tau$.
Later it was argued by Finkelstein~\cite{finkel2} and shown (within
semiclassical approach) by Levitov and Shytov~\cite{LS}
that the first correction must be exponentiated to get a result
valid in the low-frequency limit as well.
Recently, Kamenev and Andreev \cite{Kam_Andr} rederived this result
microscopically using the Keldysh $\sigma$-model approach
and separating the Coulomb phase $\roarrow{K}$ according
to Eq.~(\ref{gauge}).

After the gauge transformation (\ref{gauge}), the boundary tunneling
term (\ref{S_gamma}) acquires the form
\be
  S_\gamma = \frac{i\pi}{4} \gamma \Tr\nolimits_\Gamma
  e^{i\tensor K_{12}\tau_z} Q^{(1)} e^{-i\tensor K_{12}\tau_z} Q^{(2)} .
\label{S_gamma2}
\ee
where $\tensor K_{12} = \tensor K^{(1)} - \tensor K^{(2)}$
is the Coulomb phase difference across the interface.
In our study of the Andreev conductance, we will assume that
the superconductor is connected to a low-impedance external environment
which keeps fixed its average electric potential,
and neglect the Coulomb phase $\tensor K_S$ in the superconductor
for the reasons explained in the last paragraph of Sec.~\ref{SS:emf}.
Hence, the ZBA is determined by the fluctuations of the normal-metal
phase $\tensor K$.

On singling out the Coulomb phase by the transformation (\ref{gauge}),
the calculation of the ZBA becomes very simple~\cite{Kam_Andr}:
one has just to average the phase factors $e^{\pm i\tensor K}$
in the boundary term (\ref{S_gamma2}) over the Gaussian fluctuations
of $\roarrow K$ with the correlator (\ref{K-corr}).
Indeed, $\corr{K_iK_j}$, being integrated over frequency and momentum,
yields the aforementioned $g^{-1}\ln^21/\Omega\tau$.
On the other hand, since the bulk action (\ref{sm2}) depends only on
derivatives of $\roarrow K$, any $\corr{K_iK_j}$ pairing in the bulk
contains an additional power of frequency or momentum, its contribution
is less singular than $\ln^21/\Omega\tau$ in the limit $\Omega\to0$,
and therefore can be neglected.

\begin{figure}
\vspace{5mm}
\epsfxsize=57mm
\centerline{\epsfbox{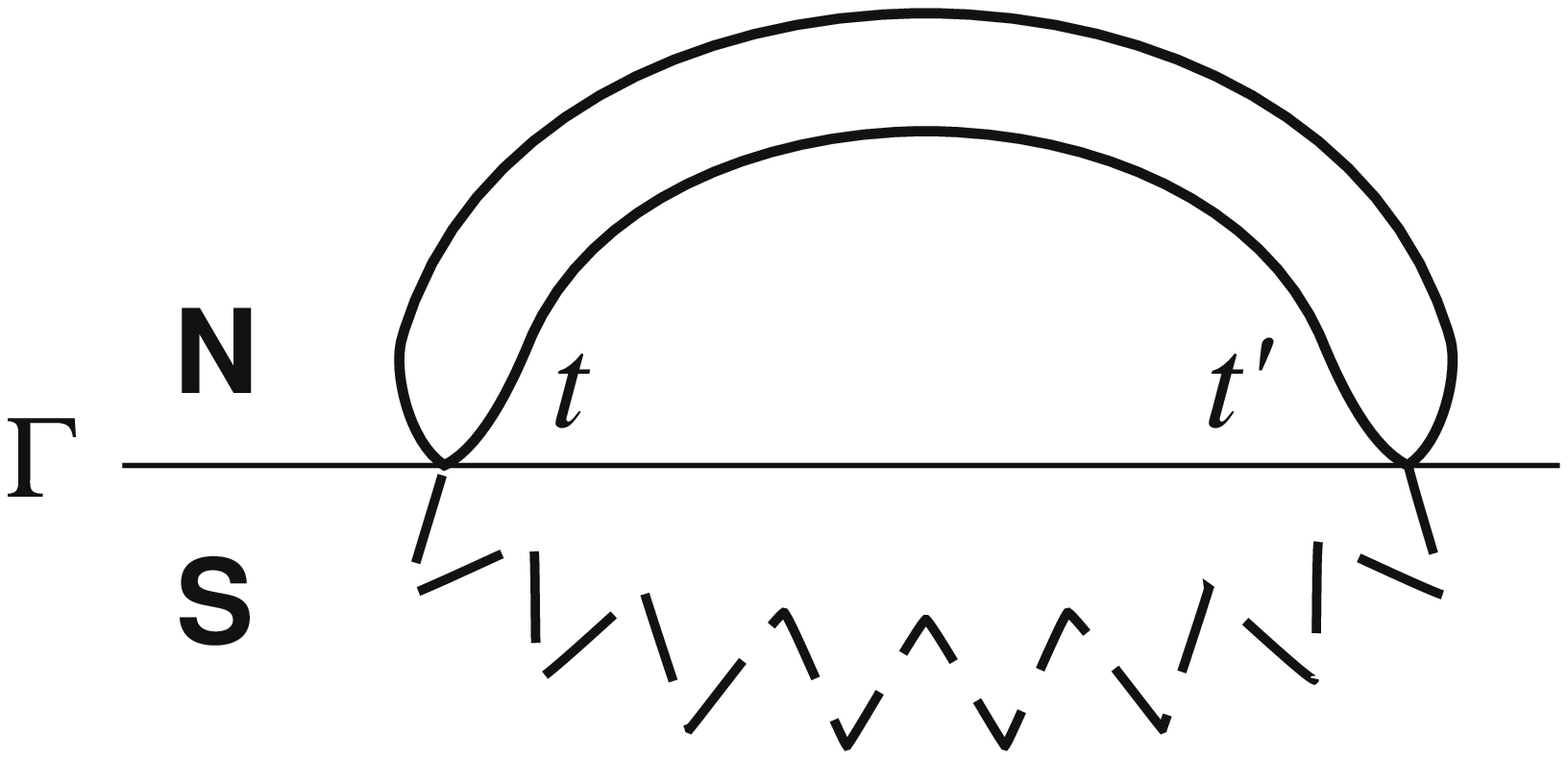}}
\caption{%
An example of factorization of the ZBA-type fluctuations for the Andreev
conductance. The dashed zigzag line denotes the correlator
$\corr{(e^{iK(t)\tau_z})_i (e^{iK(t')\tau_z})_j}$.
}
\label{F:factor}
\end{figure}

{}From this general observation it follows that in calculating the
Andreev conductance the effect of the interaction in the Cooper channel
is factorized from the ZBA effect. In other words, the full Cooperon
which determines the subgap conductance is a product (in time domain)
of the Cooperon without the ZBA (which was studied in subsections
\ref{SS:NSbar} and \ref{SS:NSisland}) and the ZBA factor $\ZBA^2(t)$
(cf.\ Eq.~(\ref{JZC}) below). Such a factorization is shown
diagrammatically in Fig.~\ref{F:factor}.
Our previous analysis implied $\ZBA(t)=1$, i.~e.\ weak ZBA at the scale
$\Omega \sim \tau^{-1} e^{2\pi\sqrt g}$ relevant for the saturation
of the RG flow for the Cooper repulsion constant $\lambda$,
cf.\ Eq.~(\ref{lambda-res}).
We will use this factorization property explicitly in the study
of the Josephson proximity coupling in Sec.~\ref{S:Josephson}.
In the calculation of the Andreev conductivity we will rather
follow another approach:
the effect of the ZBA will be taken into account within the RG scheme
together with the Cooper channel renormalization described above.

\subsubsection{Rectangular N-S contact} \label{SSS:Zbar}

High-frequency ZBA-type fluctuations of the Coulomb phase $\roarrow K$
in Eq.~(\ref{S_gamma2}) lead to an additional suppression of the effective
transparency $\gamma$ of the interface.
This effect is different in the metallic ($\Delta<\Omega<\tau^{-1}$)
and superconducting ($\Omega<\Delta$) regions.
In the metallic case, the ZBA is governed by a single-electron tunneling,
which is the only source of the transparency suppression,
whereas in the superconducting case, the reduction of $\gamma$
becomes more pronounced due to a coherent tunneling of Cooper pairs,
carrying the double charge.

The renormalized value of $\gamma$ can be extracted from the boundary action
(\ref{S_gamma2}) after averaging over fast fluctuations of the Coulomb
phase $\roarrow K'$ in the metal:
\be
  \corr{S_\gamma} = \frac{i\pi}{4} \gamma \Tr\nolimits_\Gamma
    \corr{ e^{-iK'_1(t_1)\tau_z} \tilde Q_S(t_1,t_2)
    e^{ iK'_1(t_2)\tau_z} } \tilde Q(t_2,t_1) ,
\label{Sg1}
\ee
where we neglected the Coulomb phase $\roarrow K_S$
in the superconductor as discussed above
and used the fact that only the correlator of classical components of the
phase $\roarrow K$ may be retained with logarithmic accuracy\cite{Kam_Andr}.
On eliminating fast degrees of freedom, the term $S_\gamma$ will
reproduce itself with the modified value of the transparency $\gamma$
(cf.~\cite{Oreg}).
Further transformation of Eq.~(\ref{Sg1}) depends on the structure of the
matrix $Q_S$ in the Nambu and time spaces and on the geometry of the
interface.

In the metallic energy range, $\Delta<\Omega$, the off-diagonal in the Nambu
space components of the superconductive $Q_S$ can be neglected
(cf.\ Eq.~(\ref{Q-SC})), so that it commutes with the phase factor
$e^{iK_1(t)\tau_z}$. Then the perturbative correction to $\gamma$
can be deduced from Eq.~(\ref{Sg1}):
\be
  \Delta\gamma = \gamma
    \left[ e^{-\frac12\corr{[K'_1(t_1)-K'_1(t_2)]^2}} - 1 \right] .
\label{dg1}
\ee
The fast phases $K_1'$ are correlated only at small times $\leq\Omega_*^{-1}$,
while the time difference in Eq.~(\ref{dg1}) is large enough,
$|t_1-t_2|\geq\Omega_*$, due to an implicit $\eps$-dependence of $Q_S$
at the metallic region.
Therefore, the cross average in Eq.~(\ref{dg1}) can be neglected,
and expanding the exponent one arrives at
\be
  \Delta\gamma = - \gamma \, \corr{[K'_1(0)]^2} .
\label{dg2}
\ee

The value of the correlator $\corr{[K'_1(0)]^2}$ is not universal
and depends on the setup considered. For the tunneling into the whole
2D plane (that corresponds to the geometry of a small SC island shown
in Fig.~\ref{F:NSisland}) one can employ Eq.~(\ref{K-corr}):
\be
  \corr{[K'_1(0)]^2}_{\rm plane}
  = \int_{\Omega_*}^{\Omega} \frac{d\omega}{\pi}
    \int \frac{d\bq}{(2\pi)^2}
      \Im \frac{V_0(q)}{(Dq^2-i\omega) (2\nu V_0(q)Dq^2+Dq^2-i\omega)} .
\ee
The leading double-log contribution to this expression comes
from the region $Dq^2 \ll \omega \ll \nu V_0(q)Dq^2$
where the integrand is given by $1/(g\omega q^2)$.
Note that in this limit of strong Coulomb interaction,
specific form of the potential $V_0(q)$ drops from
the integrand and enters the result only through the cutoff
of logarithmic $q$-integration.
For the case of the bare 2D Coulomb potential (\ref{V0}),
one obtains
\be
\label{KK-RG-plane}
  \corr{[K'_1(0)]^2}_{\rm plane}
  = \frac{1}{2\pi^2g}
    \int_{\Omega_*}^{\Omega} \frac{d\omega}{\omega}
    \int_{\omega/2\pi\sigma}^{\sqrt{\omega/D}} \frac{dq}{q}
  = \frac{1}{8\pi^2g}
    \left(
      \ln^2\frac{\omega_0}{\Omega_*} -
      \ln^2\frac{\omega_0}{\Omega}
    \right) ,
\ee
where $\omega_0 = (2\pi\sigma)^2/D$.

In the rectangular geometry of Fig.~\ref{F:NSbar},
a charge tunnels into the edge of the half-plane.
A similar problem for the half-space was considered in Ref.~\cite{AAZ}.
Though a complete treatment of such problems is involved,
a double-log asymptotics can be easily derived.
To find it, one should use the fact that in the relevant
region, $Dq^2 \ll \omega \ll 2\pi\sigma q$, the screened Coulomb
interaction is determined solely by the inverse density-density
correlator\cite{AAZ} whereas the bare Coulomb potential drops
from equations as we have seen above. The density-density correlator
can be obtained with the help of the eigenfunctions of the diffusion
equation with the proper boundary conditions. For the rectangular
geometry such eigenfunctions are given by Eq.~(\ref{eigenpsi}).
Then the value of $\corr{[K'_1(0)]^2}$ can be obtained analogously
to Eq.~(\ref{KK-RG-plane}); it depends on the distance $r$ from the
edge of the half-plane and is given by
\be
\label{KK-RG-half}
  \corr{[K'_1(0)]^2}_{\mbox{\scriptsize half-plane}}
  = \frac{1}{\pi^2g}
    \int_{\Omega_*}^{\Omega} \frac{d\omega}{\omega}
    \int_{\omega/2\pi\sigma}^{\sqrt{\omega/D}}
      \frac{d\bq}{2\pi q^2} \cos^2(q_xr) ,
\ee
where the infinite half-plane ($L_x$, $L_y\to\infty$) is implied.
For large $r$, the cosine squared in Eq.~(\ref{KK-RG-half}) can be
substituted by its average value, 1/2, that leads to
the infinite-plane result (\ref{KK-RG-plane}).
For the tunneling directly into the edge ($d=r$) the correlator
(\ref{KK-RG-half}) appears to be two times larger.

Hence the RG equation for the transparency $\gamma$ of a flat
N-S boundary in the metallic energy range ($\zeta<\zeta_\Delta$) reads
\be
\label{gamma_rg1}
  \frac{\partial\gamma}{\partial\zeta} =
    - \frac{\zeta-\zeta_0}{2\pi^2g} \gamma ,
\ee
where $\zeta_0$ is defined at the scale $\omega_0$,
$\zeta_0 = \ln 1/\omega_0\tau$.
Being integrated over $\zeta$ from $0$ to $\ln\frac{1}{\Omega\tau}$,
it yields exponential reduction of the tunneling
conductance (which is two times stronger than the result
\cite{finkel2,LS,Kam_Andr} for the infinite plane):
\be
\label{gamma-res1}
  \gamma(\zeta) = \gamma_0 \ZBA_M(\zeta),
\ee
where the factor $\ZBA_M(\zeta)$ is given by
\be
\label{ZM}
  \ZBA_M(\zeta)
  = \exp \left( - \frac{(\zeta-\zeta_0)^2-\zeta_0^2}{4\pi^2g} \right)
  = \exp \left( - \frac{1}{4\pi^2g}
      \ln\frac{1}{\Omega\tau} \cdot
      \ln\frac{\omega_0^2\tau}{\Omega}
  \right) .
\ee

When the RG cutoff $\Omega$ becomes smaller than $\Delta$,
situation changes. First of all, $Q_S$ acquires an $\eps$-independent
form (\ref{QS}), i.~e.\ it becomes a $\delta$-function in time domain,
$t_1=t_2$. It also anticommutes with $e^{iK_1(t)\tau_z}$ due to its
structure in the Nambu space.
 In this limit, the Coulomb phase $\roarrow K$ can be considered
as an additive correction to the SC phase $\roarrow\theta$ on the island,
with the total effective phase
$\roarrow\theta_{\rm eff} = \roarrow\theta - 2\roarrow K$.
Then the effect of the ZBA in the sub-gap limit can be attributed
to the high-frequency fluctuations of $\roarrow\theta_{\rm eff}(t)$,
destroying the Cooper pair coherence at large time scales.
Repeating the steps that lead to Eqs.~(\ref{dg1}) and (\ref{dg2}),
one has instead
\be
  \Delta\gamma = \gamma
    \left[ e^{-\frac12\corr{[2K'_1(t)]^2}} - 1 \right]
  = - 2 \gamma \, \corr{[K'_1(0)]^2} .
\label{dg3}
\ee

The effect of the ZBA anomaly should be taken into account together with
the renormalization of the transparency due to interactions in the Cooper
channel, cf.\ Eq.~(\ref{gamma_rg}).
Thus, we get the following RG equation for $\gamma$ valid in the sub-gap
limit $\zeta>\zeta_\Delta$:
\be
\label{gamma_rg2}
  \frac{\partial\gamma}{\partial\zeta}
  = - \lambda \gamma
    - \frac{\zeta-\zeta_0}{\pi^2g} \gamma .
\ee
Solving this equation for $\zeta>\zeta_\Delta$,
we obtain a modification of Eq.~(\ref{gamma-res}):
\be
\label{gamma-res2}
  \gamma(\zeta) =
    \frac{\gamma_\Delta \ZBA_S(\zeta)}
      {\displaystyle
      \left(
        1+\frac{\lambda_n}{\lambda_g}\tanh\lambda_g (\zeta-\zeta_\Delta)
      \right)
      \cosh\lambda_g(\zeta-\zeta_\Delta)} ,
\ee
where $\gamma_\Delta \equiv \gamma(\zeta_\Delta) = \gamma_0 \ZBA_M(\zeta_\Delta)$
is the transparency renormalized by the ZBA
in the metallic energy region $\Delta<\Omega<\tau^{-1}$,
and the multiplicative factor $\ZBA_S(\zeta)$
accounts for the sub-gap ZBA effect:
\be
\label{ZS}
  \ZBA_S(\zeta)
  = \exp \left(
      - \frac{(\zeta-\zeta_0)^2-(\zeta_\Delta-\zeta_0)^2}{2\pi^2g}
    \right)
  = \exp \left( - \frac{1}{2\pi^2g}
      \ln\frac{\Delta}{\Omega} \cdot
      \ln\frac{\omega_0^2}{\Delta\Omega}
\right) .
\ee

So far we considered an infinite half-plane with $L_x$, $L_y\to\infty$.
In analogy with the treatment in section \ref{SSS:NSbarC},
it might be tempting to substitute the renormalized transparency
$\gamma(\ln\frac{1}{\Omega\tau})$ given by Eq.~(\ref{gamma-res2})
into Eq.~(\ref{ga2}) to get the Andreev conductance $G_A(\Omega)$
in the rectangular geometry. This however would be wrong.
The point is that the expression (\ref{KK-RG-half}) for
$\corr{[K'(0)]^2}$ contains two coupled integrations over
$\omega$ and $q$, with {\em two} frequency-dependent length scales,
$\sqrt{D/\omega}$ and $2\pi\sigma/\omega$. The former is the usual
diffusive length that shows how far a charge spreads during time $\omega^{-1}$.
The latter is associated with the electric field propagation in a
conducting medium. One logarithm in the double-log ZBA expression
comes from the spacial region $\sqrt{D/\omega} < R < 2\pi\sigma/\omega$.
However for small enough $\omega$,
the length $2\pi\sigma/\omega$ becomes
larger than the system size $L_x$ (it is assumed that $L_x \leq L_y$).
This finite-size effect can be accounted in Eq.~(\ref{KK-RG-half})
by substituting $\max(\omega/2\pi\sigma, 1/L_x)$ as a lower limit of
$q$-integration. Physically this procedure means that we neglect
external impedance of the circuit connected to our dirty
film in comparison with the effective ``spreading resistance" of the film
(given by $\pi r(\zeta)$, where function $r(\zeta)$ is determined
 below in Eq.~(\ref{r})).
It is then straightforward to generalize the RG equations
(\ref{gamma_rg1}) and (\ref{gamma_rg2}) for finite systems.
In the metallic energy region the transparency obeys
\be
\label{gamma_rg1'}
  \frac{\partial\gamma}{\partial\zeta}
  = - r(\zeta) \gamma ,
\ee
while in the sub-gap limit one has
\be
\label{gamma_rg2'}
  \frac{\partial\gamma}{\partial\zeta}
  = - \lambda \gamma - 2 r(\zeta) \gamma .
\ee
The function $r(\zeta)$ is defined as
\be
\label{r}
  r(\zeta) = \frac{1}{2\pi^2g} \times
  \cases{
    \zeta-\zeta_0, & if $\Omega > \omega_\sigma$;\cr
    \zeta_{Th}-\zeta, & if $\omega_\sigma > \Omega > E_{Th}$;\cr
    0, & if $E_{Th} > \Omega$,
  }
\ee
where $\omega_\sigma=2\pi\sigma/L_x$ and
$\zeta_{Th} = \ln 1/E_{Th}\tau = 2\ln (L_x/l)$.

The solutions of Eqs.~(\ref{gamma_rg1'}), (\ref{gamma_rg2'})
can also be represented in the form (\ref{gamma-res1}), (\ref{gamma-res2})
with the modified functions $\ZBA_M(\zeta)$ and $\ZBA_S(\zeta)$.
We will not present here the complete list of formulae for
arbitrary values of $\omega_\sigma/\Delta$ and $\omega_\sigma\tau$
but will focus instead on $\ZBA_S(\zeta)$ in the experimentally relevant case.
If $L_x$ is measured in $\mu$m then the energy $\omega_\sigma$
(in Kelvins) is given by $100g/L_x$, that appears to be greater
than $\Delta$ for a reasonable experimental setup.
Hence, in studying the sub-gap frequency region,
the lower $q$-cutoff is given by the inverse system size, $L^{-1}$,
and any information about the Coulomb potential $V_0(q)$
drops from the resulting expression.
Solving then Eq.~(\ref{gamma_rg2'}) one obtains
\be
\label{ZS'}
  \ZBA_S(\zeta)
  = \exp \left(
      - \frac{(\zeta-\zeta_\Delta) [\zeta_{Th}-(\zeta+\zeta_\Delta)/2]}{2\pi^2g}
    \right)
  = \exp \left( - \frac{1}{2\pi^2g}
      \ln\frac{\Delta}{\Omega} \cdot
      \ln\frac{\Delta\Omega}{E_{Th}^2}
\right) .
\ee
Note that $\Omega = \max(E_{Th},\omega,eV,T,\tau_\phi^{-1})$
and cannot be smaller then $E_{Th}$.

Using Eq.~(\ref{ZS'}) and substituting $\gamma(\zeta)$
from Eq.~(\ref{gamma-res2}) into Eq.~(\ref{ga2}), we obtain
the resulting expression:
\be
\label{Ga-all}
  G_A =
    \frac{G_{T\Delta}^2}{g} \, \frac{L_{\rm eff}(L_x,\omega,T)}{L_y} \,
    \frac{ 4 (\Omega/\Delta)^{2\lambda_g} }
      { \left[ (1 + \lambda_n/\lambda_g)
      + (1 - \lambda_n/\lambda_g) (\Omega/\Delta)^{2\lambda_g} \right]^2 }
\cdot
  \exp \left( - \frac{1}{\pi^2g}
    \ln\frac{\Delta}{\Omega} \cdot
    \ln\frac{\Delta\Omega}{E_{Th}^2}
  \right) ,
\ee
where $G_{T\Delta} \equiv G_T \ZBA_M(\zeta_\Delta)$ stands for
the single-particle tunneling conductance at the scale $\Delta$
renormalized by the normal-metal ZBA.
Eq.~(\ref{Ga-all}) is one of the main results of this paper,
it shows that the original growth of
$G_A(\Omega) \propto \Omega^{-x_A}$ with the $\Omega$
decrease (the exponent $x_A$ is defined in (\ref{x_A}))
stops at $\ln(\Delta/\Omega) \sim \sqrt{g}$, and
at lower $\Omega$ the Andreev conductance decreases
due to the zero-bias anomaly.
In the intermediate frequency range
$\ln(\Delta/\Omega) \sim \sqrt{g}$
both the Finkelstein's effect and the ZBA effect are of the same importance,
whereas in the infrared limit the ZBA effect is the most important one.
The influence of the last effect upon the Andreev conductance was predicted by
Huck, Hekking and Kramer~\cite{G_A_Hekking} on phenomenological grounds.
They considered  N-S junction coupled to the model dissipative environment
characterized by some impedance ${\cal Z}(\omega)$.
We provide here a microscopic calculation of this effective impedance:
 ${\cal Z}(\omega) = \pi r(\ln\frac{1}{\Omega\tau})$, with the
function $r(\zeta)$ defined in Eqs.~(\ref{r}) and (\ref{r2}) below.

Until now we considered the case of
unscreened bare Coulomb potential $V_0(q) = \frac{2\pi e^2}{q}$.
In the presence of an additional (clean) metal gate, the Coulomb potential
in the dirty metal layer changes according to Eq.~(\ref{Vscreen}).
As a result, $V^{\rm scr}_0(0) = 4\pi e^2 b$,
that modifies the law of propagation of the electric field,
which now becomes diffusive with the effective diffusion coefficient
$D_* = 8\pi\nu e^2 b \cdot D$. The ratio $D_*/D = 2\kappa_2b \gg 1$,
where $\kappa_2 = 4\pi\nu e^2$ is the inverse 2D screening radius.
In a finite-size system, Eq.~(\ref{r}) should be modified as
\be
\label{r-scr}
  r^{\rm scr}(\zeta) = \frac{1}{2\pi^2g} \times
  \cases{
    \ln\frac{D_*}{D}, & if $\frac{D}{D_*} \Omega > E_{Th}$;\cr
    \zeta_{Th}-\zeta, & if $\Omega > E_{Th} > \frac{D}{D_*} \Omega$;\cr
    0, & if $E_{Th} > \Omega$.
  }
\ee
In an effectively infinite system (for $\Omega > \frac{D}{D_*} E_{Th}$,
when the propagating field does not have enough time to reach the edge
of the system), one of the two logarithms entering
Eq.~(\ref{KK-RG-plane}) becomes $\omega$-independent, cf.~\cite{LS}.
Then $G_A$ is given by Eq.~(\ref{Ga-all}), provided that the
last exponential factor is replaced by the power-law factor
\be
  [\ZBA_S^{\rm scr}]^2 = \left( \frac\Omega\Delta \right)^{x_z},
\label{xzba}
\ee
where
\be
  x_z = \frac{2}{\pi^2g} \ln(8\pi \nu e^2 b).
\label{xz}
\ee

\subsubsection{SC island} \label{SSS:Zisl}

Now let us consider ZBA effects in the small island geometry
shown in Fig.~\ref{F:NSisland}.
Let us first study the renormalization of the junction
transparency $\gamma$. Depending on the relation between
momentum and the inverse island size $d^{-1}$, the geometry
of the interface can be either flat, for $qd \gg 1$, or
point-like, for $qd \ll 1$. Then the renormalization of $\gamma$
can be derived from Eqs.~(\ref{dg2}), (\ref{dg3}), where
the correlator $\corr{[K'(0)]^2}$ is given by Eq.~(\ref{KK-RG-half}),
with $\cos^2(q_xr)$ being formally replaced by $[1+\theta(qd-1)]/2$.
The resulting RG equations can be written in the form
(\ref{gamma_rg1'}), (\ref{gamma_rg2'})
with the function $r(\zeta)$ defined as
\be
\label{r2}
  r(\zeta) = \frac{1}{4\pi^2g} \times
  \cases{
    2(\zeta-\zeta_0), & if $\Omega > 2\pi\sigma/d$;\cr
    \zeta_{Th}-\zeta_0,
      & if $2\pi\sigma/d > \Omega > \max(\omega_\sigma, \omega_d)$;\cr
    \zeta-\zeta_0, & if $\omega_d > \Omega > \omega_\sigma$;\cr
    \zeta_{Th}-\zeta_d, & if $\omega_\sigma > \Omega > \omega_d$;\cr
    \zeta_{Th}-\zeta,
      & if $\min(\omega_\sigma, \omega_d) > \Omega > E_{Th}$;\cr
    0, & if $E_{Th} > \Omega$.
  }
\ee

The RG equations for $\gamma(\zeta)$ can be easily solved.
Again, we will not present here a general solution depending
on the relations between various energy scales, $E_{Th}(L)=D/L^2$,
$\omega_\sigma=2\pi\sigma/L$, $\omega_d$, $\Omega$ and $2\pi\sigma/d$.
Rather we concentrate on the solution in the superconducting region,
$\Omega<\Delta$, assuming that $\omega_d \sim \Delta < \omega_\sigma$.
It can be written in the form (\ref{gamma-res2}) with $\ZBA_S(\zeta)$
given by
\be
\label{ZS"}
  \ZBA_S(\zeta)
  = \exp \left(
      - \frac{(\zeta-\zeta_\Delta) [\zeta_{Th}-(\zeta+\zeta_\Delta)/2]}{4\pi^2g}
    \right)
  = \exp \left( - \frac{1}{4\pi^2g}
      \ln\frac{\Delta}{\Omega} \cdot
      \ln\frac{\Delta\Omega}{E_{Th}^2(L)}
    \right) .
\ee
The difference by 2 compared to Eq.~(\ref{ZS'}) accounts for
the difference between spreading over the whole plane and the half-plane.

According to section~\ref{SS:NSisland}, in the island geometry,
$G_A$ entering the action as a parameter in Eq.~(\ref{SA'}) gets
renormalized after eliminating fast degrees of freedom.
In the absence of the ZBA, its renormalization comes from the
averaging of two vertices $S_\gamma$, cf.\ Eq.~(\ref{gar1}).
With the ZBA effect taken into account, the Andreev action $S_A$
will renormalize itself similar to the term $S_\gamma$
in section~\ref{SSS:Zbar}. The corresponding expression reads
\be
  \corr{S_A} = \frac{i\pi}{16} G_A \Tr
    \corr{
      Q_S(t_1) e^{ 2iK'_1(t_1)\tau_z} \tilde Q(t_1,t_2)
      Q_S(t_2) e^{ 2iK'_1(t_2)\tau_z} \tilde Q(t_2,t_1)
    } ,
\label{Sg2}
\ee
that is written for $\Omega<\Delta$ when
$Q_S$ given by Eq.~(\ref{QS}) is local in time.
On averaging over fast $K'_1$, one finds for the correction to $G_A$:
\be
  \Delta G_A = G_A
    \left[ e^{-\frac12\corr{[2K'_1(t_1)-2K'_1(t_2)]^2}} - 1 \right] .
\ee
Expanding the exponent and omitting the cross averages, one arrives at
\be
  \Delta G_A = - 4G_A \, \corr{[K'_1(0)]^2} .
\ee
As a result, we obtain the following modification of the RG equation
(\ref{GAdot}):
\be
  \frac{\partial G_A}{\partial \zeta}
    = \frac{{\cal A}^2\gamma^2}{4\pi g} - 4 r(\zeta) G_A .
\label{GA-all}
\ee
Taking $\gamma(\zeta)$ from Eq.~(\ref{gamma-res2})
and integrating this differential equation,
one obtains the solution for the Andreev conductance:
\be
  G_A(\Omega) =
  \frac{G_{T\Delta}^2}{4\pi g} \,\frac{1-(\Omega/\Delta)^{2\lambda_g}}
{ (\lambda_g + \lambda_n)
+ (\lambda_g - \lambda_n)
  (\Omega/\Delta)^{2\lambda_g}}
\cdot
  \exp \left(
    - \frac{1}{2\pi^2g}
    \ln\frac{\Delta}{\Omega} \cdot
    \ln\frac{\Delta\Omega}{E_{Th}^2}
  \right)
\label{GAdot-res2}
\ee
where $G_{T\Delta}$ is the normal-state tunneling conductance
at the scale $\Delta$ renormalized by the ZBA.
The coefficient in front of the double logarithm in the ZBA exponent
is four times larger than the one for the single electron tunneling,
due to the doubled charge of a Cooper pair.
Expression (\ref{GAdot-res2}) again shows that the ZBA effect
upon $G_A$ is described by the separate factor $\ZBA_S^2(\zeta)$
as it should be due to the factorization property discussed above.
With the frequency decrease, $G_A(\Omega)$ first grows logarithmically
and then decreases as in the case of the rectangular
N-S contact.
In the presence of a screening metal gate, the ZBA factor
$[\ZBA_S^{\rm scr}]^2 = (\Omega/\Delta)^{x_z/2}$
(cf.\ Eq.~(\ref{xzba})).

\section{Josephson proximity coupling}
\label{S:Josephson}

\subsection{General treatment and the RG equation}

In this section we first rederive semiclassical expressions for the proximity
coupling between superconductors, separated by dirty normal metal,
and then generalize them, taking into account quantum fluctuations,
similar to the way it was done above for the Andreev conductance.
The term in the effective action, which is responsible for this coupling,
can be written in the form
\be
  S_J = \frac{1}{2}E_J \Tr ( \check{Q}_S^{(1)}\check{Q}_S^{(2)} \sigma_x ) ,
\label{SJ}
\ee
where superscripts $^{(1)}$ and $^{(2)}$ refer
to two superconductive banks or islands.
Using the low-energy representation (\ref{QS}) for $Q_S$,
and neglecting the Coulomb phase factors $\exp(i\tensor{K}(t)\tau_z)$,
one rewrites Eq.~(\ref{SJ}) as
\be
  S_J = - 2 E_J \int \sin\Theta_1(t) \sin\Theta_2(t) dt ,
\label{SJ'}
\ee
with $\Theta_1$ and $\Theta_2$ being the classical and quantum
components of the phase difference between the superconductors,
$
  \roarrow\Theta = \roarrow{\theta}^{(1)} - \roarrow{\theta}^{(2)}
$.
The meaning of the term (\ref{SJ'}) becomes transparent
in the initial basis before the Keldysh rotation (\ref{vecrot}):
it yields the usual expression for the Josephson coupling energy,
$-E_J \cos\Theta$.
Employing Eq.~(\ref{I}) one obtains the standard relation
between the Josephson current and (the classical component of)
the phase difference (with dimensional units restored):
\be
  I_J = \frac{2e E_J}{\hbar} \sin\Theta_1 .
\ee

The term (\ref{SJ}) can be derived in the way very similar to the derivation
of the expression (\ref{SA1}) in section~\ref{SS:Andreev}:
one should consider the cross term in the perturbative correction
to the action,
$\frac{i}{2}\langle (S_\gamma^{(1)} + S_\gamma^{(2)})^2 \rangle$,
where $S_\gamma^{(j)}$ is the boundary action (\ref{S_gamma})
at the interface with the superconductor $(j)$,
and the average is taken over fluctuations of the normal-metal matrix $Q$.

\begin{figure}
\vspace{8mm}
\epsfxsize=102pt
\centerline{\epsfbox{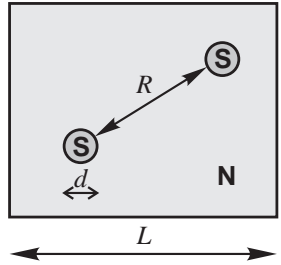}}
\vspace{-5mm}
\caption{%
Two superconductive islands (S) of size $d$ with separation $R \gg d$
coupled via a dirty normal film (N) of size $L$.
}
\label{F:Josephson}
\end{figure}

\begin{figure}
\vspace{4mm}
\epsfxsize=130pt
\centerline{\epsfbox{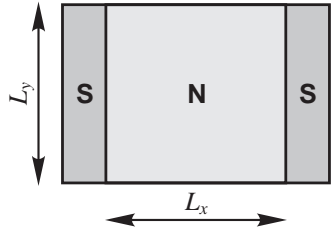}}
\vspace{-5mm}
\caption{%
Two superconductive terminals (S) coupled via a dirty normal film (N)
of sizes $L_x$ and $L_y$.
}
\label{F:Josephson2}
\end{figure}

The Josephson coupling energy $E_J$ depends on the geometry of the system,
and below we will find $E_J$ for two geometries shown in
Figs.~\ref{F:Josephson} and \ref{F:Josephson2},
that are natural counterparts of Figs.~\ref{F:NSisland} and \ref{F:NSbar}
discussed above with respect to the Andreev conductance.
In both cases $E_J$ is given by the Fourier transform
(cf.\ Eqs.~(\ref{J1}), (\ref{J4})) of the zero-frequency
Cooperon amplitude.
The latter, $\C(q,\omega) \equiv \J(\zeta)$, logarithmically
depends on $\Omega=\max(Dq^2,\omega,T)$.
Within the semiclassical approximation~\cite{ALO} it is given by
an integral over the ``center-of-mass'' Cooperon energy:
\be
  \J(\zeta)
  = \frac{G_T^2}{4\nu} \int_0^\Delta \frac{[F(E_+)+F(E_-)] E dE}{(Dq^2)^2+4E^2}
  = \frac{G_T^2}{8\nu}\ln\frac{\Delta}{\Omega} .
\label{Jclass}
\ee
To take into account the effect of interaction in the Cooper channel
that makes $G_T = {\cal A} \gamma$ scale-dependent, we replace
Eq.~(\ref{Jclass}) by the following differential RG equation:
\be
\label{Jrg1}
  \frac{\partial\J(\zeta)}{\partial\zeta} =
  {\cal A}^2 \frac{\gamma^2(\zeta)}{8\nu} .
\ee
This RG equation coincides (up to a numerical factor) with the RG equation
(\ref{GAdot}) for $G_A$ in the small island geometry.
In the study of the Andreev conductance, such an RG equation appeared
as a result of the $q$-integration and thereby hold only for the geometry
of Fig.~\ref{F:NSisland}. In the case of the Josephson coupling,
Eq.~(\ref{Jrg1}) comes from the $E$-integration and thus is
insensitive to a particular geometry holding for both setups shown
in Figs.~\ref{F:Josephson}, \ref{F:Josephson2}.

Although the RG equations (\ref{GAdot}) and (\ref{Jrg1}) look similar,
the answers for $G_A$ and $E_J$ are quite different. The main difference
is that $\J$ is not a final quantity, it should be Fourier transformed
to get the Josephson energy $E_J$. As a result, only the lowest spacial modes
effectively contribute to $E_J$, which appears to be exponentially
suppressed\cite{ALO} due to loss of coherence if temperature or frequency
of phase fluctuations exceed the Thouless energy.
Note, that in the case shown in Fig.~\ref{F:Josephson},
the relevant Thouless energy scale is determined by the inter-island distance
$R$; we denote this energy as $\omega_R = D/R^2$ to discern it from
$E_{Th}(L) = D/L^2$ which will also enter results for the ZBA factor
for the two-island setup.
Below we will assume that both $T$ and $\omega$ are much lower than
$E_{Th}$ for the  rectangular geometry, and much lower than $\omega_R$
for the island geometry.

Eq.~(\ref{Jrg1}) is sufficient to find the proximity coupling
energy in the absence of the zero-bias anomaly effects (which will
be incorporated in the next subsection).
We start from an example of two small SC islands of size $d$,
separated by the distance $R \gg d$ (see Fig.~\ref{F:Josephson}).
In this case
\be
  E_J(R)
  = \int \frac{d\bq}{(2\pi)^2} e^{i \bf qR} \, \C(q)
  = \frac{1}{2\pi} \int_0^\infty q\, dq J_0(qR) \J(\zeta_q) ,
\label{J1}
\ee
where $\zeta_q = \ln(1/Dq^2\tau)$,
and the use of the continuous Fourier transform implies that the
distance between islands is much shorter that the total size of the
film, $L \gg R$.
Using the identity
$J_0(x) = \frac{1}{x} \frac{\partial}{\partial x} [ x J_1(x) ]$,
and integrating by parts we obtain
\be
  E_J(R)
  = \frac{1}{\pi R} \int_0^\infty dq J_1(qR)
    \frac{\partial \J(\zeta_q)}{\partial\zeta_q} .
\label{J2}
\ee
Since $\J(\zeta_q)$ is a very slow function of $q$ (provided that $g\gg1$),
the $q$-integration converges rapidly near $q\sim R^{-1}$.
Then we may take $\J$ at $\zeta_q = \ln\frac{R^2}{D\tau}$
and perform the remaining integration that yields
\be
  E_J(R)
  = \frac{1}{\pi R^2}
      \left. \frac{\partial\J(\zeta)}{\partial\zeta}
      \right|_{\zeta = \ln\frac{R^2}{D\tau}} .
\label{J3}
\ee
Combining Eqs.~(\ref{J3}) and (\ref{Jrg1}) with Eq.~(\ref{gamma-res}),
we obtain
\be
  E_J(R)
  = \frac{G_T^2}{8\pi\nu R^2} \,
  \frac{4 (\xi_\Delta/R)^{4\lambda_g}}
      { \left[ (1 + \lambda_n/\lambda_g)
      + (1 - \lambda_n/\lambda_g) (\xi_\Delta/R)^{4\lambda_g} \right]^2} .
\label{EJ1}
\ee
Thus, the effect of repulsive interaction in the Cooper channel is
to produce an anomalous power-law exponent $x_J = 2 + 4\lambda_g$,
describing decay of the Josephson coupling at long distances:
$E_J \sim R^{-x_J}$.

Now we proceed to the case of the rectangular geometry shown
in Fig.~\ref{F:Josephson2}. In this case the set of eigenfunctions
with the proper boundary conditions reads (cf.~(\ref{eigenpsi}))
\be
\label{eigenpsi'}
  \psi_{mn}(x,y) =
  \frac{2 \cos q_mx \cos k_ny}{\sqrt{(1+\delta_{m,0})(1+\delta_{n,0})L_xL_y}},
  \qquad q_m=\frac{\pi}{L_x} m,
  \quad  k_n=\frac{\pi}{L_y} n,
  \quad m,n=0,1,\dots
\ee
The difference with the case of two islands is that now one has
to sum over the single component $q_m$ of the wavevector only
(cf.~section~\ref{SS:NSbar}):
\be
  E_J(L_x,L_y) = \frac{1}{L_xL_y} \sum_{m=-\infty}^\infty (-1)^m \, \C(q_m) .
\label{J4}
\ee
Eq.~(\ref{Jrg1}) for $\J = \mbox{const} \cdot G_A$ has been already solved
in section~\ref{SS:NSisland};
substituting the solution into Eq.~(\ref{J4}) and using the equality
$\sum_{m=-\infty}^\infty (-1)^m = 0$ we transform $E_J$ to the form
\be
  E_J(L_x,L_y)
  = \frac{G_T^2}{4\nu} \,
    \frac{1}{L_xL_y} \,
    \frac{\lambda_g}{\lambda_g-\lambda_n} \,
    \sum_{m=-\infty}^\infty
    \frac{(-1)^m}
      { (\lambda_g + \lambda_n)
      + (\lambda_g - \lambda_n) (E_{Th}m^2/\Delta)^{2\lambda_g} } ,
\ee
where $E_{Th} = D/L_x^2$.
Consider this sum and write it as $\sum_{m=-\infty}^\infty (-1)^m f_m$.
For $\lambda_g\ll1$, $f_m$ is a very slow function for $m\neq0$
but has a cusp at $m=0$. Then the sum can be well estimated as
$f_0-f_1$ and is equal to
\be
  \frac{ (\lambda_g - \lambda_n) (E_{Th}/\Delta)^{2\lambda_g} }
    { ( \lambda_g + \lambda_n ) [ (\lambda_g + \lambda_n)
    + (\lambda_g - \lambda_n) (E_{Th}/\Delta)^{2\lambda_g} ]} .
\ee
As a result, we obtain
\be
  E_J(L_x,L_y)
  = \frac{G_T^2}{8\nu (\lambda_g+\lambda_n) L_xL_y} \,
    \frac{2 (\xi_\Delta/L_x)^{4\lambda_g}}
      { (1 + \lambda_n/\lambda_g)
      + (1 - \lambda_n/\lambda_g) (\xi_\Delta/L_x)^{4\lambda_g} } .
\label{EJ2}
\ee
In the above expression $G_T$ is the total normal-state tunneling
conductance proportional to the width $L_y$ of the contact.
In the rectangular geometry a non-trivial exponent
$x_J' = 1 + 4\lambda_g$  enters the $E_J(L_x)$ dependence:
$E_J \propto L_x^{-x_J'}$.
In the case of weak disorder, $\lambda_g \ln(L_x/\xi_{\Delta}) \ll 1$,
the result (\ref{EJ2}) reduces to the one obtained in~\cite{ALO}, whereas
Eq.~(\ref{EJ1}) reduces to the form used in~\cite{FeigLar}
(an extra factor $4/\pi^2$ in the Eq.~(18) of~\cite{FeigLar} is due to
the difference in the definition
of the dimensionless tunneling conductance $G_T$).
Both of the above results (\ref{EJ1}), (\ref{EJ2}) were obtained
neglecting the zero-bias anomaly effects.
Now we proceed to take them into account.

\subsection{Proximity coupling in the presence of the zero-bias anomaly}

As it was explained in Sec.~\ref{SSS:ZBA0}, the effect of the Coulomb
phase $\roarrow{K}(t)$ fluctuations (which are due to the long-range
Coulomb potential) can be separated (factorized) from the effect
of the interaction in the Cooper channel.
Below we will use this factorization explicitly.

Consider first the two island geometry of Fig.~\ref{F:Josephson}.
Using the gauge-transformed form (\ref{S_gamma2}) of the boundary
action term, and repeating the steps that lead to Eqs.~(\ref{Jrg1}) and
(\ref{J1}), we obtain now the Josephson coupling energy $E_J(R)$
as a time-domain convolution of the Cooperon and the ZBA exponent:
\be
  E_J(R) = \int dt \ZBA_J^2(t,R)
  \int \frac{d\bq}{(2\pi)^2} e^{i \bf qR} \,\C(q,t) ,
\label{JZC}
\ee
where $C(q,t) = \int\frac{d\omega}{2\pi} \J(\zeta) e^{-i\omega t}$
is the time-domain Cooperon amplitude, and $\zeta=\ln1/\Omega\tau$
with $\Omega = \max(Dq^2, \omega)$.
The ZBA factor $\ZBA_J(t,R)$ is defined as (cf.~Eq.~(\ref{KK-RG-plane}))
\be
  - \ln \ZBA_J(t,R) = \langle [K'_1(t,R)-K'_1(0,0)]^2 \rangle =
  \frac{1}{\pi^2g}
    \int_0^{\Delta} \frac{d\omega}{\omega}
    \int_{\max(\omega/2\pi\sigma,1/L)}^{\sqrt{\omega/D}} \frac{dq}{q}
      \left( 1 - \cos\omega t J_0(qR) \right) ,
\label{JZ1}
\ee
where $L \gg R$ is the size of the metal film.
Here we assume that $\omega_d \sim \Delta$, and include all ZBA
fluctuations with frequency $\omega\gg\Delta$ into redefinition
of the normal-state tunneling conductance $G_{T\Delta}$ at the scale $\Delta$.
In the absence of the ZBA, $\ZBA_J=1$ and Eq.~(\ref{JZC}) reduces
to Eq.~(\ref{J1}), with $\C(q) \equiv \C(q,\omega=0)$.
The lower limit of the $q$-integration depends on the relation
between the field spreading distance, $2\pi\sigma/\omega$,
and the sheet size, $1/L$.
To be more specific, we will assume, following section~\ref{SS:ZBA},
that $\Delta/2\pi\sigma \ll L$.
Then the ZBA exponent is equal to
\be
  \ZBA_J(t,R)
  = \exp \left( - \frac{1}{4\pi^2g}
      \ln\frac{\Delta}{\omega_R} \cdot
      \ln\frac{\Delta\,\omega_R}{E_{Th}^2(L)}
    \right) ,
\label{JZ2}
\ee
where $\omega_R=D/R^2$, $E_{Th}(L)=D/L^2$, and it is assumed that
$t \sim \omega_R^{-1}$.
Now substituting $\J(\zeta)$ from Eq.~(\ref{Jrg1}) into Eq.~(\ref{JZC}),
using the trick that led to Eq.~(\ref{J2}) and the identity
$\partial\J(\zeta)/\partial\zeta_q
= \theta(Dq^2-|\omega|) \partial\J(\zeta_q)/\partial\zeta_q$,
and performing a trivial integration over $\omega$, we obtain
\be
  E_J(R)
  = \frac{1}{\pi^2 R}
    \int dt \ZBA_J^2(t,R)
    \int_0^\infty dq J_1(qR) \frac{\partial\J(\zeta_q)}{\partial\zeta_q} \,
    \frac{\sin Dq^2t}{t} .
\ee
Both momentum and time integrals converge fairly well in the vicinity
of $q\sim R^{-1}$ and $t \sim \omega_R^{-1}$ respectively.
As a result, to logarithmic accuracy we get
\be
  E_J(R)
  = \frac{1}{\pi R^2}
      \left. \ZBA_J^2(\zeta) \frac{\partial\J(\zeta)}{\partial\zeta}
      \right|_{\zeta = \ln\frac{R^2}{D\tau}} .
\ee
Note that the same result could be obtained within the RG approach
with the help of the equation similar to Eq.~(\ref{GA-all}):
\be
  \frac{\partial \J}{\partial \zeta}
    = \frac{{\cal A}^2\gamma^2}{8\nu} - 4 r(\zeta) \J .
\label{J-all}
\ee
with $r(\zeta) = -(1/2)\, \partial\ln \ZBA_J/ \partial\zeta$.

Finally, we obtain the Josephson coupling energy for two islands:
\be
  E_J(R)
  = \frac{G_{T\Delta}^2}{8\pi\nu R^2} \,
  \frac{4 (\xi_\Delta/R)^{4\lambda_g}}
      { \left[ (1 + \lambda_n/\lambda_g)
      + (1 - \lambda_n/\lambda_g) (\xi_\Delta/R)^{4\lambda_g} \right]^2}
  \cdot
  \exp \left( - \frac{2}{\pi^2g}
    \ln\frac{R}{\xi_\Delta} \cdot
    \ln\frac{L^2}{\xi_\Delta R}
  \right) .
\label{Jf1}
\ee
The presence of the total size $L$ of the film in the above expression
for the inter-island proximity coupling may appear to be unexpected.
It originates from the fact that electric field propagates much faster
than electron density: during  diffusion time $R^2/D$ corresponding to the
inter-island distance, the electric field propagates the distance
$L(R) = \kappa_2 R^2$, where $\kappa_2 = 4\pi\nu e^2$ is the inverse
2D screening length. The effective electric impedance that determines the
ZBA factor does depend on the  distance $L(R)$ as long as it is shorter
than the total size $L$. If the dirty film is of metallic origin,
$\kappa_2^{-1}$ is of the order of Angstr\"oms, so $L(R)$ easily exceeds
any reasonable size of the film even for $R$ in submicron range.
In this case the effective impedance is determined by the film size $L$.
All these considerations do not hold for the case of screened Coulomb
potential (\ref{Vscreen}); in that case the exponential factor
in Eq.~(\ref{Jf1}) should be replaced by $(\xi_\Delta/R)^{x_z}$,
cf.\ the end of Section VI.

Consider now the ZBA effect for the rectangular geometry
(see Fig.~\ref{F:Josephson2}). In this case one has
\be
  E_J(L_x,L_y) = \frac{1}{L_xL_y} \int dt \ZBA_J^2(t,L_x)
  \sum_{m=-\infty}^\infty (-1)^m \, \C(q_m,t) ,
\label{JZC2}
\ee
cf.~Eqs.~(\ref{J4}) and (\ref{JZC}).
Again, the relevant scale for the $t$-integration is given
by $E_{Th}^{-1} = R^2/D$.
Then one can integrate over $t$ keeping $\ZBA_J(t)=\ZBA_J(E_{Th}^{-1})$ constant
that amounts to multiplying the previous result (\ref{EJ2})
by $\ZBA_J^2(E_{Th}^{-1})$, provided that $G_T$ is substituted by $G_{T\Delta}$.
The ZBA factor for a flat N-S interface is given by
\be
  \ZBA_J(E_{Th}^{-1},L_x)
  = \exp \left( - \frac{1}{2\pi^2g}
      \ln^2\frac{\Delta}{E_{Th}}
    \right) ,
\label{JZ3}
\ee
where $E_{Th}=D/L_x^2$.
Therefore we obtain
\be
  E_J(L_x,L_y)
  = \frac{G_{T\Delta}^2}{8\nu (\lambda_g+\lambda_n) L_xL_y} \,
    \frac{2 (\xi_\Delta/L_x)^{4\lambda_g}}
      { (1 + \lambda_n/\lambda_g)
      + (1 - \lambda_n/\lambda_g) (\xi_\Delta/L_x)^{4\lambda_g} }
  \cdot
  \exp \left( - \frac{4}{\pi^2g} \ln^2\frac{L_x}{\xi_\Delta} \right) .
\label{Jf2}
\ee
For the screened Coulomb potential (parallel metal gate), the last
factor in (\ref{Jf2}) is replaced by $(\xi_{\Delta}/L_x)^{2x_z}$.
Note, finally, that both expressions (\ref{Jf1}) and (\ref{Jf2}) refer
to the zero-temperature limit $T \ll \hbar D/R^2$, $\hbar D/L_x^2$.


\section{Discussion}
\label{S:Discussion}

We have developed field theory functional formalism of the Keldysh type
for disordered superconductors.
This approach provides a regular method for the treatment of all kinds of
quantum fluctuations beyond the standard
semiclassical theory of superconductivity~\cite{LO1,rammer,lambert}
which can be understood as the saddle-point approximation of our theory.
The theory is formulated in terms of the local (in space) matrix order
parameter whose components corresponds to the retarded/advanced and Keldysh
Green functions.
The main advantage of our approach with respect to the standard
Matsubara replica formalism~\cite{finkel2} is that
it provides a possibility to treat nonequilibrium problems.

General formulation of the theory involves $8\times 8$
matrices $Q_{t,t'}$, composed as the direct product
of the $2\times 2$ blocks corresponding to the Nambu, time-reversal and
Keldysh spaces. However, in order to present our approach in the most
transparent form, we restricted here our
specific calculations to the case  of spin-independent interactions,
and neglected usual weak-localization conductivity
corrections, that makes it possible to use reduced $4\times 4$ matrices.
We focused here on the fluctuational effects specific for the
dirty superconductive films, those relative magnitude is
known~\cite{finkel1,finkel2} to be of the order of
$(1/g)\ln^2(L/l)$ (in the two-dimensional limit),
and thus can be considered independently of the normal-metal
weak-localization and interaction corrections~\cite{GLK,AA}
which are $\propto (1/g)\ln(L/l)$.
There are two physically different types of these effects:
i) Coulomb-induced
repulsive contribution to the electron-electron interaction in the
Cooper channel (coming from the ``diffusive" frequency
region $\omega \ll Dq^2$), and ii) reduction of the
averaged single-particle density of states
due to high-frequency ($\omega \gg Dq^2$)
 fluctuations of electric potential.
We have shown, following the approach developed in~\cite{Kam_Andr},
that these two effects can be separated at the non-perturbative level,
by ``gauging away"  phase factors induced by the fluctuating long-range
 electric potential.
The first effect was treated by the
 renormalization group method within the single-loop approximation.
Being applied to the case of uniformly disordered superconductive films,
this method essentially reproduces the well-known results by
Finkelstein~\cite{finkel1,finkel2,finkel3} for the Coulomb-induced
 suppression of the superconductive transition temperature $T_c$.
The second effect is formally similar to the ``infrared
catastrophe" of quantum electrodynamics; in our technique it is
 taken into account in an essentially exact way, similar to
the normal-metal case treated in~\cite{Kam_Andr}.
This second effect (analogous to the zero-bias anomaly of the
tunneling conductance), contrary to the first one, does depend on the
long-range part of the bare Coulomb potential, and, thus, can be
modified by the presence of an additional external screening
(provided, e.~g., by a metallic gate nearby the dirty film).

It was argued previously (cf.~\cite{finkel3}) that the fluctuations
responsible for the zero-bias anomaly effect are specific
for the single-particle density of states, due to
gauge-noninvariance of the single-particle Green function, whereas
the same fluctuations do not contribute to other (gauge-invariant) physical
quantities like $T_c$. We have calculated two
important characteristics of mesoscopic S/N structures,
the Andreev subgap conductance and the Josephson proximity coupling,
and found that they are affected by the zero-biased anomaly effect
as well as by the Finkelstein's effect.
To be more specific, we have calculated linear Andreev
conductance $G_A(\Omega)$ in the presence
of the tunneling barrier (N-I-S structure), for the two different
geometries: i) small SC island ``sitting" on a large-area dirty metal
(Fig.~\ref{F:NSisland}), and ii) rectangular normal film between
superconductive and normal reservoirs (Fig.~\ref{F:NSbar}).
 Our results
for those two cases are given by Eqs.~(\ref{GAdot-res2}) and (\ref{Ga-all})
correspondingly.
In these formulae $\Omega$ stands for the infrared cutoff of renormalization
procedure, i.~e.\ $\Omega = {\rm max}(T,\omega,\tau_{\phi},D/L^2)$.
A detailed generalization of these results for the steady-state
 nonlinear Andreev current
$I_A(V)$ can be obtained with the use of formulae from Appendix B;
qualitatively the behaviour of $G_A(V) = dI_A/dV$ can be found by
substituting $2eV$ for $\Omega$
in Eqs.~(\ref{GAdot-res2}) and (\ref{Ga-all}).

The second new physical quantity we have studied is
the Josephson proximity coupling $E_J$ as a function of the size of
the normal region between two SC contacts. It was calculated, in the
low-temperature limit $T \ll E_{Th}$, for the case
of two small islands in contact with dirty normal metal
(Fig.~\ref{F:Josephson}), as well as for the rectangular film between
two SC banks (Fig.~\ref{F:Josephson2}). The electron-electron
repulsion in the Cooper channel leads to the appearance of an anomalous
scaling for $E_J$. In particular, for two small SC islands
$E_J(R) \propto R^{-2-2/\pi\sqrt{g}}$, whereas for the rectangular
contact the Josephson energy decays with the film length as
$E_J(L_x) \propto L_x^{-1-2/\pi\sqrt{g}}$. In addition, the ``ZBA effect"
adds to the above behaviour a log-normal suppression factor
$\exp[-(4/\pi^2 g)\ln^2(L_x/\xi_{\Delta})]$ for the rectangular contact,
and a similar factor for the island geometry.
Full results for both geometries are
given, at the low-temperature limit $T\to 0$, by Eqs.~(\ref{Jf1}) and
(\ref{Jf2}). Screening of the long-range Coulomb interaction in the film
by an external gate makes the ZBA suppression weaker, it reduces then to
the power-law factor $(\xi_\Delta/L_x)^{2x_z}$ and
$(\xi_\Delta/R)^{x_z}$ for the rectangular and island geometries
correspondingly, where $x_z$ is defined in Eq.~(\ref{xz}).

In the present calculations
we did not took into account the existence of a finite decoherence
time $\tau_{\phi}$ in normal metals at $T > 0$.
Qualitatively, the effect
of nonzero $\tau_{\phi}^{-1}$ is to suppress the  central peak in $I(V)$
predicted within the semiclassical theory, i.~e.\ it acts in the way
similar to the effect of quantum corrections we have studied in this paper
(we emphasize that these quantum  effects have nothing to do with
``zero-temperature decoherence").
That is why the decoherence rate $\tau_{\phi}^{-1}$ extracted from the data
on the Andreev conductance  might be overestimated
unless quantum fluctuations are taken into account in the data analysis.
Another unsolved theoretical
problem is to go beyond the lowest-order expansion over the tunneling
conductance $G_T$, that becomes necessary if $G_T$ approaches or
exceeds the conductance of the metal region.

\acknowledgements

This research was started during ICTP Trieste Extended Workshop
in August 1998; we are grateful to the Workshop organizers for a kind
hospitality and excellent working conditions.
We gained a lot from many discussions with I.~L.~Aleiner,
A.~Kamenev, A.~S.~Ioselevich, G.~B.~Lesovik, Yu.~V.~Nazarov, Y.~Oreg,
A.~V.~Shytov, B.~Z.~Spivak and F.~Zhou.
We thank financial support from the Swiss NSF collaboration
grant \# 7SUP J048531, INTAS-RFBR grant \# 95-0302,
RFBR grant \# 98-02-19252, Program ``Statistical Physics" of the Russian
Ministry of Science (M.~V.~F. and M.~A.~S.), DGA grant \# 94-1189
(M.~V.~F.), and from the NSF grant DMR-9812340 (A.~I.~L.).


\appendix

\sbox{\sigmaa}{\bf Eq.~(\ref{sigmaA})}

\section{Alternative proof of Eq.~(\protect\ref{AAA})}

In this Appendix we present an alternative proof of Eq.~(\ref{sigmaA})
with the help of the fluctuation-dissipation theorem. Expanding the
action (\ref{SA3}) to the second order in the phase variables, we obtain
\be
\label{A:SA1'}
  S_A^{(2)}[\roarrow\theta] = \frac{i}{4} G_A
    \int \frac{d\omega}{2\pi} \,
      \roarrow{\theta}^T(-\omega)
        \omega \Pi^{-1}_\omega
      \roarrow{\theta}(\omega) .
\ee
Hence, in the Gaussian approximation, the correlator of phases is given by
\be
  \corr{\theta_i(\omega) \theta_j(-\omega)}
  = \frac{2(\Pi_\omega)^{ij}}{G_A\omega} .
\ee
Employing Eq.~(\ref{2eV}) relating the phase of the island with the applied
voltage, we get for the spectral voltage correlator:
\be
  \corr{V_\omega V_{-\omega}}
  = \frac{1}{e^2G_A} \omega \coth \frac{\omega}{2T} ,
\ee
that, according to the Nyquist formula, results in Eq.~(\ref{sigmaA}).

\section{Nonlinear Andreev conductance}

To derive nonlinear Andreev current $I_A(V)$ we start from
Eq.~(\ref{SA22}). Tracing over the Nambu space we reduce
the expression for the action to the form similar to Eq.~(\ref{SA3}),
with $G_A$ being replaced by $G_A(\omega)$.
Calculating the current following Eq.~(\ref{I}), we find
\bea
I_A(t) = \frac{e}{4}\int dt_1 \int \frac{d\omega}{2\pi}
       \omega G_A(\omega) e^{-i\omega(t_1-t)}
       \biggl\{\langle e^{i\theta_1(t_1) - i\theta_1(t)} \rangle -
       \langle e^{-i\theta_1(t_1) + i\theta_1(t)} \rangle -
\nonumber \\ {}
2 i\,\coth\frac{\omega}{2T}\left[
      \langle e^{i\theta_1(t_1)} \theta_2(t_1) e^{- i\theta_1(t)} \rangle +
\langle e^{-i\theta_1(t_1)} \theta_2(t_1) e^{ i\theta_1(t)} \rangle
                       \right] \biggr\} .
\label{a21}
\eea
Angular brackets in Eq.~(\ref{a21}) mean averaging over quantum dynamics
of the phase $\theta(t)$.  In the case when fluctuations of
$\theta$ can be neglected,  the second line in Eq.~(\ref{a21})
should be omitted, whereas in the first line we put
$e^{i\theta_1(t_1) -i\theta_1(t)} = e^{2i\,eV (t_1-t)}$.
As a result, integration over $t_1$ produces
$2\pi\delta(\omega-2eV)$, and one obtains
\be
  I_A = e^2\, V\, G_A(2eV) .
\label{a22}
\ee
The function $G(\omega)$ in Eq.~(\ref{a22}) should
be determined with the account of fluctuational corrections due to
the interaction in the normal metal, as discussed in Secs.~\ref{SS:NSbar},
\ref{SS:NSisland}.
The effect of the zero-bias anomaly also can be taken into account along
the same lines, cf.\ Sec.~\ref{SS:ZBA}.
However, the RG method employed there
does not determine the dependence of the ZBA factor $\ZBA_S(\Omega)$
(cf.\ Eq.~(\ref{ZS}) upon the ratio $2eV/T$.
To find this dependence, we use the factorization
property discussed in Sec.~\ref{SSS:ZBA0}
and note that the low-energy ZBA effect
can be accounted for by the replacement $\roarrow\theta(t) \to
\roarrow\theta_{\rm eff}(t) = \roarrow\theta(t) - 2\roarrow K(t)$,
cf.\ discussion after Eq.~(\ref{ZM}).
After such a replacement the correlation functions entering Eq.~(\ref{a23})
can be expressed via the correlation matrix
\be
  {\cal D}(\omega)
  = \left( \begin{array}{cc}
    {\cal D}^K(\omega) & {\cal D}^R(\omega) \\ {\cal D}^A(\omega) & 0
  \end{array} \right) ,
\ee
where, similar to Eqs.~(\ref{phi-corr}) and (\ref{K-corr}),
the matrix elements
${\cal D}_{ij}(\omega) = - 2i \corr{ \Psi_i(\omega) \Psi^+_j (-\omega) }$,
and $\roarrow\Psi(t) = e^{-2i\roarrow K(t)}$.
As a result, the whole
expression for the current can be written as
\be
  I_A(V) =  \frac{e}{4}\int \frac{d\omega}{2\pi}\omega G_A(\omega)
  \left[ i {\cal D}^K(2eV-\omega) +
  2\coth\frac{\omega}{2T} \Im {\cal D}^R(2eV-\omega) \right] .
\label{a23}
\ee
Note that $i {\cal D}^K(\omega)$ is a real positive function.
Below it will be convenient to rewrite Eq.~(\ref{a23}) via the
 forward/backward correlation functions ${\cal D}^{>(<)} (\omega)$
defined as
\be
  {\cal D}^K = {\cal D}^> + {\cal D}^< ,
    \quad
  {\cal D}^R - {\cal D}^A = {\cal D}^> - {\cal D}^< .
\label{a23a}
\ee
In the equilibrium state,  the detailed balance relation
${\cal D}^<(\omega) = e^{-\omega/T} {\cal D}^>(\omega)$ is valid
for bosonic correlation functions, in addition to the general relations
(\ref{a23a}).
In terms of the original (before the Keldysh rotation (\ref{vecrot}))
bosonic variables, ${\cal D}^> (t) = - i \langle \Psi (t)\Psi^+ (0)\rangle$.
The correlators labeled by the index ``$>$'' contain fields taken
at the forward branch of the Keldysh contour $\cal C$
(for the sake of brevity we will omit the corresponding index ``1'').

If correction to $G_A(\omega)$ from the interaction vertex $\lambda$ in the
Cooper channel can be neglected, it can written as
\be
  G_A(\omega) = \frac{G_T^2}{2\nu}\int \frac{d E}{\omega}
  \left[\tanh\frac{E_+}{2T}-\tanh\frac{E_-}{2T}\right] C(E) ,
\label{a24}
\ee
where $E_\pm = E\pm \omega/2$, and
\be
  C(E) = \frac{1}{L_xL_y} \Re\sum_{\bf q}\frac{1}{-2i E + Dq^2}
\label{a25}
\ee
is the Cooperon amplitude. The sum in Eq.~(\ref{a25}) goes over
Cooperon eigenmodes (cf.\ derivation in Sec.~\ref{SSS:NSbar}).

Combining Eqs.~(\ref{a23})--(\ref{a25}), we obtain finally
\be
  I_A(V) =
  i\frac{e G_T^2}{4\nu}\int_{-\infty}^{\infty} C(E) dE
  \int_{-\infty}^{\infty}\frac{dE'}{2\pi}
  \left[
    \tanh\frac{E-E'/2+eV}{2T} - \tanh\frac{E+E'/2-eV}{2T}
  \right]
  {\cal D}^>(E') \frac{1-e^{-2eV/T}}{1-e^{(E'-2eV)/T}} .
\label{a26}
\ee
The result (\ref{a26}) is equivalent to the one obtained
in~\cite{G_A_Hekking} by the method of analytic continuation.
In the $T\to 0$ limit expression (\ref{a26}) simplifies and can be
transformed to the following form for the differential conductance:
\be
  \frac{dI_A}{dV} = \frac{e^2G_T^2}{\nu}\int_{-\infty}^{2eV}
  dE\, {\cal P}(E)\, C(eV-E/2) ,
\label{a27}
\ee
where ${\cal P}(E) = (i/2\pi) {\cal D}^>(E)$.

Note finally, that the results (\ref{a26}), (\ref{a27}) are valid in
 the situation when the order parameter phase $\theta(t)$
is subject to quantum fluctuations as well;
in this case the correlation function ${\cal D}^>$ is defined as
${\cal D}^> (t) = -i\langle e^{i\tilde\theta(t)} \Psi (t)
e^{-i\tilde\theta(0)}\Psi^+(0) \rangle$, where
$\tilde\theta (t) = \theta(t) - 2eVt$.


\end{document}